\documentclass[a4paper,11pt]{article}
\pdfoutput=1 

\usepackage{jcappub} 

\usepackage[T1]{fontenc} 
\usepackage{multirow}

\begin{document}

\title{Possibility of stable thin-shell around wormholes within the string cloud and quintessential field via the van der Waals and polytropic EOS in General relativity}                      

\author[a]{G. Mustafa,}
\author[b]{Faisal Javed,}
\author[c]{S.K. Maurya,}
\author[d]{Saibal Ray}

\affiliation[a]{Department of Physics, Zhejiang Normal University, Jinhua 321004, People's Republic of China}
\affiliation[b]{Department of Physics, Zhejiang Normal University, Jinhua 321004, People's Republic of China}
\affiliation[c]{Department of Mathematical and Physical Sciences, College of Arts and Sciences, University of Nizwa, Nizwa, Sultanate of Oman}
\affiliation[d]{Centre for Cosmology, Astrophysics, and Space Science (CCASS), GLA University, Mathura 281406, Uttar Pradesh, India}

\emailAdd{gmustafa3828@gmail.com}
\emailAdd{faisaljaved.math@gmail.com}
\emailAdd{sunil@unizwa.edu.om}
\emailAdd{saibal.ray@gla.ac.in}

\date{\today}

\abstract{In this work, we present the Einstein field equations (EFE) in the framework of a modified matter source and thus find out the solutions for the Wormhole. To obtain characteristic solutions set, we employ two kinds of equations of states (EOS), i.e., the van der Waals and polytropic EOS to the EFE. We adopt embedding class as a general technique for the system and discuss several physical attributes of the stellar system under embedded wormhole solutions. Detailed discussion about the matter contents of the thin-shell developed around the obtained wormhole solutions in the background of Schwarzschild BH surrounded
by cloud and quintessence-type fluid distribution by using a well-known cut and paste approach. We have also performed stability analysis using the linearized radial perturbation method for the van der Waals EOS and polytropic EOS. Several interesting points have evolved from the entire investigation along with the features of the thin shell around the wormhole.}


\maketitle
\flushbottom

\section{Introduction}  \label{sec1}

Let us start with the present article with an appropriate quotation from Albert Einstein: ``However, we select from nature a complex [of phenomena] using the criterion of simplicity, in no case will its theoretical treatment turn out to be forever appropriate (sufficient). I do not doubt that the day will come when [general relativity], too, will have to yield to another one, for reasons which at present we do not yet surmise. I believe that this process of deepening theory has no limits.'' So, following Einstein's far-reaching comment, eventually, we have arrived at a new avenue of modern cosmology which draws our attention to look into various observational evidence, like late-time acceleration of the Universe~\cite{Riess1998, Perlmutter1999,Bernardis2000,Knop2003,Bennett2003}, Cosmic Microwave Background Radiation (CMBR)~\cite{Spergel2003,Spergel2007}, baryon acoustic oscillations~\cite{Percival2010}, Planck data~\cite{Ade2014}. These results, therefore, invite a serious as well as an awkward challenge to Einstein's most successful theory of gravitation, i.e., the general theory of relativity (GR). 

How to reconcile with the conflict thus aroused in the modern-day scenario of science became a much debatable topic and scientists became divided into two groups: conventional as well as revolutionary. The conventional scientists tried to tackle the situation by addressing the problem within the framework of GR where they introduce the presence of {\it dark energy}, a kind of exotic entity~\cite{Caldwell2002,Nojiri2003a,Odinstov2003,Padmanabhan2002,Kamenshchik2001,Bento2002}. On the other hand, the revolutionary group which now belongs to some conventional candidates also, came out with a proposal of modified theories to coup up with the confronting situation where the standard Einstein-Hilbert action have been replaced by an arbitrary function of Ricci scalar $R$, i.e., an extension of GR via $f(R)$ gravity theory which could successfully explain most of the issues, such as the late time cosmic acceleration, unified inflation with dark energy, galactic dynamics of massive test particles~\cite{Capozziello2002,Nojiri2003b,Carroll2004,Nojiri2007,Bertolami2007,Nojiri2008,Cognola2008,Lobo2008,Sotiriou2010,Capozziello2010,Cognola2011,Nojiri2011,GPS2020}. However, later on several modified gravity theories take part with definite role to confront with the observational results in the field of cosmology as well as astrophysics, e.g., $f(R,T)$~\cite{harko2011,SC2013,shabani13,Moraes2015,Amit2016,Amit2017,Deb2018a,Hulke2020,Mishra2021a}, $f\left(G\right)$ gravity~\cite{Bamba2010,Rodrigues2014}, $f\left(R,G\right)$ gravity~\cite{Nojiri2005}, $f\left(\mathbb{T}\right)$ gravity~\cite{Bengochea2009,Linder2010,Bohmer2011}, Brans-Dicke (BD) gravity~\cite{Avilez2014,Bhattacharya2015,Singh2020,Singh2021} and so on, where $T$, $G$ and $\mathbb{T}$ are the energy-momentum tensor, Gauss-Bonnet scalar and the torsion scalar, respectively.

Following the above-mentioned modified/extended/alternative theories of gravity, several investigators have proposed compact stellar models, however, within the 4-dimensional spacetime. It is therefore obviously a wider need to understand the inherent geometry of the underlying spacetime in a more effective manner and hence the physical process should be considered to embed a $4D$ spacetime. To facilitate the purpose, in the present work, we have employed the Karmarkar condition~\cite{Karmarkar1948} keeping in mind that he used the technique of embedding 4-dimensional spacetime into 5-dimensional Euclidean space. This very embedding technique can simplify the process of solving the Einstein field equations in an easier way. In this context, it is to note that different types of manifolds are interlinked by embedding 4-dimensional Einstein's field equations into 5-dimensional flat Riemannian space~\cite{Rippl95,Lidsey97}. However, in the framework of GR Maurya et al.~\cite{Maurya16} provided an exact generalized model for anisotropic compact stars embedding Class I with viable physical conditions. Physically acceptable solution for compact star under $f(R,T)$ gravity theory also obtained by Waheed et al.~\cite{Waheed20} where they have used the Karmarkar condition~\cite{Karmarkar1948} whereas Ahmed and Abbas ~\cite{Ahmed20} have investigated the gravitational collapse by exploiting the Karmarkar condition~\cite{Karmarkar1948} to the case of the spherically symmetric non-static radiating star. 

In the present investigation, our target theme is rather the wormhole (WH), especially the geometry of the so-called tunnel in between two sources (white hole) and sink (black hole) in the remote spacetime within the observable universe or in another universe. Historically, the concept of a wormhole comes into the limelight through the naive research by Flamm~\cite{Flamm1916} which was then got physical detailing via the suggestion of a kind of ``bridge" through spacetime, known as the Einstein-Rosen bridge~\cite{ER1935}. Later on, it was structured mathematically and explained physically by several scientists~\cite{FH1962,Morris1988,Garfinkle1991,Visser1991,Bueno2018,Jusufi2019b,Collas2012,Cataldo2017}. In a series of papers Capozziello and coworkers~\cite{Capo1,Capo2,Capo3,Capo4,Capo5} have discussed wormhole solutions in different physical contexts, e.g., in the background of hybrid metric-Palatini gravity by specifying the redshift function, a scalar field nonminimally coupled with torsion and a boundary term to construct the Lorentzian wormholes under Noether symmetries, calculated the equations for the motion of a test particle in the equatorial plane around a wormhole geometry by considering the Poynting-Robertson effect and the Epicyclic frequencies in static and spherically symmetric wormhole geometries. There are several other noteworthy investigations related to WH astrophysics as well as cosmology can be obtained in the following Refs.~\cite{Ellis1973,Bronnikov1973,Visser1987,Hochberg1993,Visser1989,Kim2001,Shinkai2002,Lemos2003,Kuhfitig2003,Arellano2006,Hayward2009,Usmani2010a,Sarbach2010,Bronnikov2013,Bejarano2017,Moraes2017,Cremona2019,Garattini2019,Bronnikov2021,Jose2021,Tripathy2021,Mishra2021b,Konoplya2022,Mishra2022,Sengupta2022,Huang2022} whereas in connection to WH detection some works with interesting theoretical suggestions and observational techniques are available in the literature~\cite{Ohgami2015,Wang2016,Shaikh2019,Dai-Stojkovic2019,Piotrovich2020,Shaikh2021}. To get a kind of different flavor regarding some interesting papers on galactic WH readers may consult the following Refs.~\cite{Kuhfittig2014,Rahaman2014a,Rahaman2014b,Rahaman2016a,Rahaman2016b,Chakraborty2021,Mustafa2022a,Mustafa2022b}.

{The time-like thin-shells in spherically symmetric static spacetimes are the most interesting cosmological object which can be constructed in general relativity. Such models of cosmological objects have been used to analyze some astrophysical phenomena such as gravitational collapse and supernova explosions. The seminal work of Israel~\cite{1s} in 1966 provided a concrete formalism for constructing the time-like shells, in general, by gluing two different manifolds at the location of the thin-shell. The self-gravitating thin-shells are the solutions of a given gravitational theory describing two regions separated by an infinitesimally thin region where the matter is confined. It is important to determine whether relevant thin-shell configurations are stable, both thermodynamically as well as dynamically.}

{Many researchers have discussed the dynamics and stable configuration of thin-shell, along with thin-shell wormholes (WHs), by using the radial perturbation with different matter distributions. Interestingly, Brady et al.~\cite{2s} investigated the linear stability of thin-shell that connects the inner flat and outer Schwarzschild BH through radial perturbation. On the other hand, Martinez~\cite{3s} studied the thermodynamical stability of such a geometrical structure. Later on, Mazharimousavi et al.~\cite{4s} explored the stable configuration of such systems by using the variable EOS. LeMaitre and Poisson~\cite{5s} observed the stability of this geometrical structure in  Newtonian and relativistic gravity. They found a link between the existence of a maximum mass along a sequence of equilibrium configurations and the onset of dynamical instability. Recently, Bergliaffa et al.~\cite{6s} examined their linear and thermodynamical stability by considering barotropic and also for EOS of the type $p=p(\sigma(a))$.}

{At this juncture we would like to mention that the well-known geometrical structure of thin-shell with the interior de Sitter (DS) and exterior BH geometry referred to as a {\it gra}vitational {\it va}cuum {\it star} (gravastar)~\cite{6ss}. Several researchers have developed such geometrical structures in the background of different BH spacetimes and also investigated their stability with different EOS~\cite{6sss}. Thin-shell has great importance and their stability problems can be resolved by choosing different equations of state (EOS) with linear perturbation.} 

{On the other hand, Poisson and Visser~\cite{1k} developed thin-shell from the Schwarzschild BH and analyzed stability through the radial perturbation that depends on the exotic matter at WH throat. Lobo \cite{2k} constructed thin-shell from the joining of the Schwarzschild BH with traversable WH and analyzed the stability of the developed structure. He also examined the expansion as well as collapsing characteristics of thin-shell using surface stresses. Garcia et al.~\cite{3k} studied thin-shell from traversable WHs through the {\it cut and paste} technique and found that the presence of the exotic
matter at the shell greatly affects the stable/unstable configuration. Forghani et al.~\cite{Forghani} formulated an asymmetric thin-shell from two traversable WHs with different redshift functions by assuming barotropic EOS. They analyzed that either thin-shell collapses to the original WH or expands along the shell’s radius.}

Motivated by all the above-mentioned investigations, in the present work our aim is to explore the possibility of stable thin-shell around wormholes within the string cloud and quintessential field via the van der Waals and polytropic EOS in General relativity.

The outline of the present investigation is as follows: The Einstein field equations (EFE) in the framework of modified matter source are shown in Section~\ref{sec2}. Two kinds of equations of states (EOS), i.e., van der Wall and Polytropic are employed to obtain solutions to the EFE in section~\ref{sec3}. In section~\ref{sec4} we adopt an embedding class and discuss some physical properties of our stellar system under embedded wormhole solutions. Energy conditions and matter contents of thin-shell around wormhole geometry have been discussed in Sections~\ref{sec5} and~\ref{sec6}, respectively. Stability analysis using Linearized Radial Perturbation has been performed in~\ref{sec7}, especially~\ref{subsec7.1} for the van der Waals EOS and~\ref{subsec7.2} polytropic EOS. Finally, in Section~\ref{sec8} we have provided a few concluding remarks.

\section{Einstein field equations in the framework of modified matter source}\label{sec2}
This section deals with the spherically symmetric static space-time that is bounded by a spherical surface $\Sigma$. With the arrangement of Schwarzschild coordinates the corresponding line element is provided as:
\begin{eqnarray}
ds^2 = e^{\mu(r)}dt^{2} - e^{\nu(r)}dr^{2}-r^{2}(\sin^{2}\theta d \phi^{2}+d\theta^{2}), \label{1}
\end{eqnarray}
where $\mu(r)$ and $\nu(r)$ represent the gravitational functions. 

The modified version of the Einstein field equations through matter sector for the metric (\ref{1}) can be written as:
\begin{eqnarray}
 &&   G_{ij}=R_{ij}-\frac{1}{2}\,g_{ij}\,R= T^{eff}_{ij}, \label{eq2}
\end{eqnarray}
where
\begin{eqnarray}
T^{eff}_{ij}= \hat{T}_{ij}+\Theta_{ij}+\hat{\Theta}_{ij}, \label{eq3}
\end{eqnarray}
where $\Theta_{ij}$ and $\hat{\Theta}_{ij}$ denote the matter due to the cloud and quintessence fields respectively. 

The Lagrangian density under the effect of string clouds is defined as~\cite{SC1}
\begin{equation}\label{S1}
L_s=k\big(-\frac{1}{2}\Sigma^{ij}\Sigma_{ij} \big),
\end{equation}
where $k$ is a constant that defines the tension of the bivector and string. Further, we have the following relation
\begin{equation}\label{S3}
\Sigma^{ij}=\epsilon^{a\beta}\frac{\partial x^i}{\partial \lambda^{a}}\frac{\partial x^j}{\partial \lambda^{\beta}}.
\end{equation}

In Eq. (\ref{S3}) $\epsilon^{a \beta}$ represent the Levi-Civita tensor and $\lambda^{a} (\lambda^{a}=\lambda^0, \lambda^1)$, is a relation for the parameterization of the world sheet, which is described by the string with the induced metric~\cite{SC1}
\begin{equation}\label{S4}
h_{a \beta}=g_{ij}\frac{\partial x^i}{\partial \lambda^{a}}\frac{\partial x^j}{\partial \lambda^{\beta}}.
\end{equation}

Further, the notation $\Sigma^{i j}$ describes the following important identities~\cite{SC1,SC2}
\begin{equation}\label{S5}
\Sigma^{i[a}\Sigma^{\beta \sigma]}=0,\quad \nabla_i\Sigma^{i a}\Sigma^{\beta \sigma]}=0, \quad \Sigma^{i a}\Sigma_{a\sigma}\Sigma^{\sigma j}={\bf h} \Sigma^{ji},
\end{equation}
where ${\bf h}$ defines the determinant of $h_{a \beta}$. 

Now, by varying the Lagrangian density with respect to $g_{ij}$, we can obtain~\cite{SC2}
\begin{equation}\label{S6}
\Theta_{ij}=\rho_s\frac{\Sigma^{i a}\Sigma^{j}_{a}}{\sqrt{-\bf h}},
\end{equation}
where $\rho_s$ in the Eq. (\ref{S6}) represents the string cloud density. 

Now, by plugging the three identities given in Eq. (\ref{S5}), one can obtain $\partial_i(\sqrt{-g}\Sigma^{i a})=0$. Therefore, the non-zero and viable components of the stress-energy-momentum tensor under the effect of string clouds are expressed as \cite{SC3}:
\begin{equation}\label{11}
\Theta_{tt}=\Theta_{rr}=-\frac{a}{r^2}\;\;\;\;\;\;\;\Theta_{\theta\theta}=\Theta_{\phi\phi}=0,
\end{equation}
where the cloud of strings corresponds to the constant $a$. 

For the matter due to quintessence, we have the following relation~\cite{Q1}
\begin{equation}\label{q}
L_q=-\frac{1}{2}g^{ij}\partial_i\Psi \partial_j\Psi-V(\Psi),
\end{equation}
where $\Psi$ denotes the quintessence field and $V(\Psi)$ represents the potential term. 

The non-zero and physically viable components of the stress-energy-momentum tensor under the effect of the quintessence field are given as~\cite{Toledoa,SC1}
\begin{equation}\label{q1}
\hat{\Theta}_{tt}=\hat{\Theta}_{rr}=\rho_{q}\;\;\;\; \hat{\Theta}_{\theta\theta}=\hat{\Theta}_{\phi\phi}=-\frac{\rho_{q}}{2}(3\omega_q+1),
\end{equation}
where $\omega_q$ and $\rho_q$ represent the quintessence field and the quintessence density respectively. 

To describe the internal structure of the system with the matter source $\hat{T}_{ij}$, we consider the physical content of the space-time, which is covered by an anisotropic matter distribution with energy density $\rho$, radial pressure $p_r$ and tangential pressure $p_t$. Further, the corresponding energy-momentum tensor $\hat{T}_{ij}$ is given as
\begin{eqnarray}
\hat{T}_{ij}=\rho\,u^i\,u_j-p\,K_{ij}+\Pi_{ij}, \label{eq4}
\end{eqnarray}
where
\begin{eqnarray}
&& p=\frac{p_r+2p_t}{3};~~~~\Pi_{ij}=\Pi \big(\xi^i \xi_j+\frac{1}{3} K_{ij}\big);\nonumber\\
&& \Pi=p_r-p_t; ~~~K_{ij}=g_{ij}-u^i u_j, \label{eq5}
\end{eqnarray}
and  $u^i$ (four-velocity vector) and $\xi^i$ are given by
\begin{eqnarray}
u^i=(e^{-\mu/2},~0,~0,~0)~~\text{and}~~\xi^i=(0,~e^{-\nu/2},~0,~0), \label{eq6}
\end{eqnarray}
such that $\xi^i u_i=0$ and $\xi^i\xi_i=-1$. 

Finally, we have the following expressions for the $\hat{T}_{ij}$ matter source:
\begin{eqnarray}
&& \hat{T}_{tt}=\rho,~~~\hat{T}_{rr}=-p_r,~~~~\hat{T}_{\theta\theta}=\hat{T}_{\phi\phi}=-p_t, \label{eq7}
\end{eqnarray}
and the modified version of the Einstein field equations (\ref{eq2}) read as
\begin{eqnarray}
\label{eq8}
T^{eff}_{tt}&&= \rho^{eff}=\rho +\rho_{q}+\frac{a }{r^2}=\frac{1}{8\pi}\bigg[\frac{1}{r^2}-e^{-\nu}\left(\frac{1}{r^2}-\frac{\nu^{\prime}}{r}\right)\bigg],
\\\label{eq9}
T^{eff}_{rr}&&= {p}^{eff}_{r}= {p}_{r}-\rho_{q}-\frac{a }{r^2}=\frac{1}{8\pi}\bigg[-\frac{1}{r^2}+e^{-\nu}\left(\frac{1}{r^2}+\frac{\mu^{\prime}}{r}\right)\bigg],
\\
\label{eq10}
T^{eff}_{\theta\theta}&&= {p}^{eff}_{t}={p}_{t}+\frac{1}{2} \rho_{q} (3 w_q+1)=\frac{1}{8\pi}\bigg[\frac{e^{-\nu}}{4}\left(2\mu^{\prime\prime}+\mu^{\prime2}-\nu^{\prime}\mu^{\prime}+2\frac{\mu^{\prime}-\nu^{\prime}}{r}\right)\bigg],~~~
\end{eqnarray}
where primes denote the derivatives with respect to $r$. 

Now by using the Schwarzschild coordinates within the framework of the Morris and Thorne metric for wormhole geometry, i.e., $e^{\mu(r)}=e^{2\Phi(r)}$ and $e^{\nu(r)}=\left(1-\frac{S(r)}{r}\right)^{-1}$~\cite{Morris1988} in the static spherically symmetric spacetime by Eq. (\ref{1}), we have the following set of field equations for wormhole geometry:
\begin{eqnarray}
\label{19}
&& \hspace{-0.8cm}\rho^{eff}=\rho +\rho_{q}+\frac{a }{r^2}=\frac{S'(r)}{r^2},\\\label{20}
 && \hspace{-0.8cm} {p}^{eff}_{r}= {p}_{r}-\rho_{q}-\frac{a }{r^2}=\frac{2 r (r-S(r)) \Phi '(r)-S(r)}{r^3},\\
\label{21}
 && \hspace{-0.8cm} {p}^{eff}_{t}={p}_{t}+\frac{1}{2} \rho_{q} (3 w_q+1)= \frac{1}{4 r^3} \Big[\left(2 r \Phi '(r)+2\right) \big(2 r^2 \Phi '(r)-r S'(r)-2 r S(r) \Phi '(r) \nonumber\\&& \hspace{4cm} +S(r)\big) +4 r^2 (r-S(r)) \Phi ''(r)\Big].
\end{eqnarray}

In the above equations, $\Phi(r)$ and $S(r)$ are the representatives of the redshift and shape function for the wormhole geometry.

\section{Equations of states for the physical system}\label{sec3}
Here, in this section, we shall consider two different equations of states to evaluate the unknown quantities, like the quintessence energy density and pressure for the current system.

\subsection{van der Waals EOS}
It is necessary to mention that the realistic fluid obeys the barotropic EOS, viz., $p_{r}=p_{r}(\rho)$. Here, we consider the interior matter distribution, which fulfills the modified van der Waals EOS~\cite{Van1,Van2,Van3,Van4} and can be defined as:
\begin{equation}\label{eq22}
p_r=\alpha  \rho ^2+\frac{\beta  \rho }{\gamma  \rho +1}.
\end{equation}

In the above Eq. (\ref{eq22}), $\alpha,\;\; \beta$ and $\gamma$ are the arbitrary parameters defining the acceleration and deceleration phases of the universe. As in our system, by Eqs. (\ref{19}-\ref{21}) we have four unknowns like $\rho$, $p_r$, $p_t$, and $\rho_{q}$, so we must plug the van der Waals EOS in the above system. After some suitable calculations, we have the following expressions for $\rho$, $p_r$, $p_t$, and $\rho_{q}$:
\begin{eqnarray}
\rho &&=\frac{1}{6 r^3 \alpha \gamma}\bigg[-2 r^3 \left(\alpha+\gamma\right)+\frac{2 \sqrt[3]{2} r^3 \Psi _1}{\sqrt[3]{\Psi _7}}+2^{2/3} \sqrt[3]{\Psi _7}\bigg],\label{22}\\
\rho_{q}&& =-\frac{1}{6 r^3}\bigg[\frac{-2 r^3 \left(\alpha+\gamma\right)+\frac{2 \sqrt[3]{2} r^3 \Psi _1}{\sqrt[3]{\Psi _7}}+2^{2/3} \sqrt[3]{\Psi _7}}{\alpha \gamma}+6 a  r-6 r S'(r)\bigg],\label{23}\\
p_{r}&& =\frac{1}{36 r^6 \alpha \gamma^2}\bigg[\left(-2 r^3 \left(\alpha+\gamma\right)+\frac{2 \sqrt[3]{2} r^3 \Psi _1}{\sqrt[3]{\Psi _7}}+2^{2/3} \sqrt[3]{\Psi _7}\right) \bigg(\frac{6 r^3 \beta \gamma}{\frac{\Psi _9}{6 r^3 \alpha}+1}-2 r^3 \left(\alpha+\gamma\right) \nonumber\\&&\hspace{0.5cm}+\frac{2 \sqrt[3]{2} r^3 \Psi _1}{\sqrt[3]{\Psi _7}}+2^{2/3} \sqrt[3]{\Psi _7}\bigg)\bigg],\label{24}\\
p_{t} &&=\frac{1}{12 r^3}\bigg[(3 w+1) \left(\frac{-2 r^3 \left(\alpha+\gamma\right)+\frac{2 \sqrt[3]{2} r^3 \Psi _1}{\sqrt[3]{\Psi _7}}+2^{2/3} \sqrt[3]{\Psi _7}}{\alpha \gamma}+6 a  r-6 r S'(r)\right)+3 \Psi _8\bigg],\label{25}
\end{eqnarray}
where
\begin{eqnarray*}
\Psi _1&&=r^3 \alpha^2+r^3 \gamma^2-\alpha \gamma \left(3 r^3 \beta+r \left(r^2-3 \gamma \left(S'(r)+2 r \Phi '(r)\right)\right)+3 \gamma S(r) \left(2 r \Phi '(r)+1\right)\right)\\
\Psi _2&&=r^3 \beta+r \left(r^2-\gamma \left(S'(r)+2 r \Phi '(r)\right)\right)+\gamma S(r) \left(2 r \Phi '(r)+1\right)\\
\Psi _3&&=2 \alpha^2-\left(9 \beta+5\right) \gamma \alpha+2 \gamma^2\\
\Psi _4&&=r^2 \left(-\alpha-\gamma\right) \Psi _3+9 \alpha \left(2 \alpha-\gamma\right) \gamma^2 S'(r)+18 r \alpha \left(2 \alpha-\gamma\right) \gamma^2 \Phi '(r)\\
\Psi _5&&=r \Psi _4-9 \alpha \left(2 \alpha-\gamma\right) \gamma^2 S(r) \left(2 r \Phi '(r)+1\right)\\
\Psi _6&&=r^{12} \Psi _5^2+4 \left(3 r^3 \alpha \gamma \Psi _2-r^6 \left(\alpha+\gamma\right){}^2\right){}^3\\
\Psi _7&&=-2 r^9 \alpha^3-2 r^9 \gamma^3+3 r^9 \alpha \gamma^2+9 r^9 \alpha \beta \gamma^2+3 r^9 \alpha^2 \gamma+9 r^9 \alpha^2 \beta \gamma-18 r^8 \alpha \gamma^3 \Phi '(r)\\&&+36 r^8 \alpha^2 \gamma^2 \Phi '(r)-9 r^7 \alpha \gamma^3 S'(r)+18 r^7 \alpha^2 \gamma^2 S'(r)+18 r^7 \alpha \gamma^3 S(r) \Phi '(r)-36 r^7 \alpha^2 \gamma^2 S(r) \Phi '(r)\\&&+9 r^6 \alpha \gamma^3 S(r)-18 r^6 \alpha^2 \gamma^2 S(r)+\sqrt{\Psi _6}\\
\Psi _8&&=2 \left(r \Phi '(r)+1\right) \left(2 r^2 \Phi '(r)-r S'(r)-2 r S(r) \Phi '(r)+S(r)\right)+4 r^2 (r-S(r)) \Phi ''(r)\\
\Psi _9&&=-2 r^3 \left(\alpha+\gamma\right)+\frac{2 \sqrt[3]{2} r^3 \Psi _1}{\sqrt[3]{\Psi _7}}+2^{2/3} \sqrt[3]{\Psi _7}.
\end{eqnarray*}

\subsection{Polytropic EOS}
In the current subsection, we consider the same basic polytropic EOS in the form $p_{r}=p_{r}(\rho)$ to solve the above system by Eqs. (\ref{19})--(\ref{21}) for the quintessence energy density. The general polytropic EOS considered here can be expressed as~\cite{Takisa,Chavanisa,Chavanisb}:
\begin{eqnarray}\label{26}
p_{r}=\omega (\rho)^\chi, \quad \chi=1+\frac{1}{n},\quad n\neq0,
\end{eqnarray}
where $\omega$ and $n$ are the constants. For the current analysis, by considering $n=2/3$ and doing some necessary calculations with Eqs. (\ref{19})--(\ref{21}), we have the five different expressions for the energy density. Here, we are going to include all the root-base solutions. The final expressions $\rho$, $p_r$, $p_t$, and $\rho_{q}$ against root-base solutions can be expressed as:
\begin{eqnarray}
&& \hspace{-0.5cm}\rho =\text{Root}[X_{i}],\label{27}\\
&& \hspace{-0.5cm} \rho_{q} =-\text{Root}[X_{i}]-\frac{\alpha }{r^2}+\frac{S'(r)}{r^2},\label{28}\\
&& \hspace{-0.5cm} p_{r} =\omega \text{Root}[X_{i}]^{5/2},\label{29}\\
&& \hspace{-0.5cm} p_{t} =\frac{1}{4 r^3}\Big[2 r^2 (r-S(r)) \Phi''(r)+\left(r \Phi'(r)+2\right) \left(r^2 \Phi'(r)-r S(r) \Phi'(r)-r S'(r)+S(r)\right)\Big]\nonumber\\&&\hspace{0.4cm}-\frac{1}{2} (3 \omega_{q}+1) \left(-\text{Root}[X_{i}]-\frac{\alpha }{r^2}+\frac{S'(r)}{r^2}\right),\label{30}
\end{eqnarray}
where
\begin{eqnarray*}
X_{i}&&=\Big[\text{$\#$1}^5 r^6 \omega^2-\text{$\#$1}^2 r^6+\text{$\#$1} \left(2 r^5 \Phi'(r)-2 r^4 S(r) \Phi'(r)+2 r^4 S'(r)-2 r^3 S(r)\right) \nonumber\\&&\hspace{0.4cm} -r^4 \Phi'(r)^2-2 r^3 \Phi'(r) S'(r)+2 r^3 S(r) \Phi'(r)^2+2 r^2 S(r) \Phi'(r) S'(r)-r^2 S(r)^2 \Phi'(r)^2 \nonumber\\&&\hspace{0.4cm}+2 r^2 S(r) \Phi'(r)-2 r S(r)^2 \Phi'(r)-r^2 S'(r)^2+2 r S(r) S'(r)-S(r)^2\&,i\Big].
\end{eqnarray*}

In the above expression, we have $i=1,2,3,4,5$. It is interesting to mention that an event horizon is not allowed in wormhole geometry. So, we must impose the constraint on the $g_{tt}$ component say $2\Phi(r)$ that should be finite everywhere within the traversable wormhole geometry. For the current analysis, we choose the specific kind of the redshift function \cite{rpop2,rpop3}, which is expressed as:
\begin{equation}\label{22gm}
\Phi=-\frac{ \zeta }{r},
\end{equation}
where $ \zeta $ is constant. For all values of radial coordinate with the scope of radial distance $l$, $\Phi$ should be nonzero and finite which sustains the nonexistence of horizons in the wormhole geometry.

\section{Embedded Wormhole Solutions}\label{sec4}
Herein we shall take a brief overview of embedded wormhole solutions by using two different approaches. In the current study, we will deal with generalized embedded wormhole solutions using Karmarkar condition (KC)~\cite{Karmarkar1948} with embedded Class-1 and Ellis-Bronniokv space-times. The basic construction of the KC depends upon the embedded Class-1 solution of Riemannian space. Eisenhart discussed the suitable requirement for the embedded Class-1 solution~\cite{Eisenhart1966}, which comes through the following equations where the Gauss equation is defined as
\begin{eqnarray}\label{rr10}
\mathcal{R}_{mnpq}=2\,\epsilon\,{b_{m\,[p}}{b_{q]n}}.
\end{eqnarray}

The Codazzi equation can be expressed as:
\begin{eqnarray}\label{rr11}
b_{m\left[n;p\right]}={\Gamma}^q_{\left[n\,p\right]}\,b_{mq}-{{\Gamma}^q_{m}}\,{}_{[n}\,b_{p]q},
\end{eqnarray}
where square brackets represent anti-symmetrization concept, $\epsilon=\pm1$, and $b_{mn}$ mentions the coefficients of the second differential form. Using Eqs. (\ref{rr10}) and Eq. (\ref{rr11}), the KC is expressed as
\begin{equation}\label{rr12}
R_{2323}R_{1414}=R_{1224}R_{1334}+ R_{1212}R_{3434},
\end{equation}
where $R_{2323}\neq R_{1414}\neq0$~\cite{Pandey1982}. 

By incorporating the appropriate Riemannian tensor in Eq. (\ref{rr12}) we get the following differential equation
\begin{equation}\label{rr13}
\frac{\mu'(r) \nu'(r)}{1-e^{\nu(r)}}-\left\{\mu'(r) \nu'(r)+\mu'(r)^2-2 \left[\mu''(r)+\mu'(r)^2\right]\right\}=0,\;\;\;\;\;\;\;e^{\nu(r)}\neq 1.
\end{equation}

By solving (\ref{rr13}) we get
\begin{equation}\label{rr14}
e^{\nu(r)}=1+Ae^{\mu(r)}\mu^{'2}(r),
\end{equation}
where $A$ is an integration constant. 

Now, by adopting the process, which is already reported in~\cite{gm1}, we get the following embedded shape function: For the current study, it should be considered as model-I
\begin{equation}\label{rr15}
S(r)=r-\frac{r^{5}}{r^{4}+S_{0}^{4}(S_{0}-\Psi)}+\Psi,\;\; 0<\Psi<S_{0},
\end{equation}
where $r_{0}$ wormhole throat radius. 

Now, we discuss the second embedded wormhole solution, which is known as the generalized Ellis-Bronnikov space-time (ultra-static wormhole model), which is defined as:
\begin{equation}\label{rr16}
d s^{2}=-d t^{2}+d l^{2}+r^{2}(l)\left(d \theta^{2}+\sin ^{2}(\theta) d \phi^{2}\right),
\end{equation}
with
\begin{equation}\label{rr17}
r(l)=\left(S_{0}^{m}+l^{m}\right)^{1 / m}.
\end{equation}

In the above equations, $l$ represents the proper radial distance or known as the tortoise coordinate, and can be used for both embedded solutions. As already mentioned in the previous case that $S_{0}$ is the throat radius of the embedded wormhole and $m$ is used for the wormhole parameter with condition $(m \geq 2)$. Now, Eq. (\ref{1}) can be rewritten as:
\begin{equation}\label{rr18}
d s^{2}=-d t^{2}+\frac{d r^{2}}{\left(1-\frac{S(r)}{r}\right)}+r^{2}\left(d \theta^{2}+\sin ^{2} \theta d \phi^{2}\right).
\end{equation}

The radial coordinate $r$ and radial distance $l$ can be combined through the following embedding relation
\begin{equation}\label{rr19}
d l^{2}=\frac{d r^{2}}{\left(1-\frac{S(r)}{r}\right)}.
\end{equation}

Finally, we have the following embedded shape function: For the current study, it should be considered model-II
\begin{equation}\label{rr20}
 S(r)=r-r^{(3-2 m)}\left(r^{m}-S_{0}^{m}\right)^{\left(2-\frac{2}{m}\right)}.
\end{equation}

By putting $m=2$ we get the Ellis-Bronniokv wormhole geometry with horizonless space-time. In order to complete the current analysis, we use two different embedded wormhole solutions by Eq. (\ref{rr15}) and Eq. (\ref{rr20}) by combining the radial coordinate $r$ and radial distance $l$ through Eq. (\ref{rr17}).

\begin{figure}[htb!]
\centering 
\includegraphics[width=7.5cm,height=6.0cm]{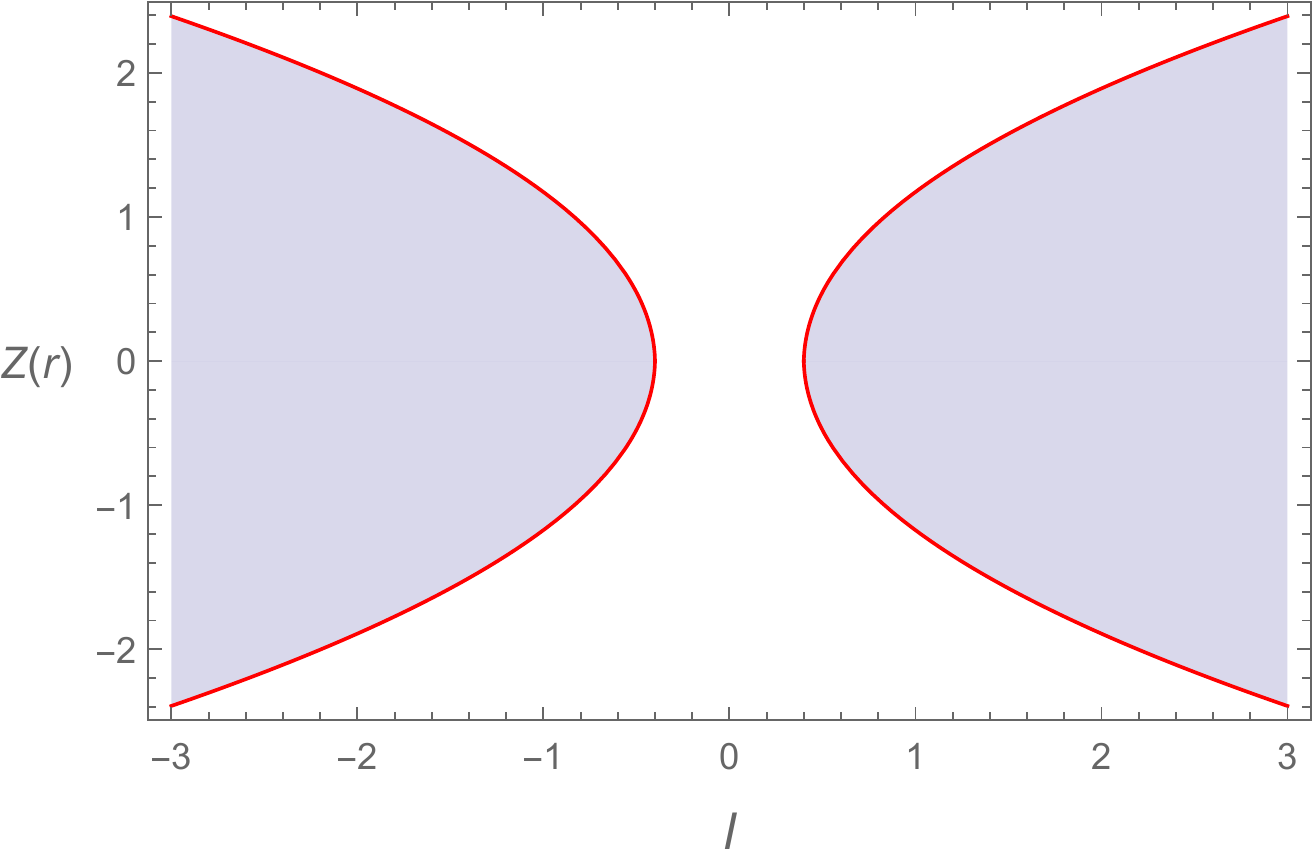}~~ \includegraphics[width=7.5cm,height=6.0cm]{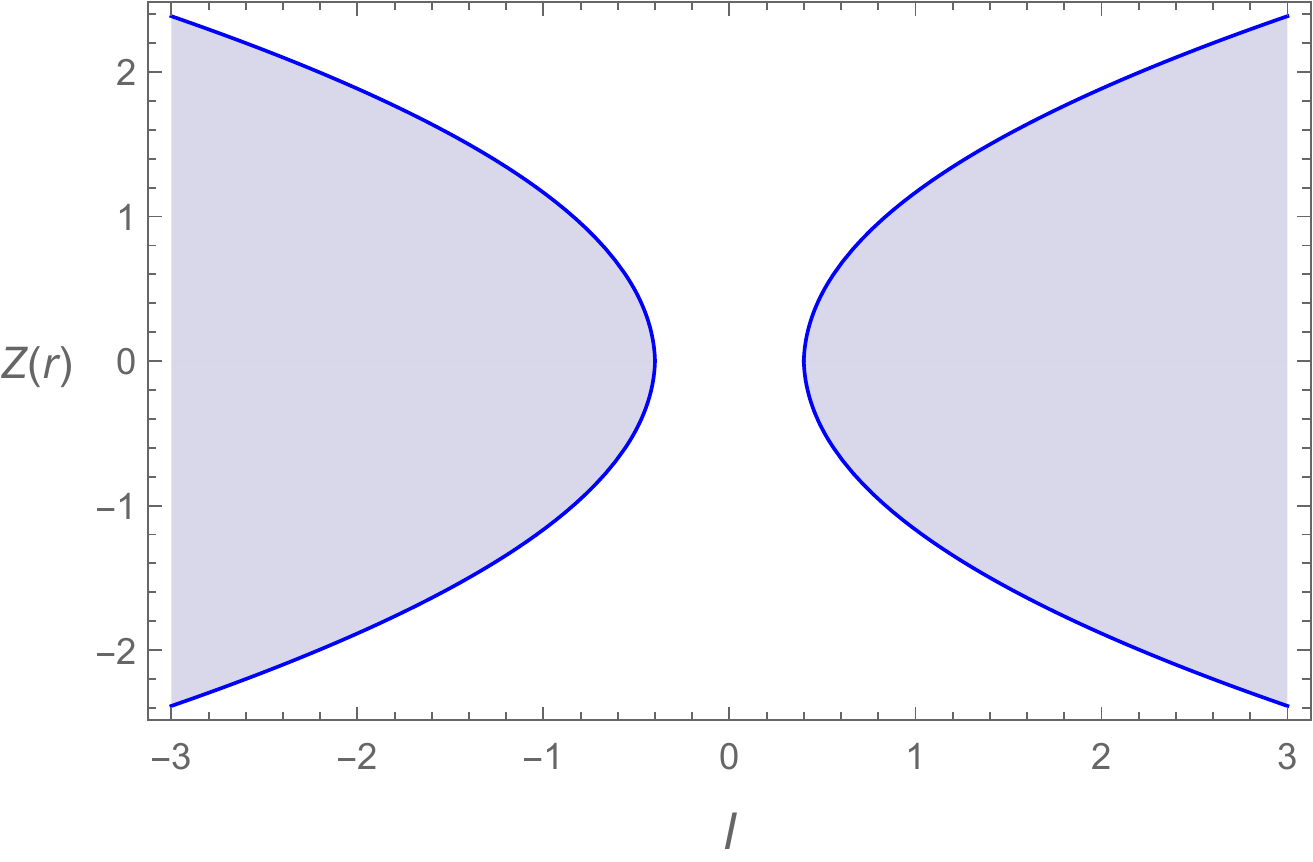}
\centering 
\includegraphics[width=7.6cm,height=6.5cm]{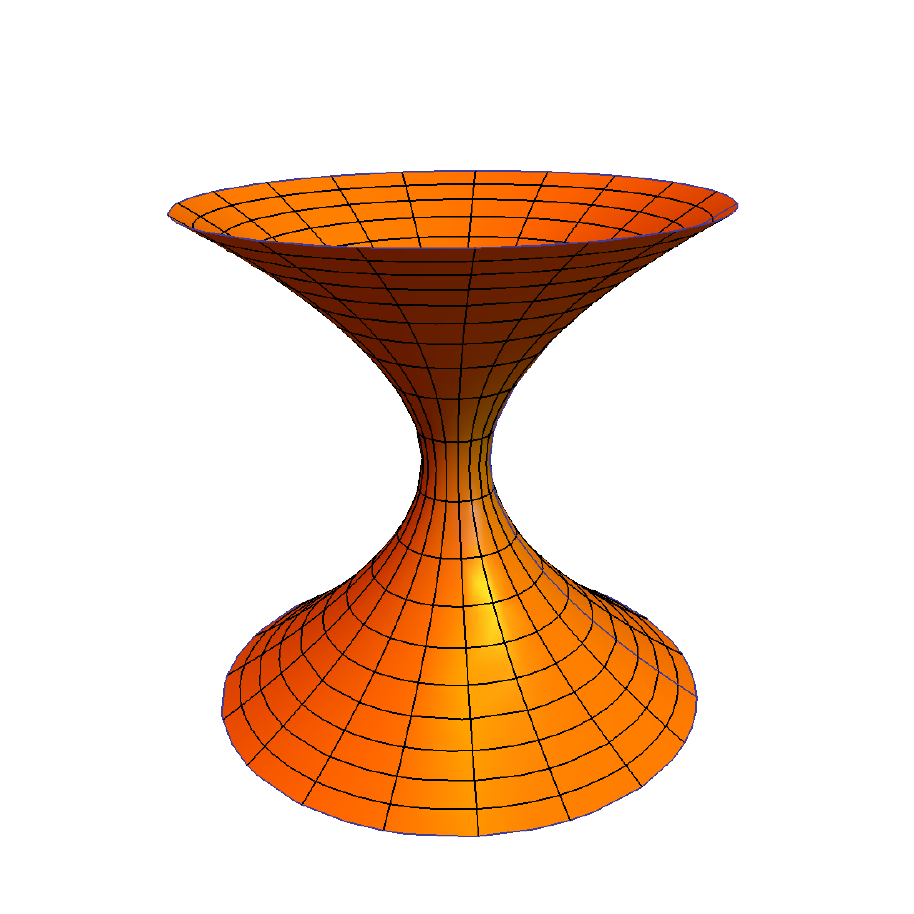} \includegraphics[width=7.6cm,height=6.5cm]{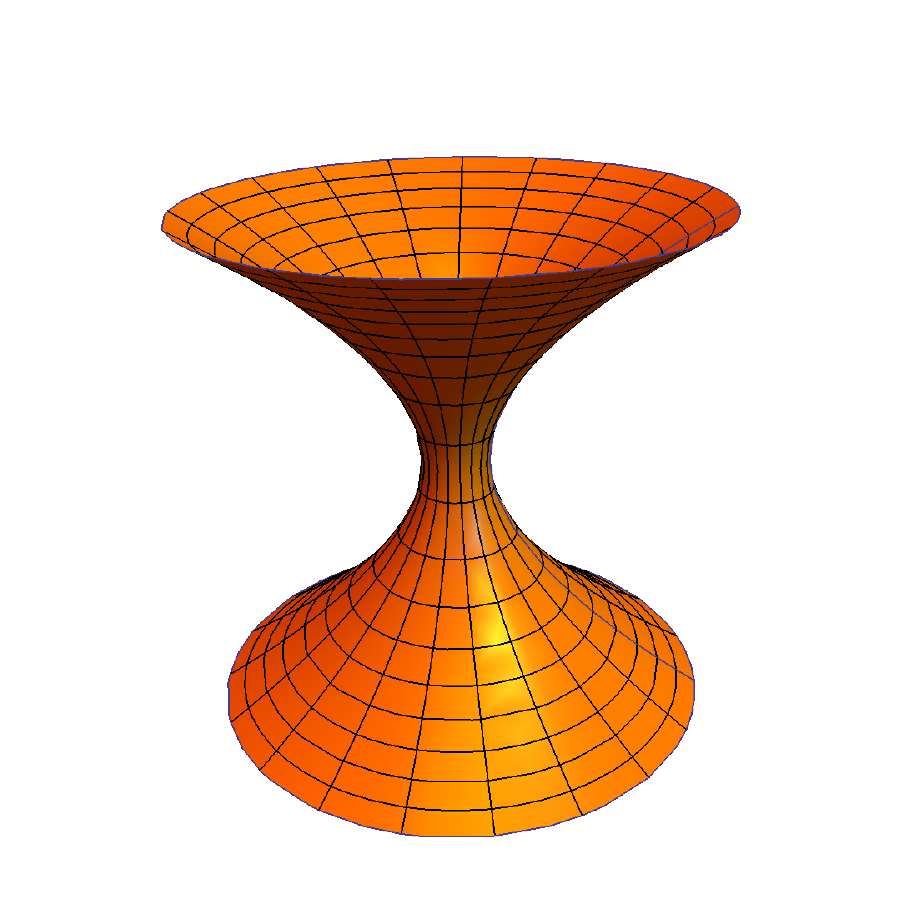}
\caption{Shows the embedding diagram with Model-I (left) and Model-II (right) for two generalized embedded wormhole solutions under $m=2$.}\label{F0}
\end{figure}

Now, we construct the physically acceptable embedding surface diagram for both embedded wormhole solutions and try to explore the required conditions for the embedded surface for wormhole configuration. For this purpose, we shall work with the equatorial plane, by plugging $\theta=\pi$ and $t$ = constant in Eq. (\ref{1}) within the Morris and Thorne wormhole geometry with $r(l)=\left(S_{0}^{m}+l^{m}\right)^{1 / m}$, which reads as:
\begin{equation} \label{r26a}
ds^{2}= \frac{dr^{2}}{\left(1-\frac{S(r)}{r}\right)}+r^{2}d\phi^{2},
\end{equation}
The above expression by Eq. (\ref{r26a}) should be embedded into 3-D Euclidean within the cylindrical symmetry approach under the effect of radial distance, which is redefined as
\begin{equation} \label{r27a}
ds^{2}_{\sum}= dZ^{2}+dr^{2}+r^{2}d\phi^{2},
\end{equation}
The above relation by Eq. (\ref{r27a}) can be recalculated as
\begin{equation} \label{r28a}
ds^{2}_{\sum}= \bigg(1+\bigg(\frac{dZ}{dr}\bigg)^{2}\bigg)dr^{2}+r^{2}d\phi^{2},
\end{equation}
By matching Eqs. (\ref{r26a}-\ref{r28a}), we get the following equation
\begin{equation} \label{r29a}
\frac{dZ}{dr}=\pm \left(\frac{r}{S(r)}-1\right)^{-1/2},
\end{equation}
From the above Eq. \ref{r26a} with $r(l)=\left(S_{0}^{m}+l^{m}\right)^{1 / m}$, the embedded surface diagram can be constructed. The embedded surface diagram construction is shown in Fig (\ref{F0}). The connection between upper universe $Z>0$ and Lower universe $Z<0$ for the radial distance at $S_{0}$ is also provided in Fig. (\ref{F0}).

\begin{figure}[htb!]
\centering 
\includegraphics[width=7.6cm,height=6.0cm]{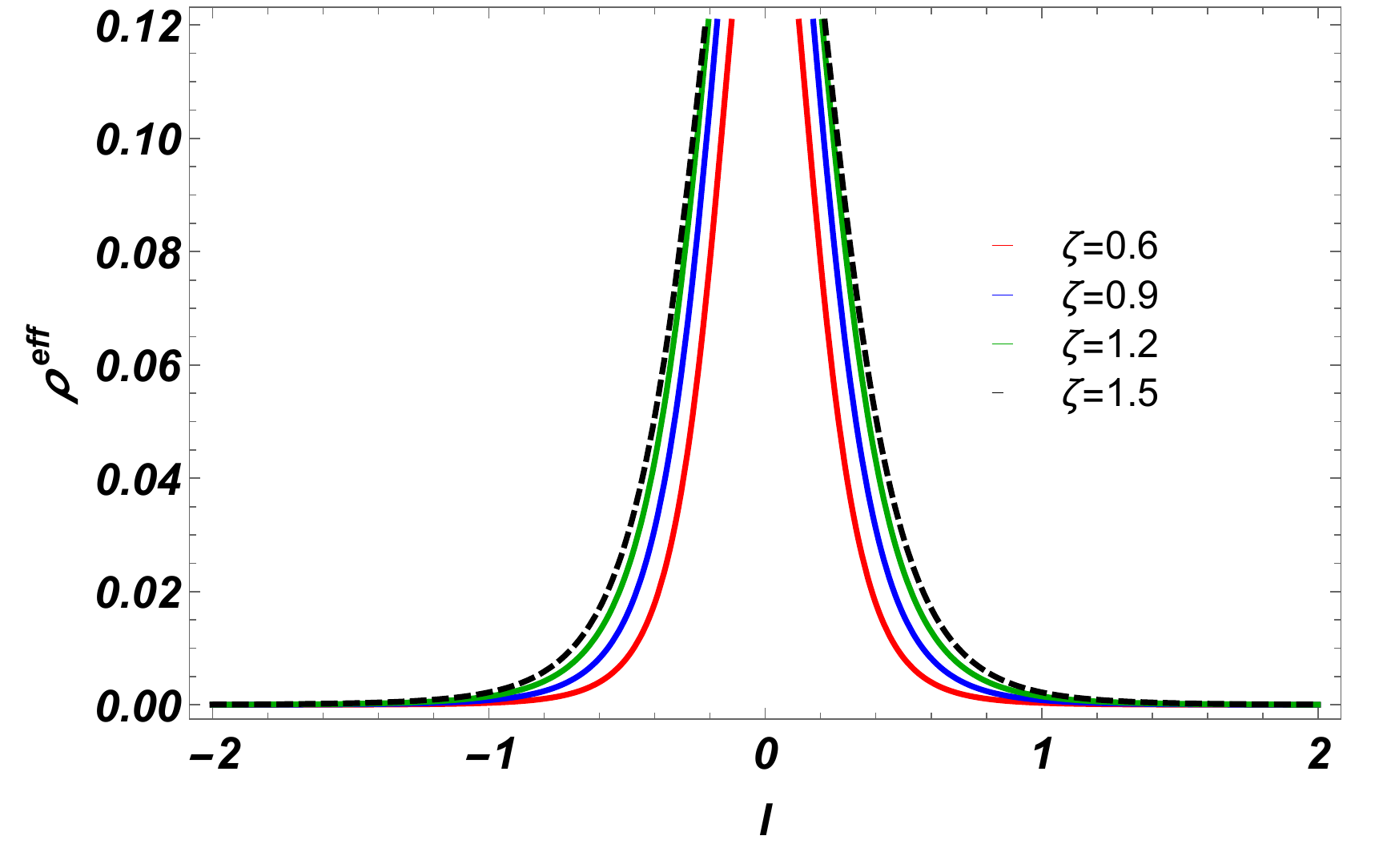} \includegraphics[width=7.6cm,height=6.0cm]{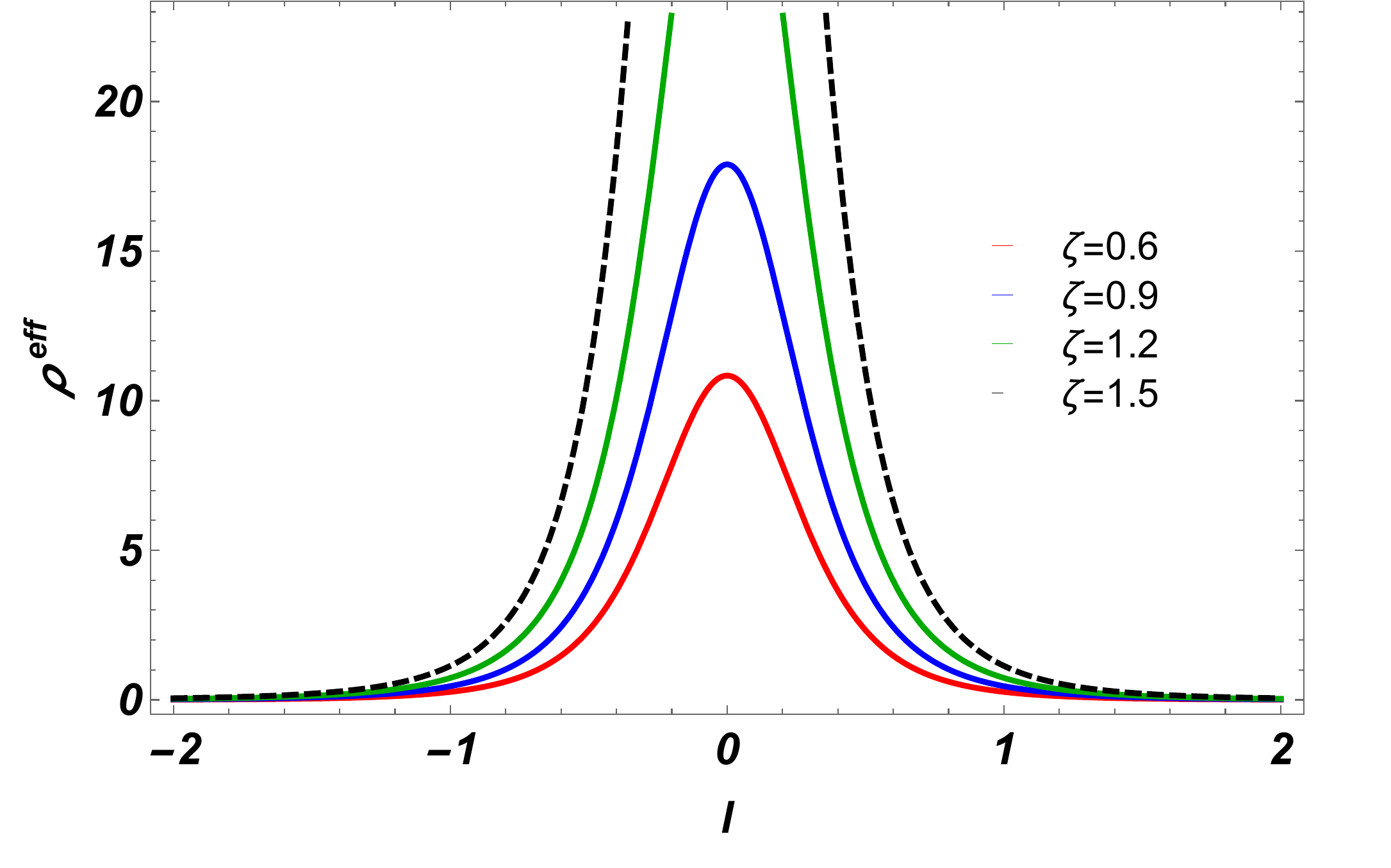}
\caption{Shows the variation of $\rho^{eff}$ with Model-I (left) and Model-II (right) for generalized embedded wormhole solutions under the effect of van der Waals EOS.}\label{F1}
\end{figure}

\begin{figure}[htb!]
\centering 
\includegraphics[width=7.6cm,height=6.0cm]{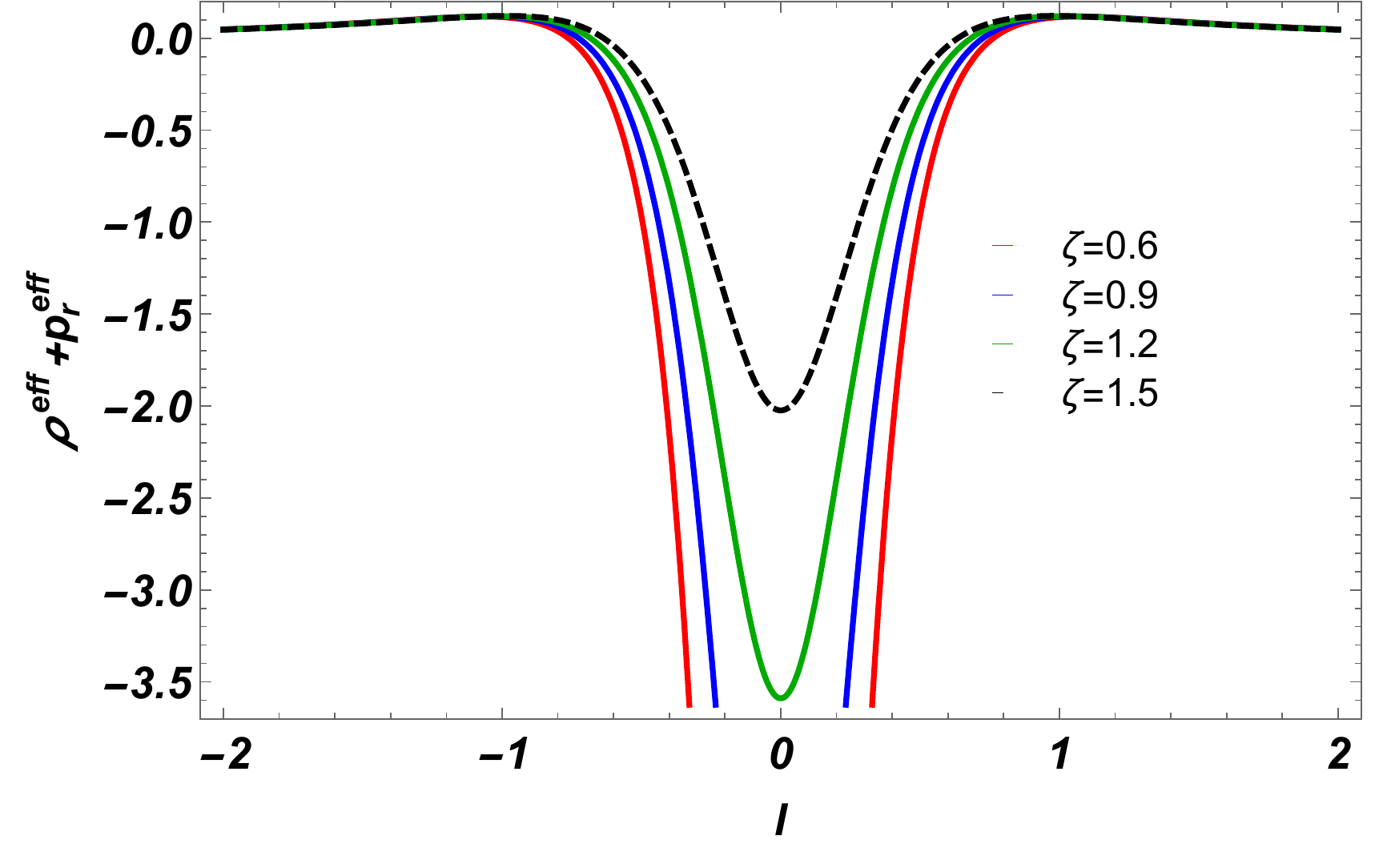} \includegraphics[width=7.6cm,height=6.0cm]{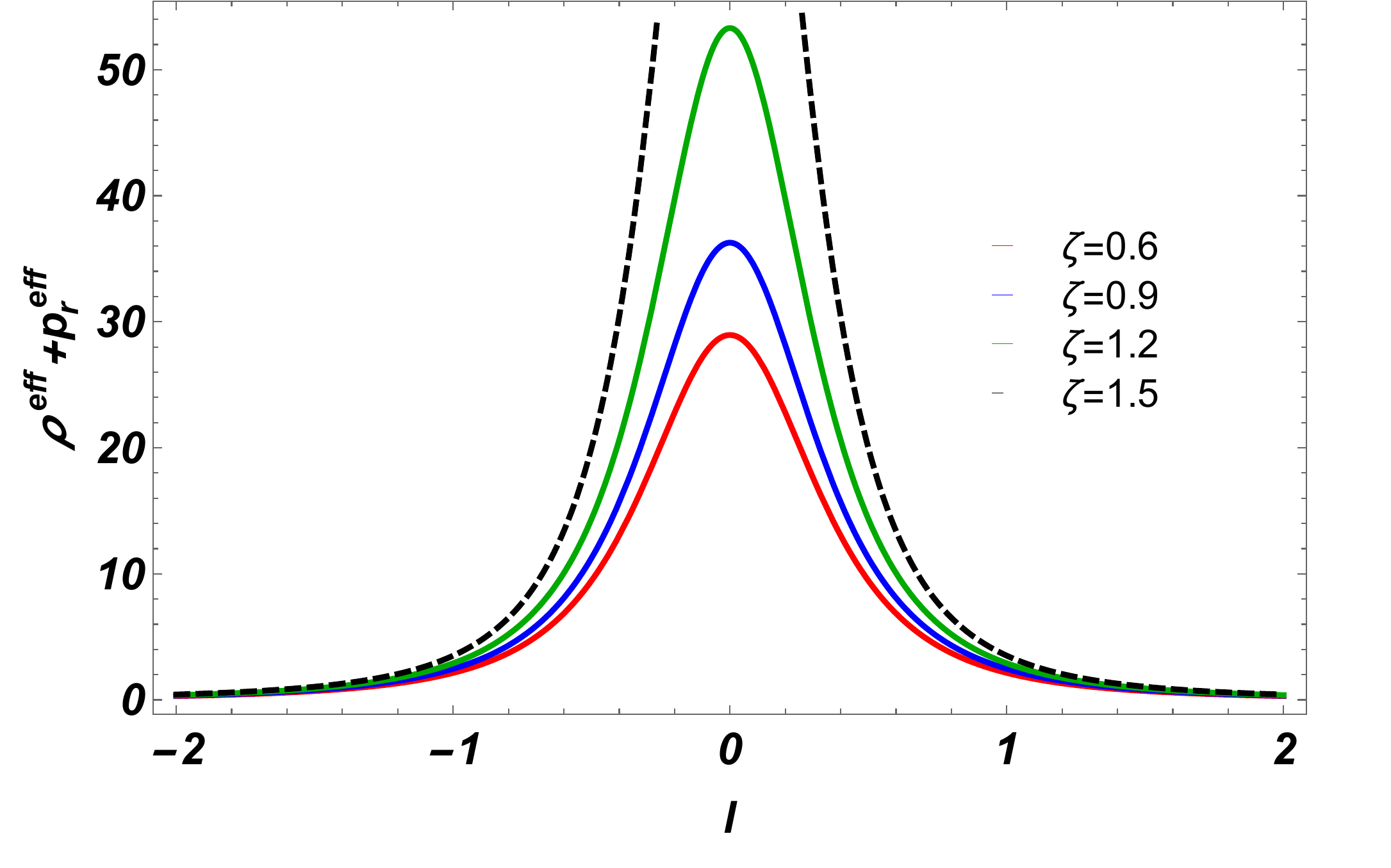}
\caption{Shows the variation of  $\rho^{eff}+p^{eff}_{r}$ with Model-I (left) and Model-II (right) generalized embedded wormhole solutions under the effect of van der Waals EOS.}\label{F2}
\end{figure}
\begin{figure}[htb!]
\centering 
\includegraphics[width=7.6cm,height=6.0cm]{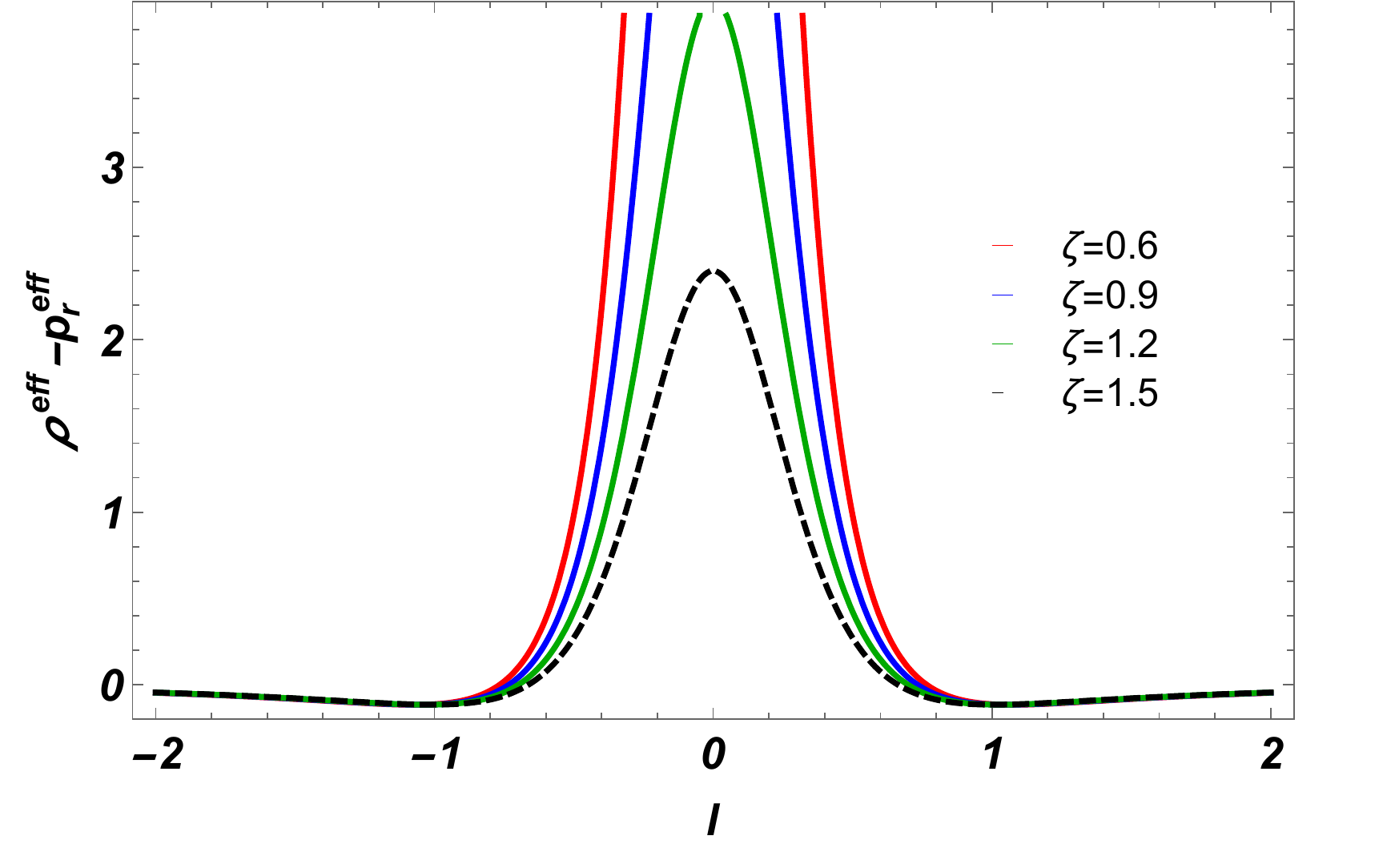} \includegraphics[width=7.6cm,height=6.0cm]{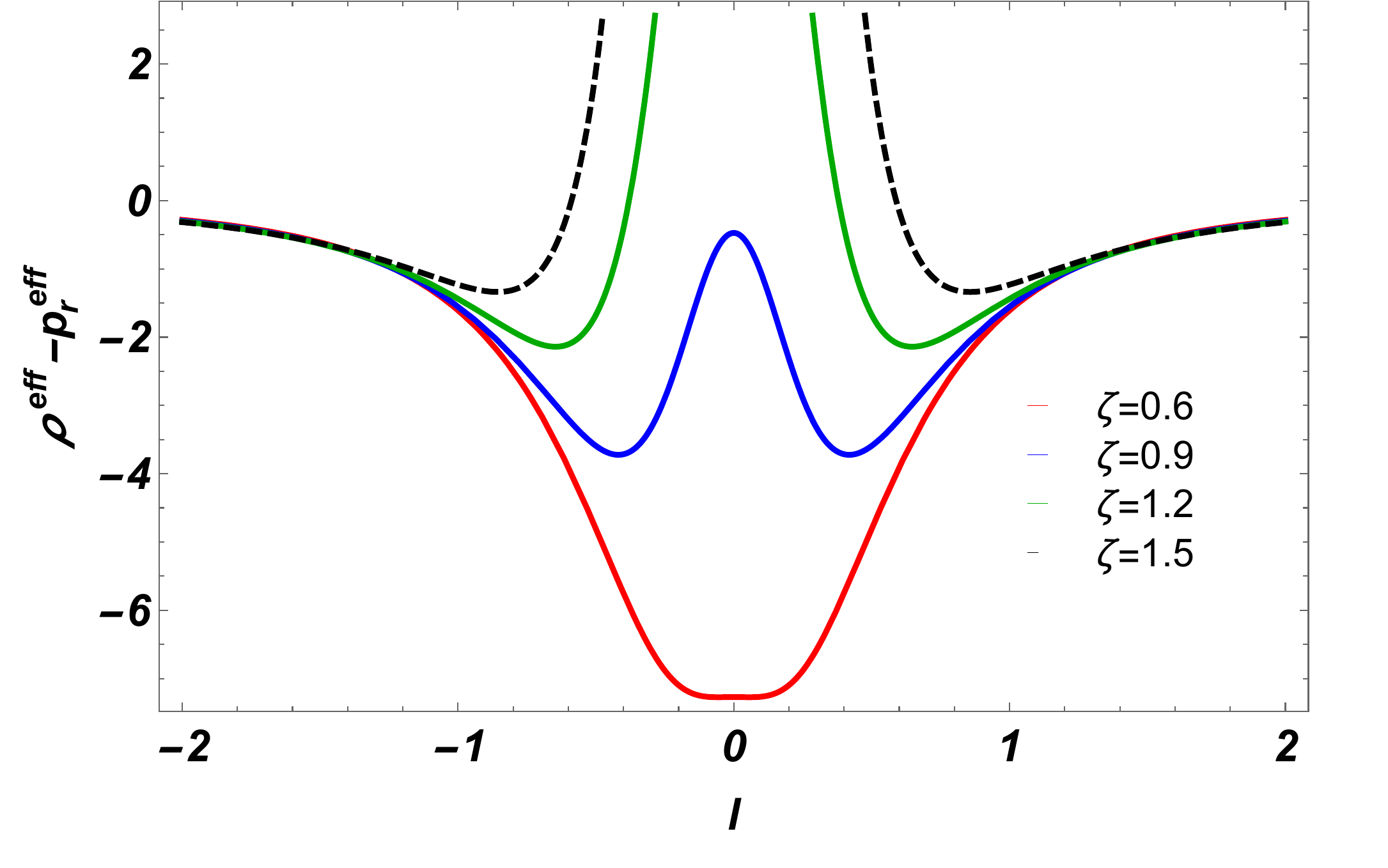}
\caption{Shows the variation of $\rho^{eff}-p^{eff}_{r}$ with Model-I (left) and Model-II (right) generalized embedded wormhole solutions under the effect of van der Waals EoS..}\label{F3}
\end{figure}

\begin{figure}[htb!]
\centering 
\includegraphics[width=7.6cm,height=6.0cm]{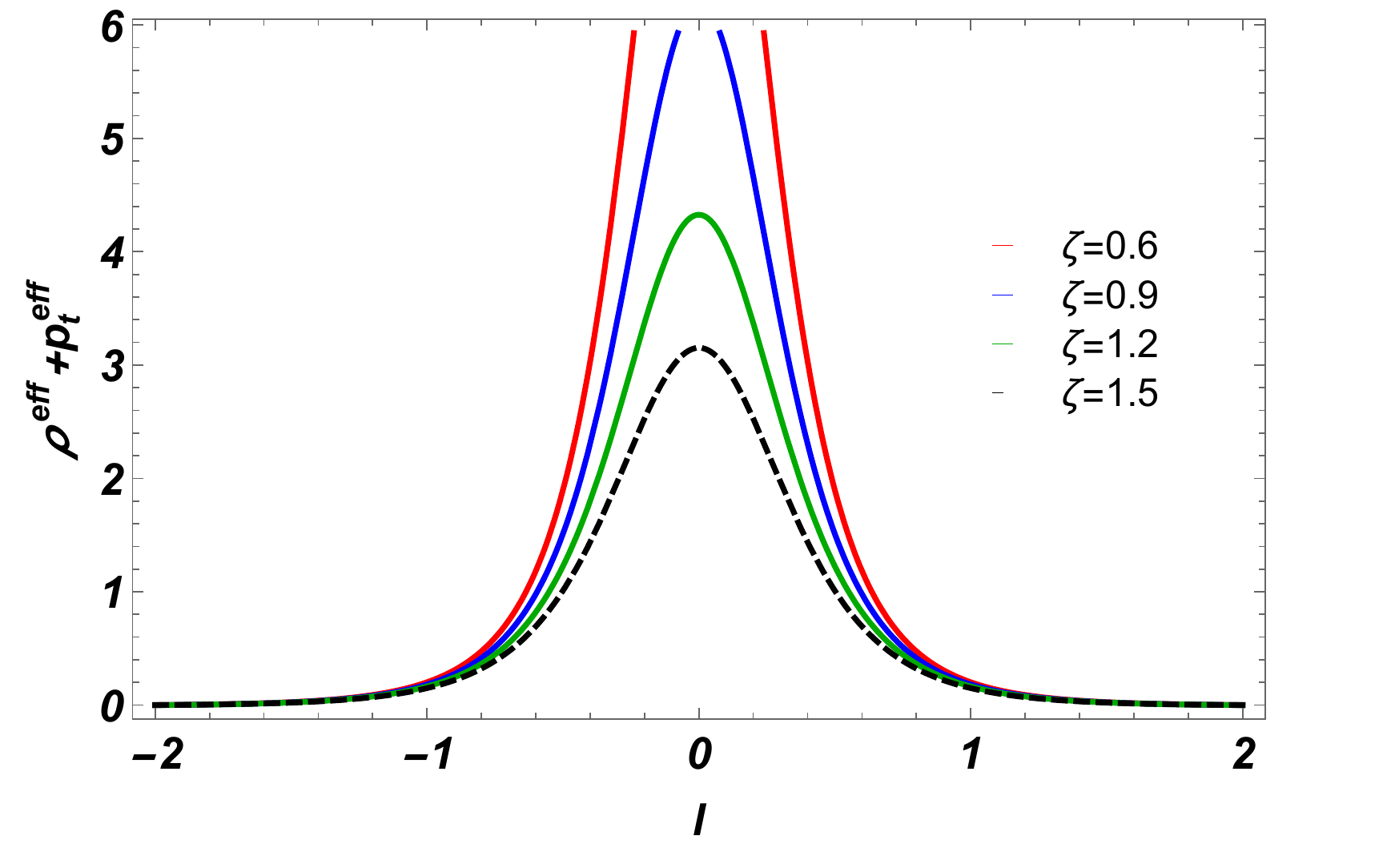} \includegraphics[width=7.6cm,height=6.0cm]{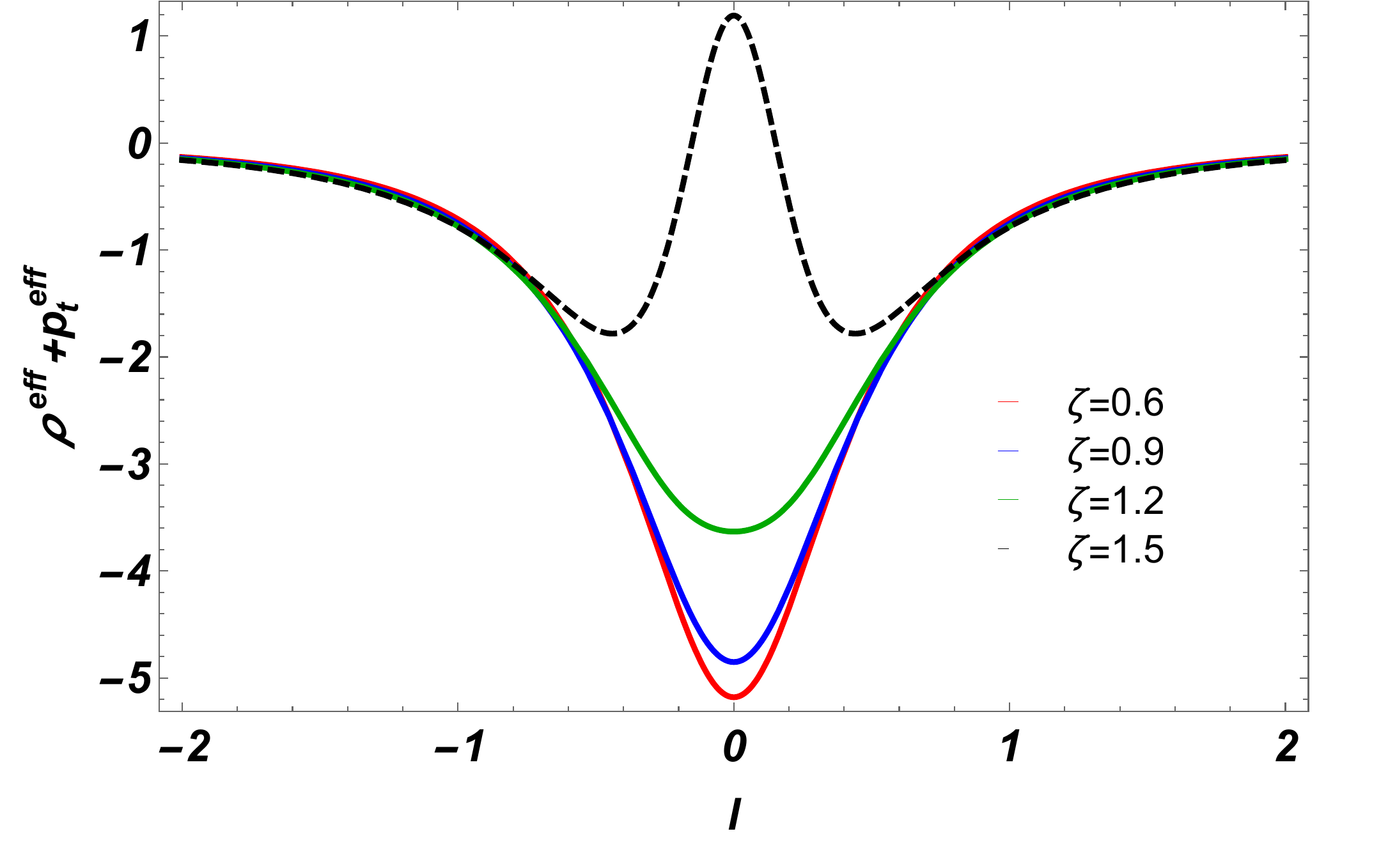}
\caption{Shows the variation of $\rho^{eff}+p^{eff}_{t}$ with Model-I (left) and Model-II (right) generalized embedded wormhole solutions under the effect of van der Waals EOS.}\label{F4}
\end{figure}

\begin{figure}[htb!]
\centering 
\includegraphics[width=7.6cm,height=6.0cm]{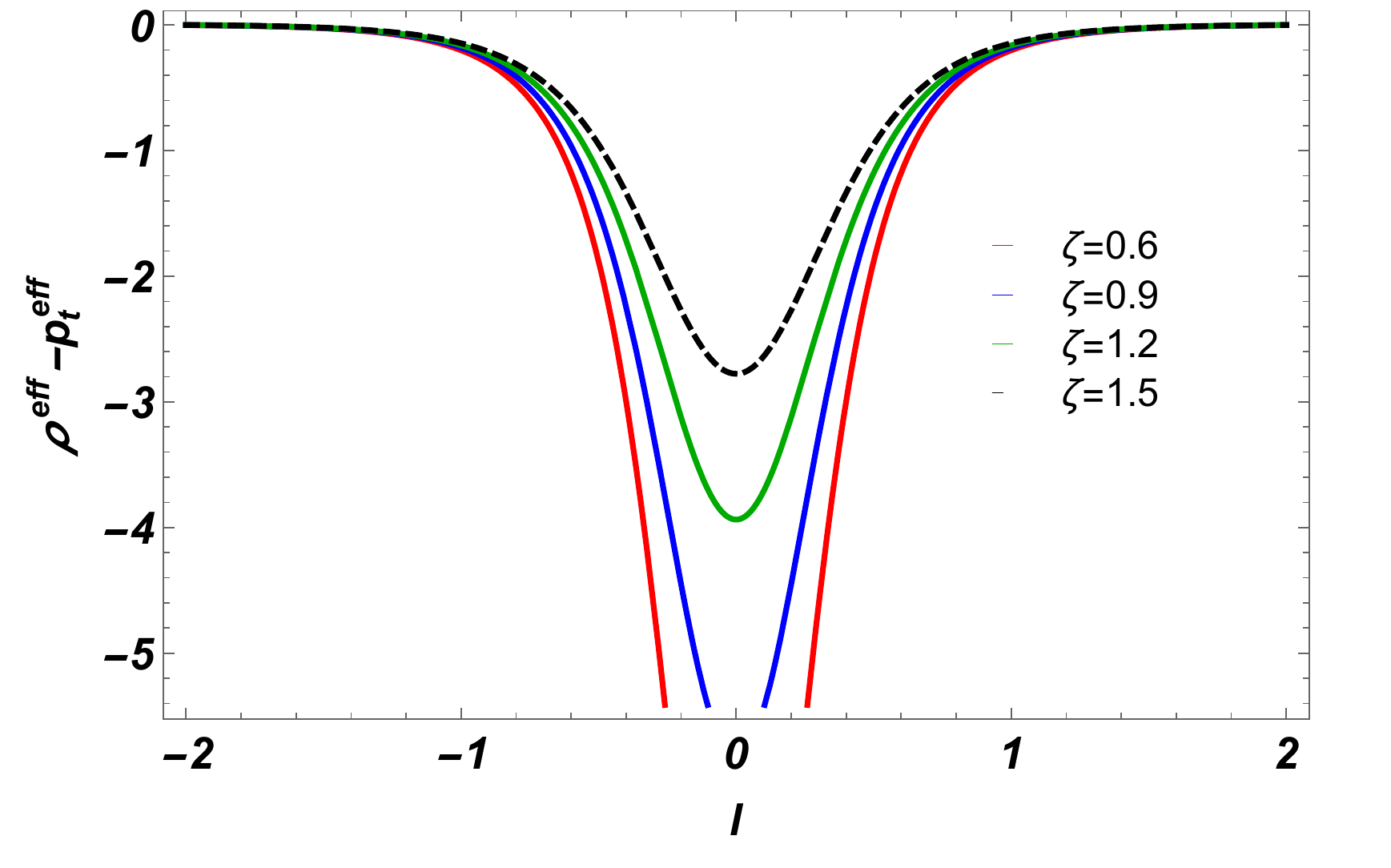} \includegraphics[width=7.6cm,height=6.0cm]{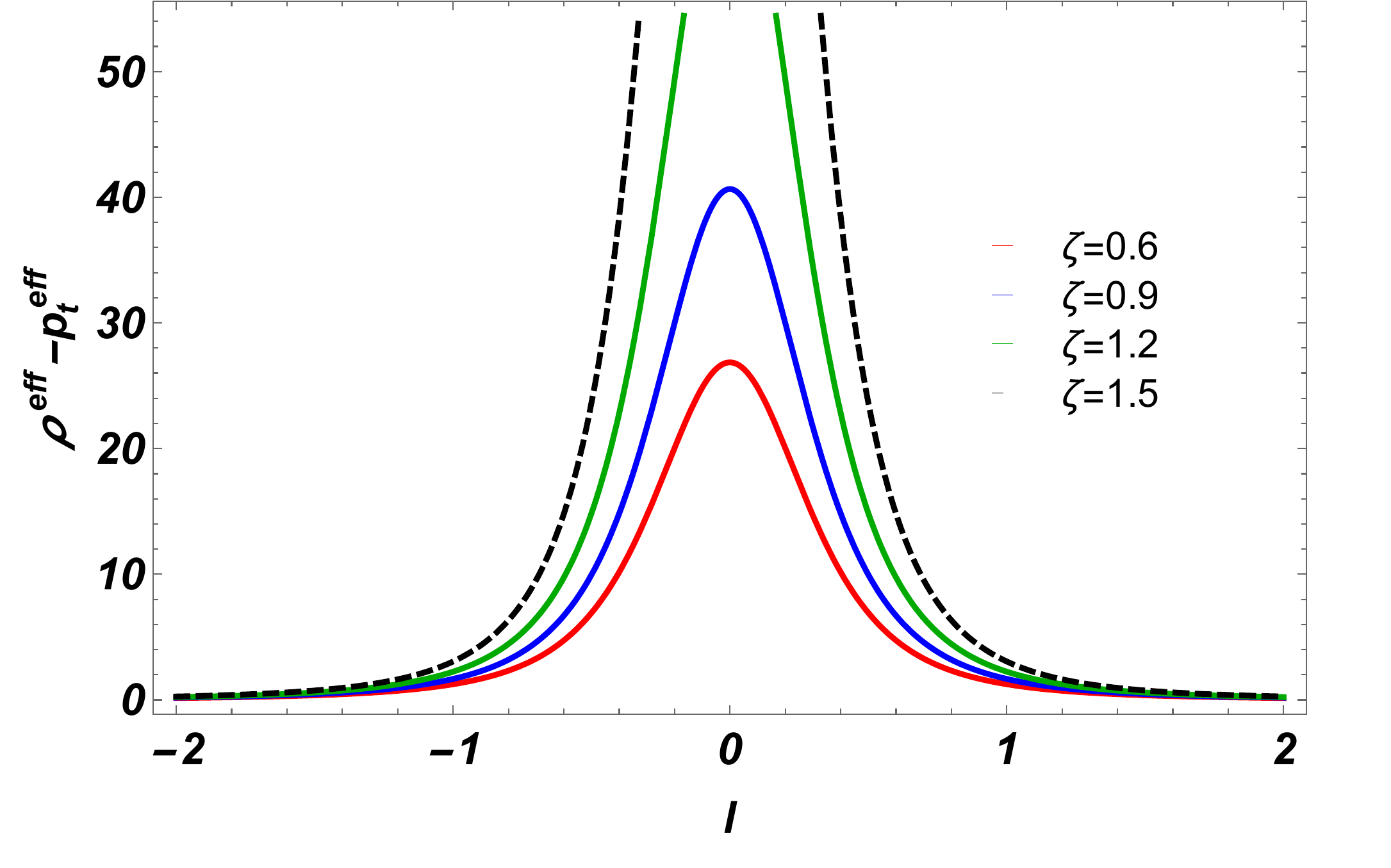}
\caption{Shows the variation of $\rho^{eff}-p^{eff}_{t}$ with Model-I (left) and Model-II (right) generalized embedded wormhole solutions under the effect of van der Waals EOS.}\label{F5}
\end{figure}

\begin{figure}[htb!]
\centering 
\includegraphics[width=7.6cm,height=6.0cm]{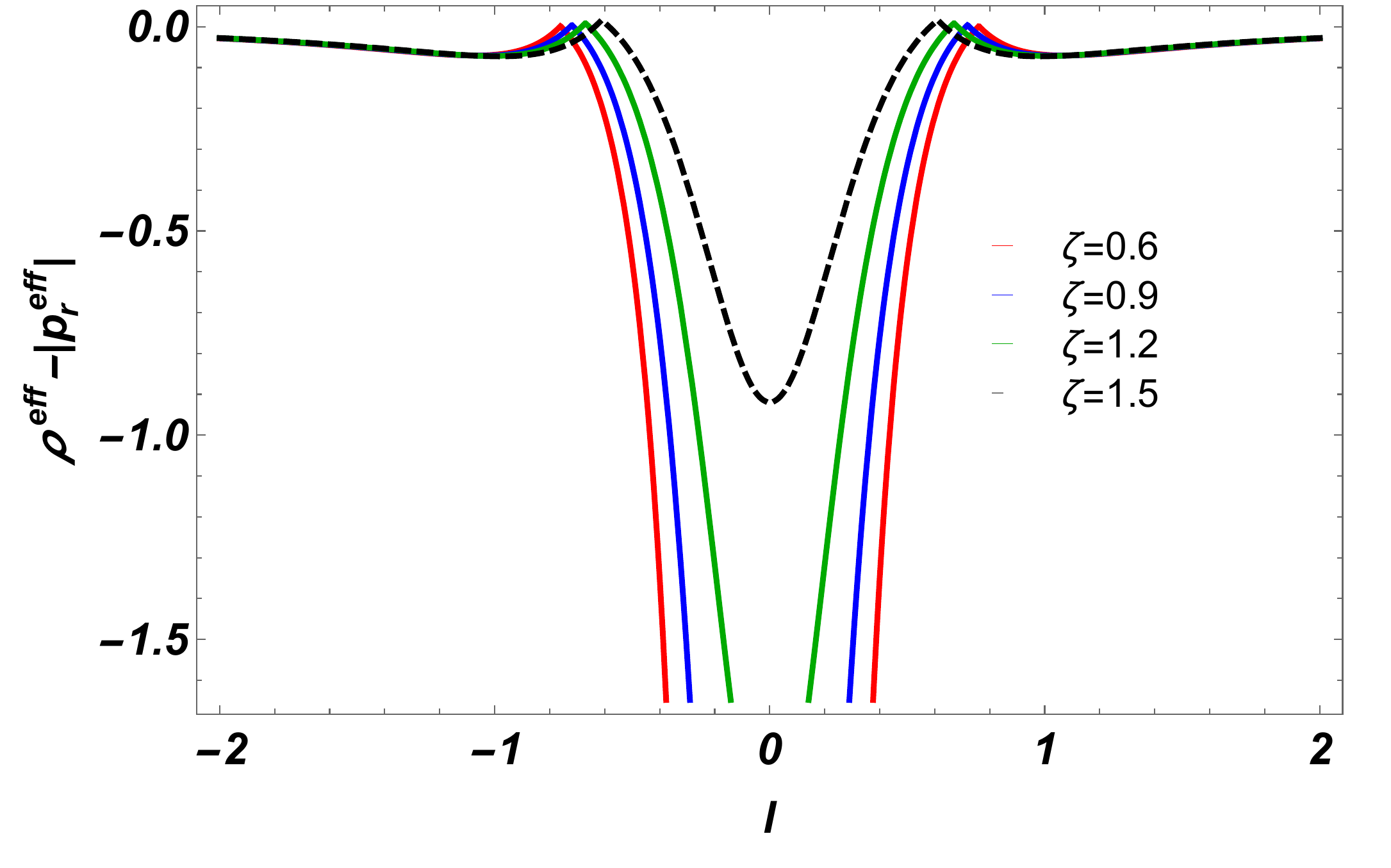} \includegraphics[width=7.6cm,height=6.0cm]{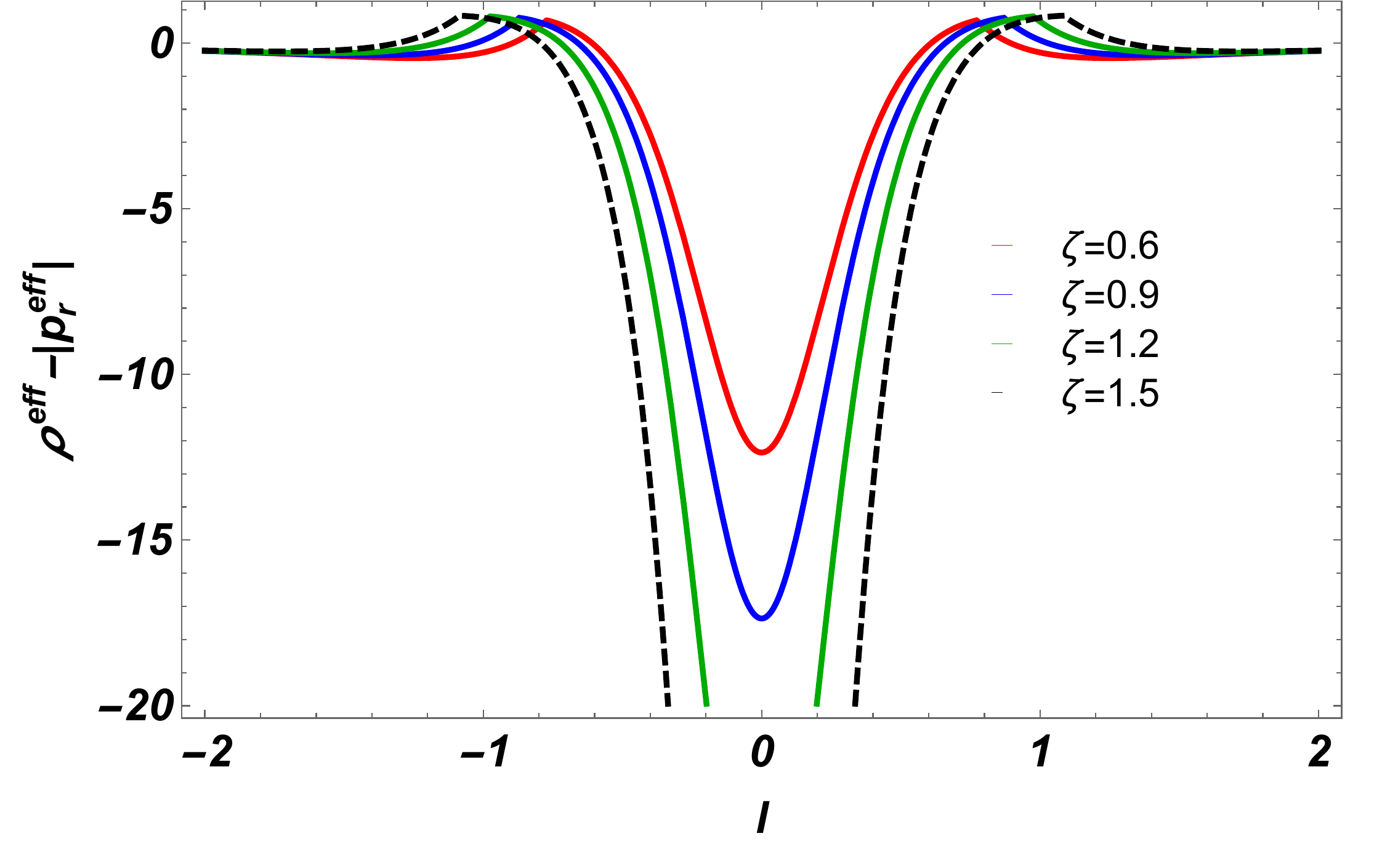}
\caption{Shows the variation of $\rho^{eff}-\mid p^{eff}_{r}\mid$ with Model-I (left) and Model-II (right) generalized embedded wormhole solutions under the effect of van der Waals EOS.}\label{rF5}
\end{figure}

\begin{figure}[htb!]
\centering 
\includegraphics[width=7.6cm,height=6.0cm]{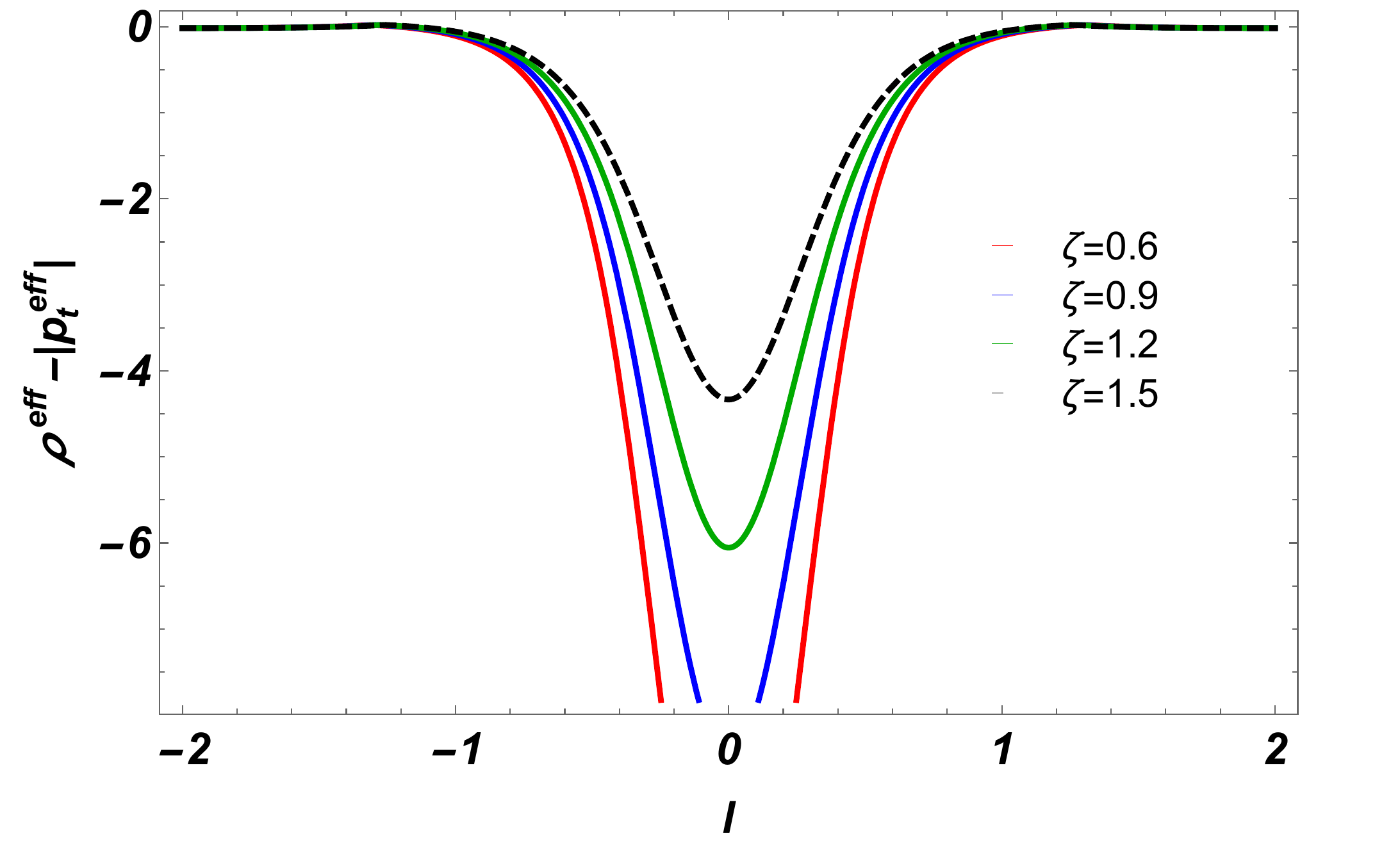} \includegraphics[width=7.6cm,height=6.0cm]{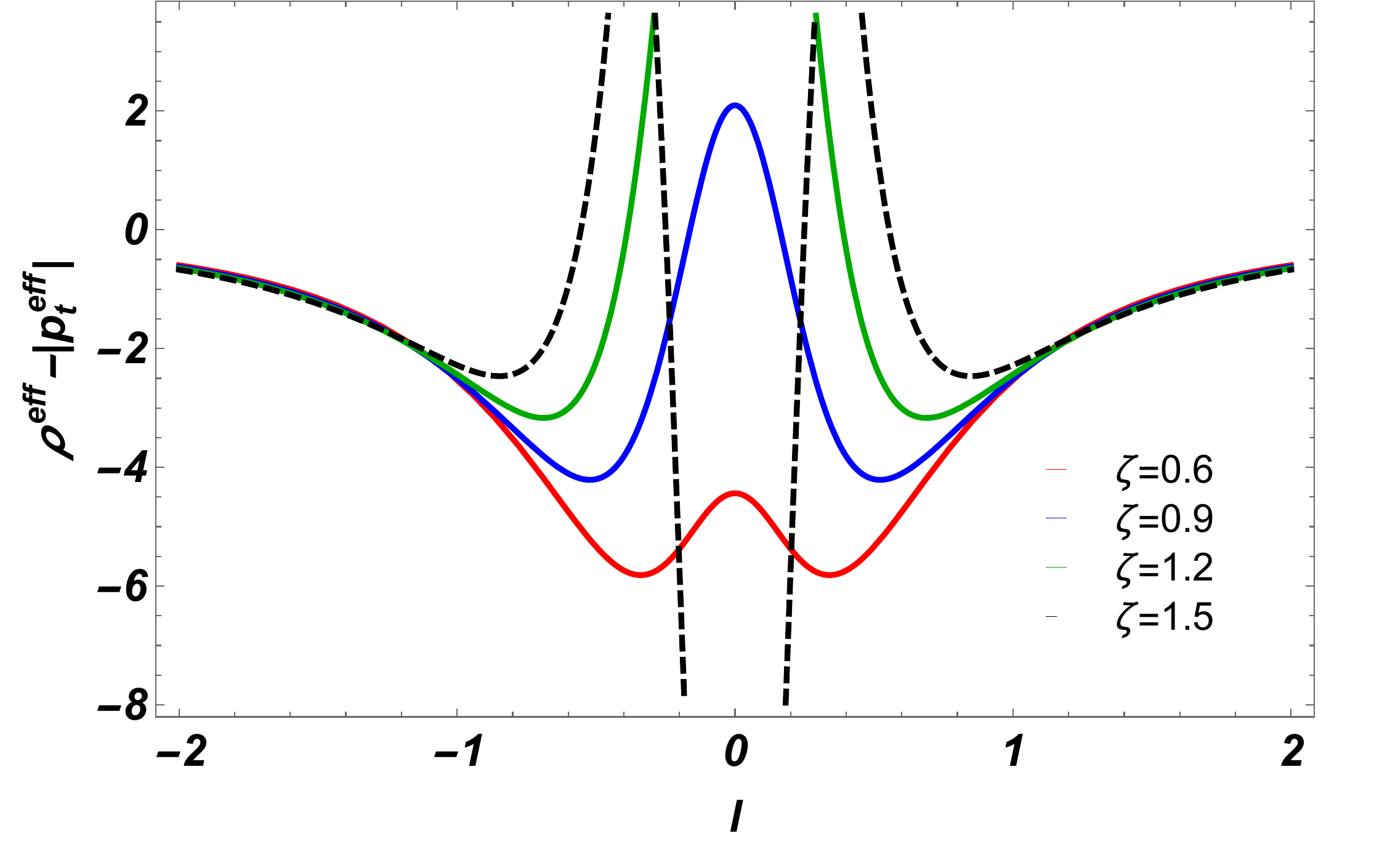}
\caption{Shows the variation of $\rho^{eff}-\mid p^{eff}_{t}\mid$ with Model-I (left) and Model-II (right) generalized embedded wormhole solutions under the effect of van der Waals EOS.}\label{rrF5}
\end{figure}

\begin{figure}[htb!]
\centering 
\includegraphics[width=7.6cm,height=6.0cm]{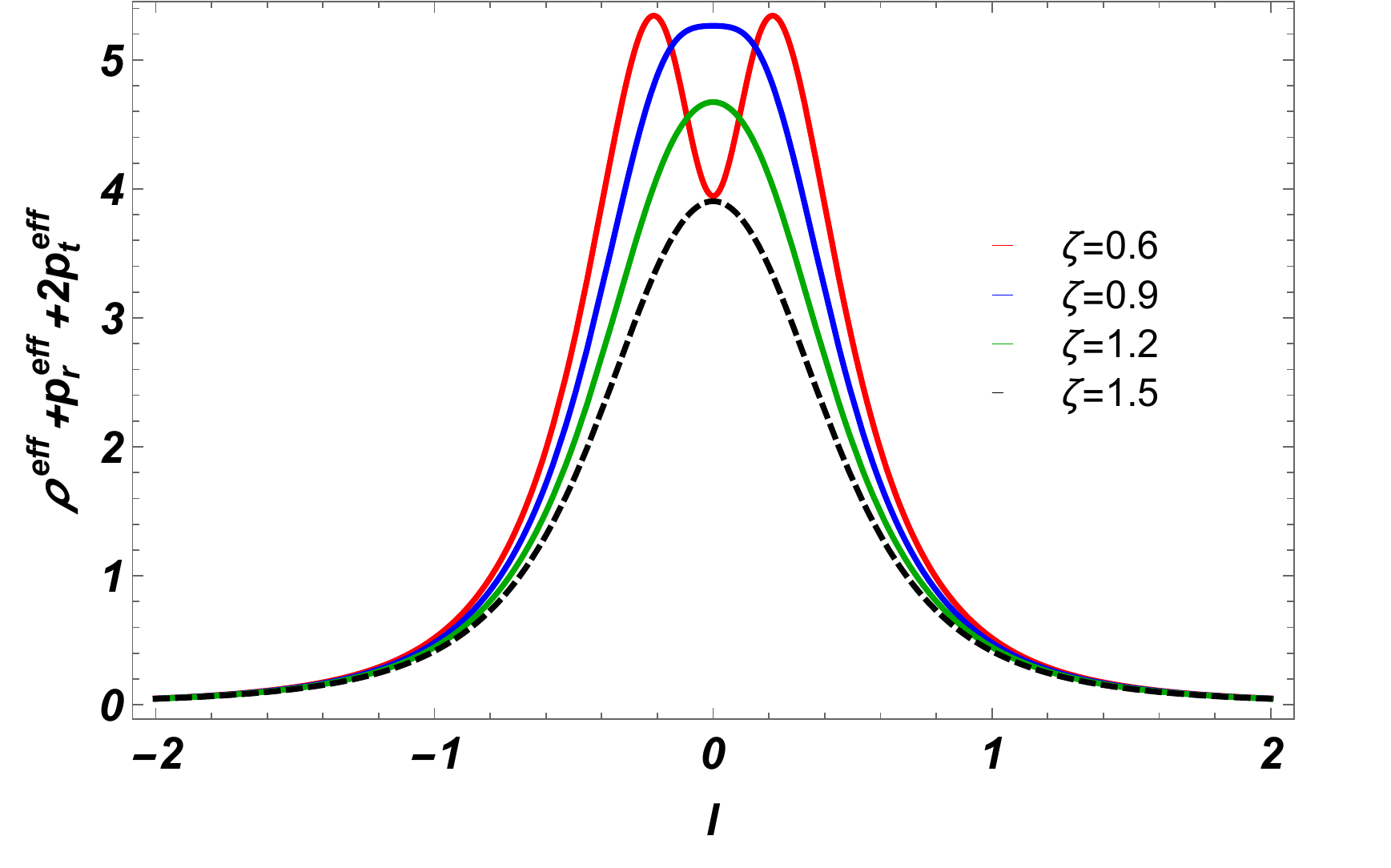} \includegraphics[width=7.6cm,height=6.0cm]{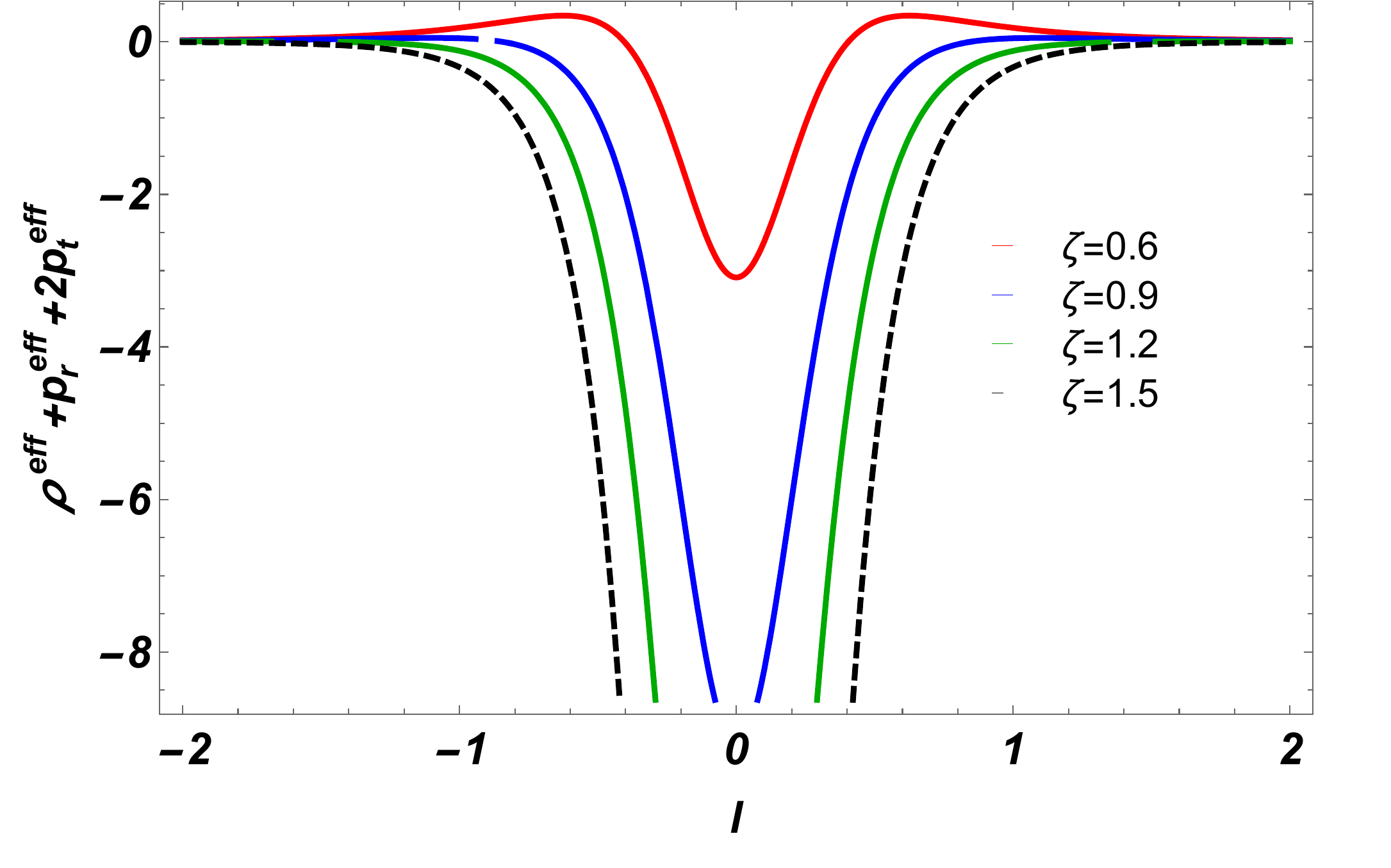}
\caption{Shows the variation of $\rho^{eff}+p^{eff}_{r}+2p^{eff}_{t}$ with Model-I (left) and Model-II (right) generalized embedded wormhole solutions under the effect of van der Waals EOS.}\label{F6}
\end{figure}

\begin{figure}[htb!]
\centering 
\includegraphics[width=7.6cm,height=6.0cm]{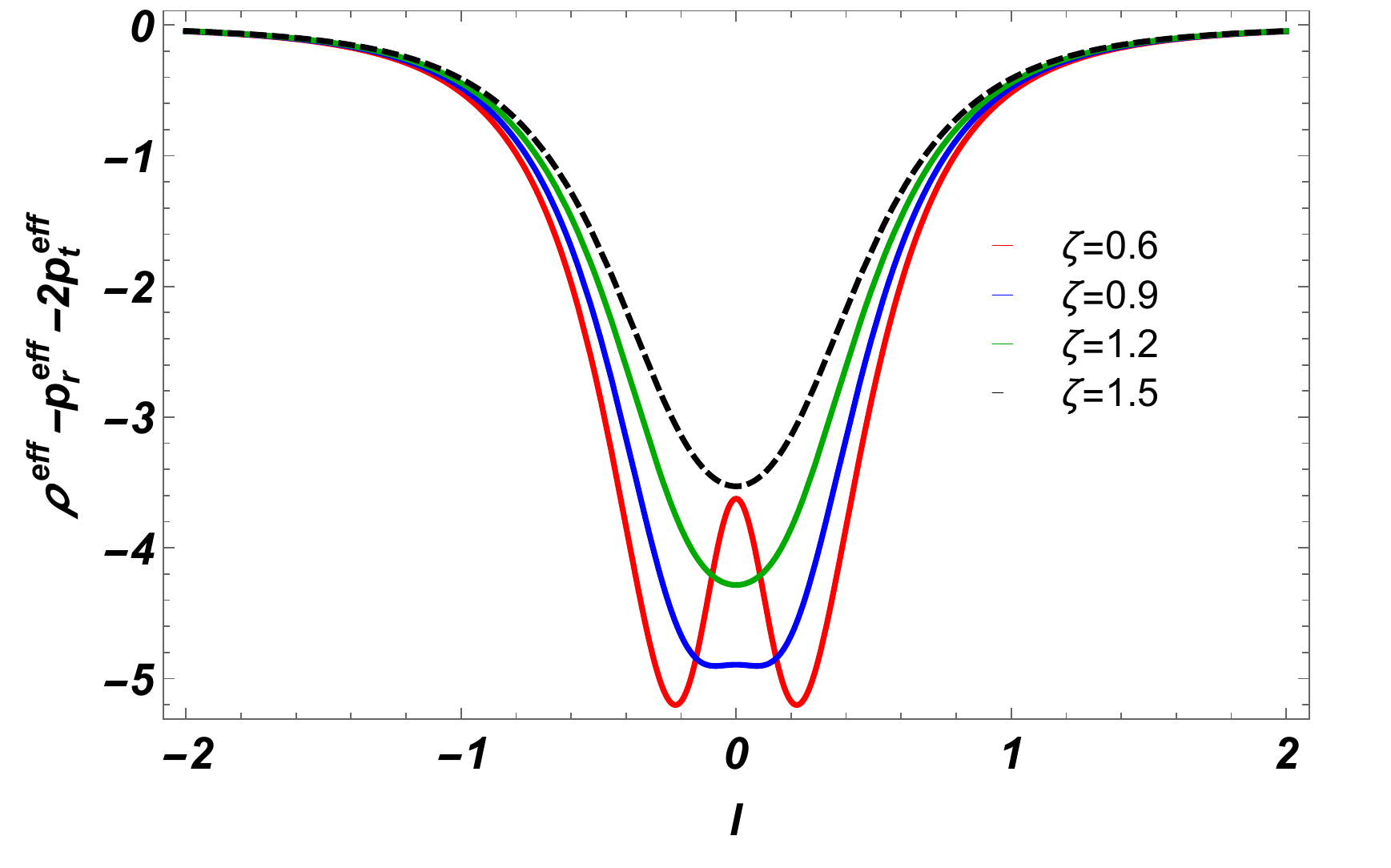} \includegraphics[width=7.6cm,height=6.0cm]{Fig7.pdf}
\caption{Shows the variation of $\rho^{eff}-p^{eff}_{r}-2p^{eff}_{t}$ with Model-I (left) and Model-II (right) generalized embedded wormhole solutions under the effect of van der Waals EOS.}\label{F7}
\end{figure}

\begin{figure}[htb!]
\centering 
\includegraphics[width=7.6cm,height=6.0cm]{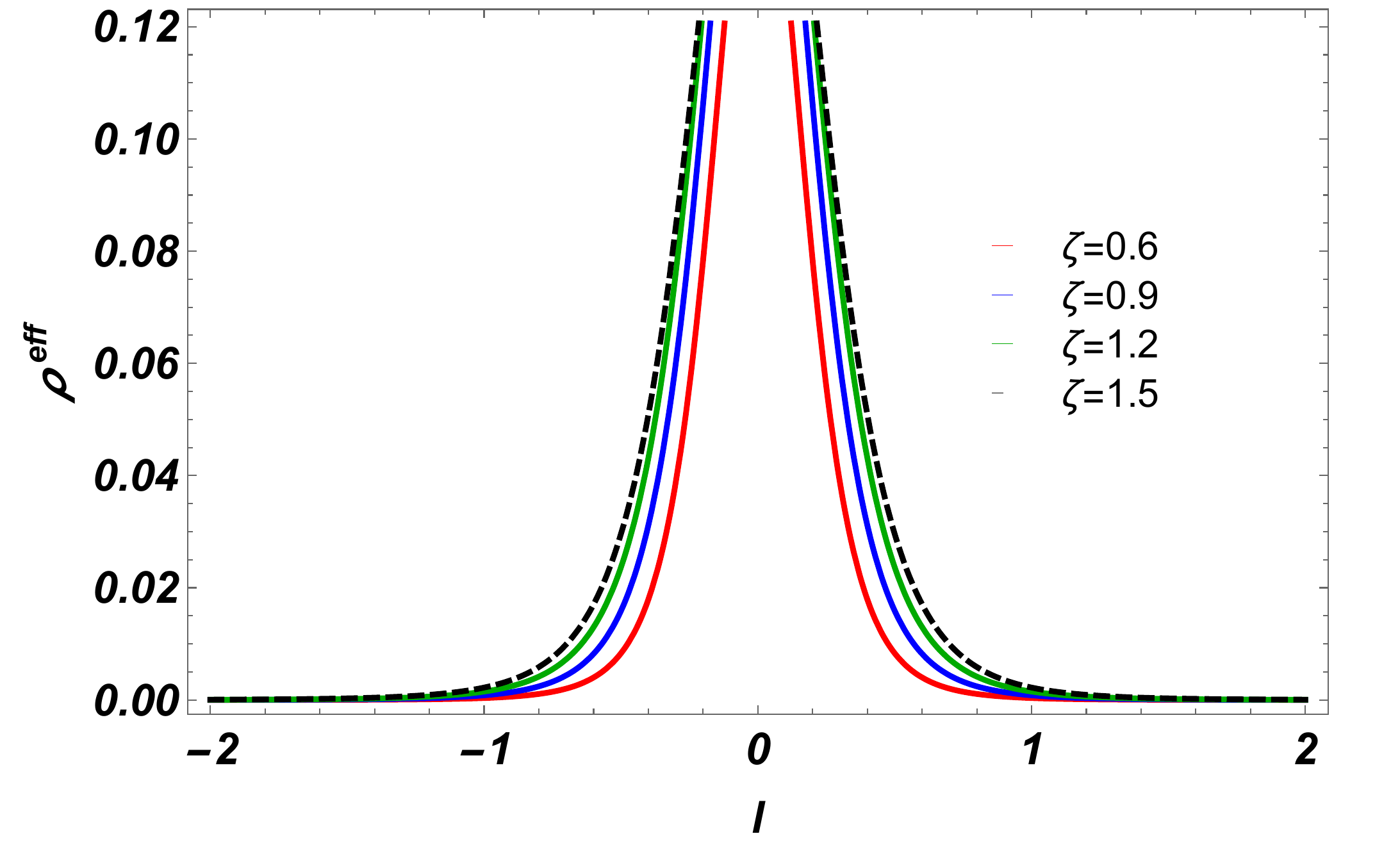} \includegraphics[width=7.6cm,height=6.0cm]{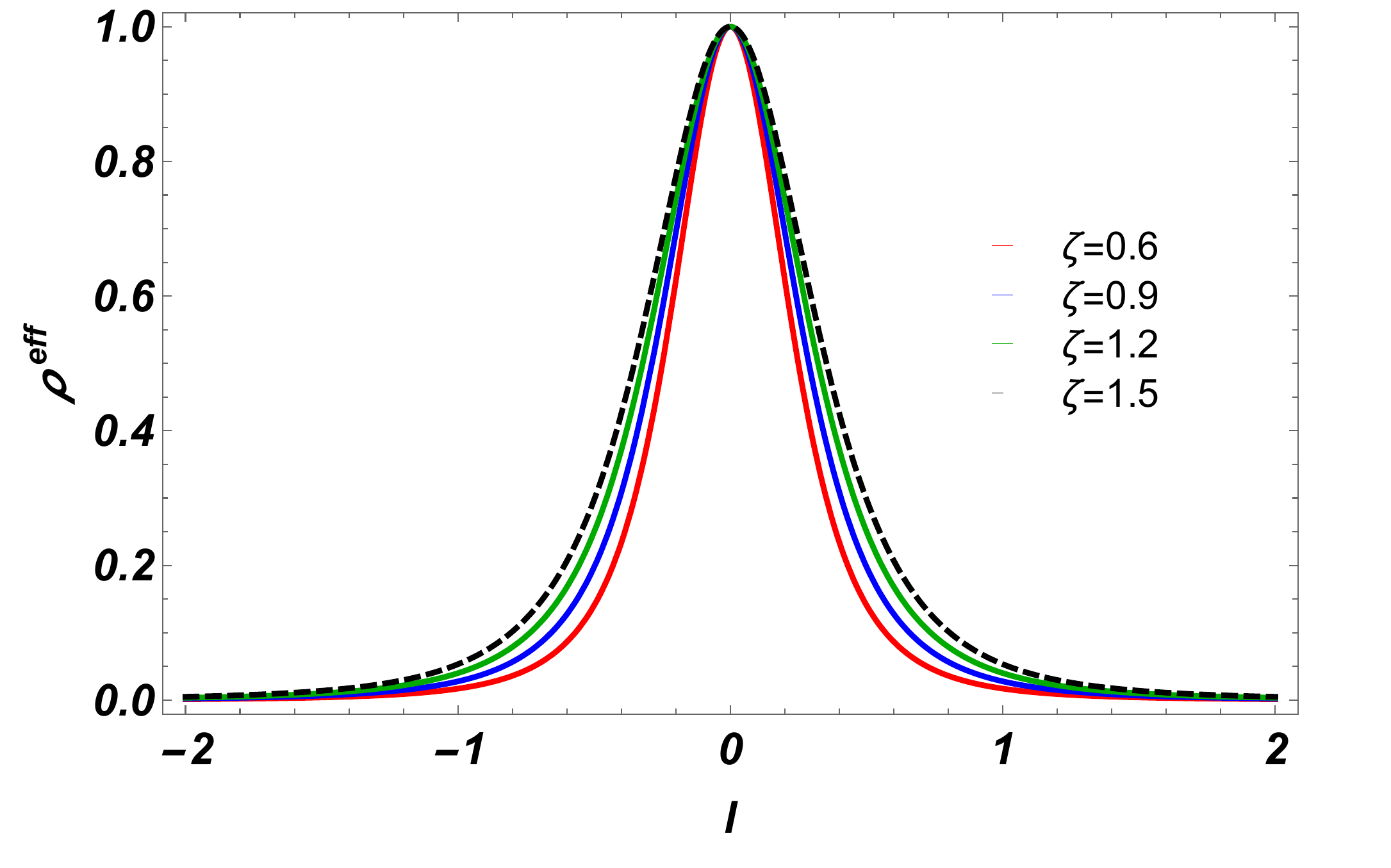}
\caption{Shows the variation of $\rho^{eff}$ with Model-I (left) and Model-II (right) generalized embedded wormhole solutions under the effect of Polytropic EOS.}\label{F8}
\end{figure}

\begin{figure}[htb!]
\centering 
\includegraphics[width=7.6cm,height=6.0cm]{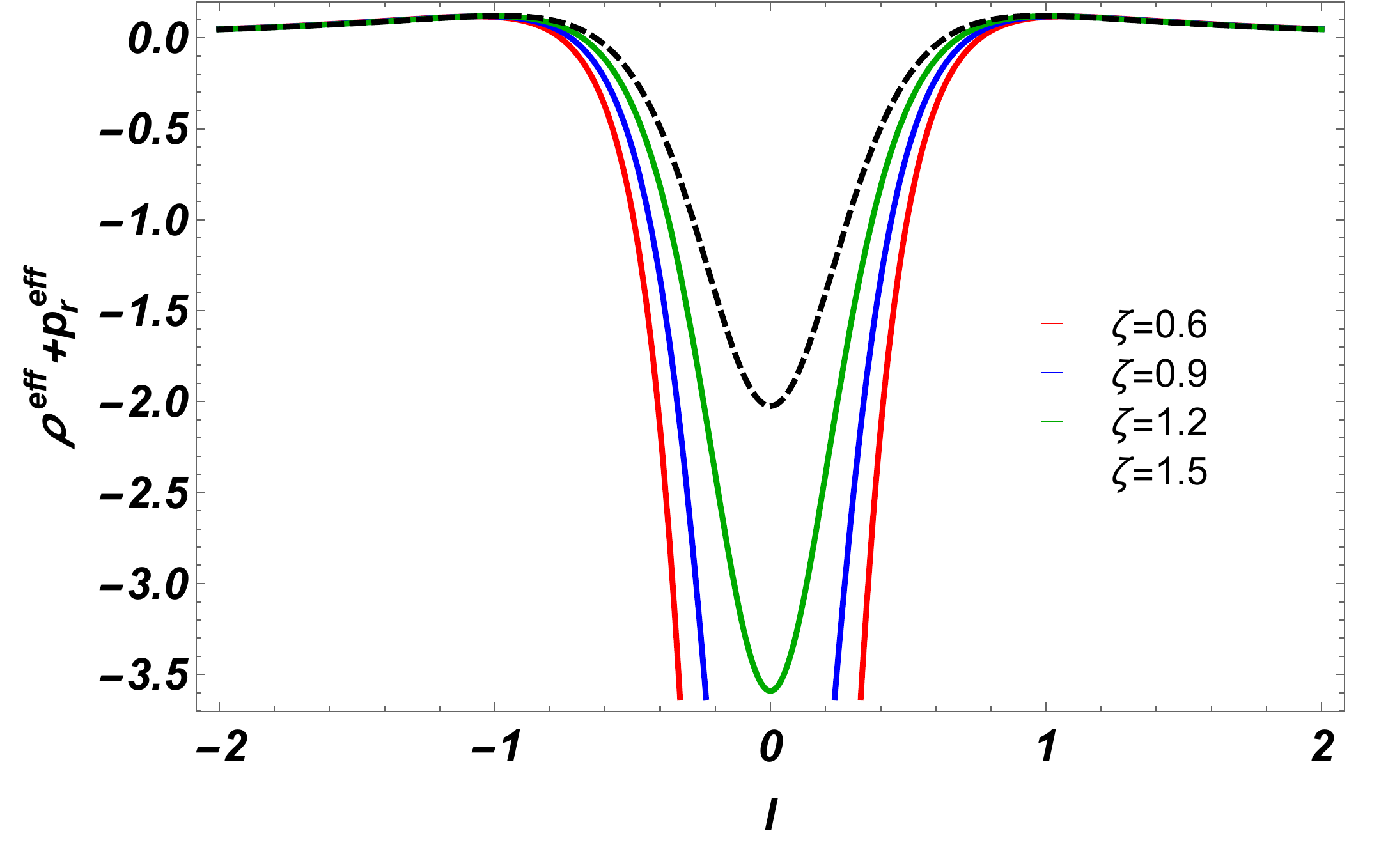} \includegraphics[width=7.6cm,height=6.0cm]{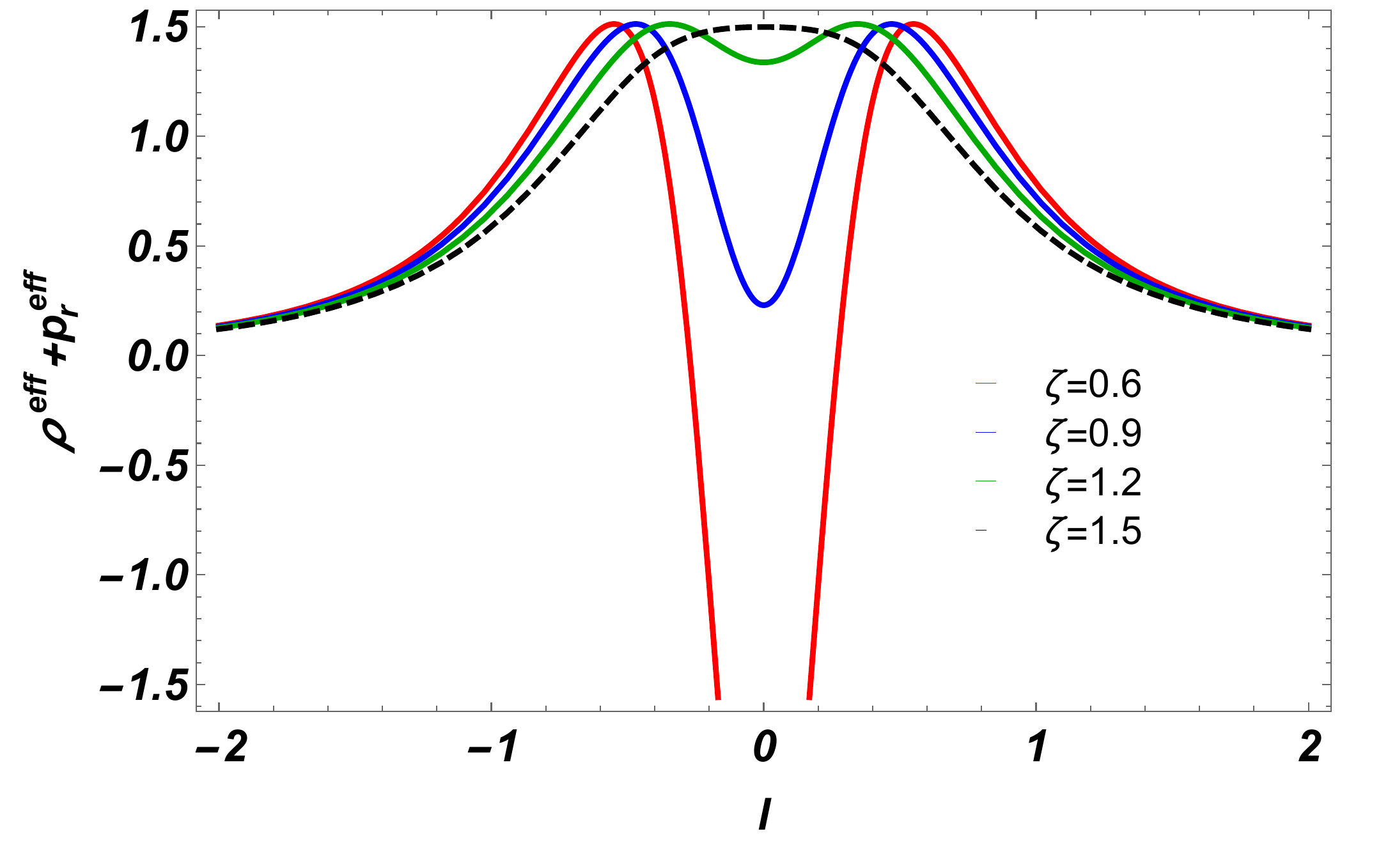}
\caption{Shows the variation of $\rho^{eff}+p^{eff}_{r}$ with Model-I (left) and Model-II (right) generalized embedded wormhole solutions under the effect of Polytropic EOS..}\label{F9}
\end{figure}

\begin{figure}[htb!]
\centering 
\includegraphics[width=7.6cm,height=6.0cm]{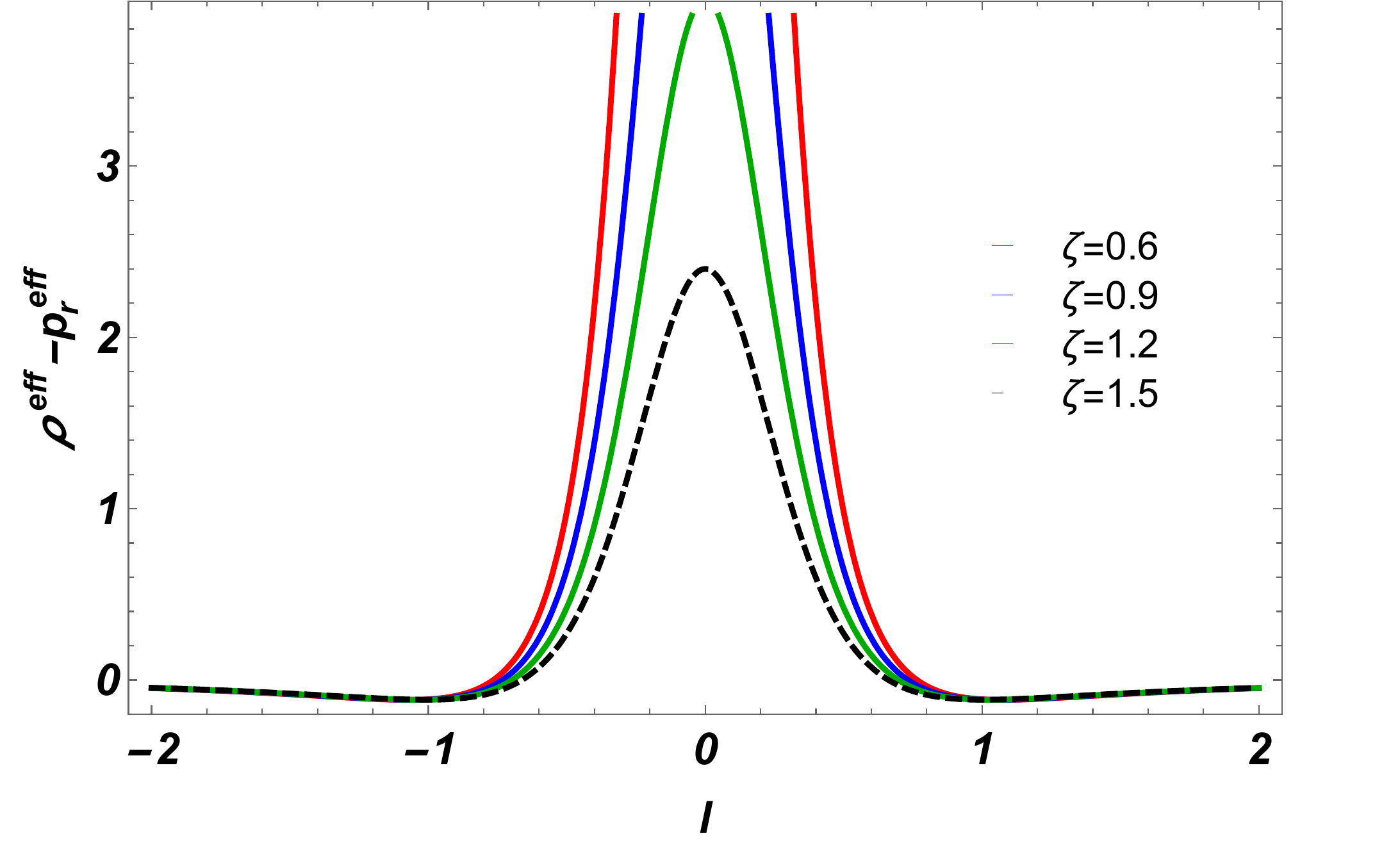} \includegraphics[width=7.6cm,height=6.0cm]{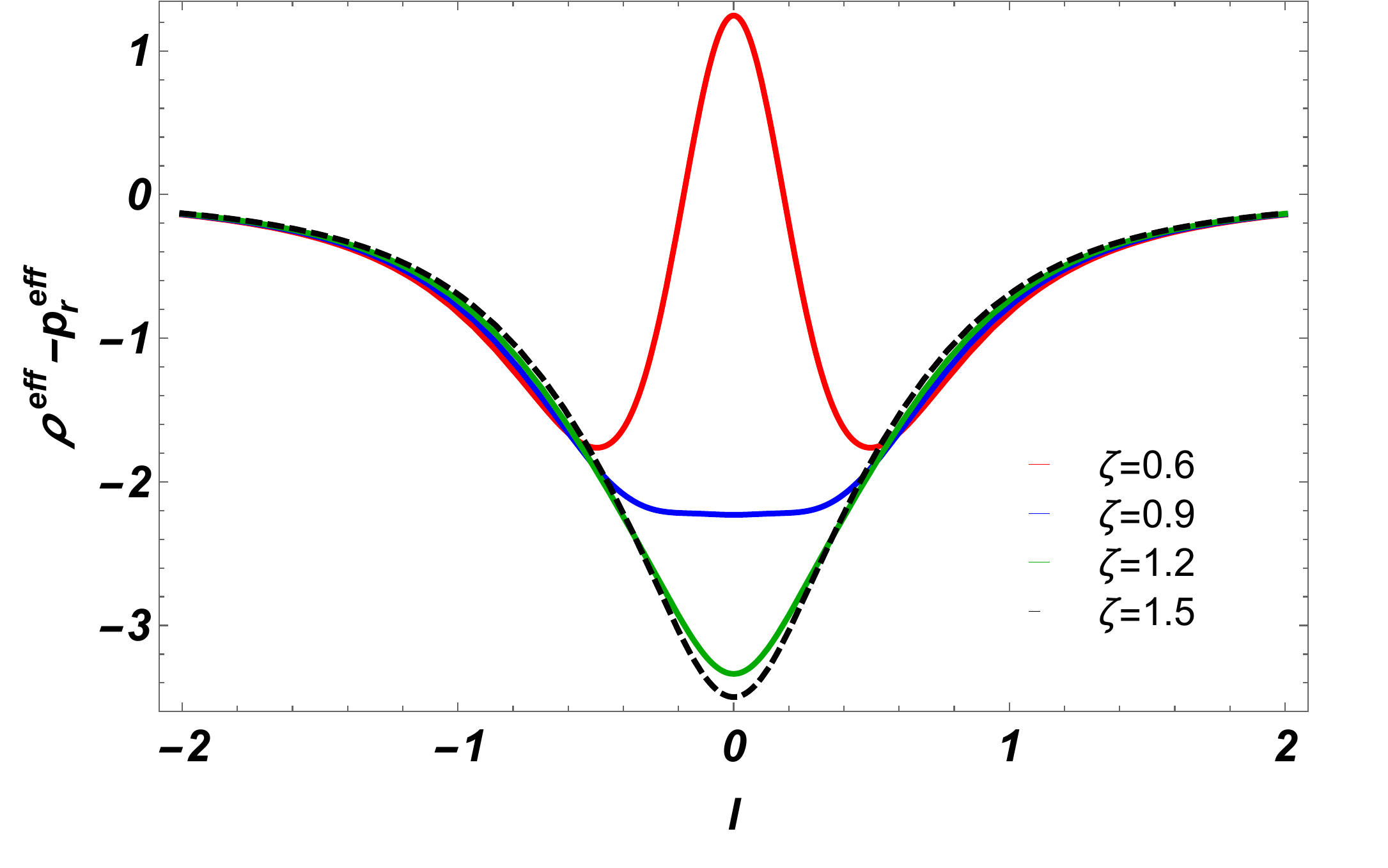}
\caption{Shows the variation of $\rho^{eff}-p^{eff}_{r}$ with Model-I (left) and Model-II (right) generalized embedded wormhole solutions under the effect of Polytropic EOS.}\label{F10}
\end{figure}

\begin{figure}[htb!]
\centering 
\includegraphics[width=7.6cm,height=6.0cm]{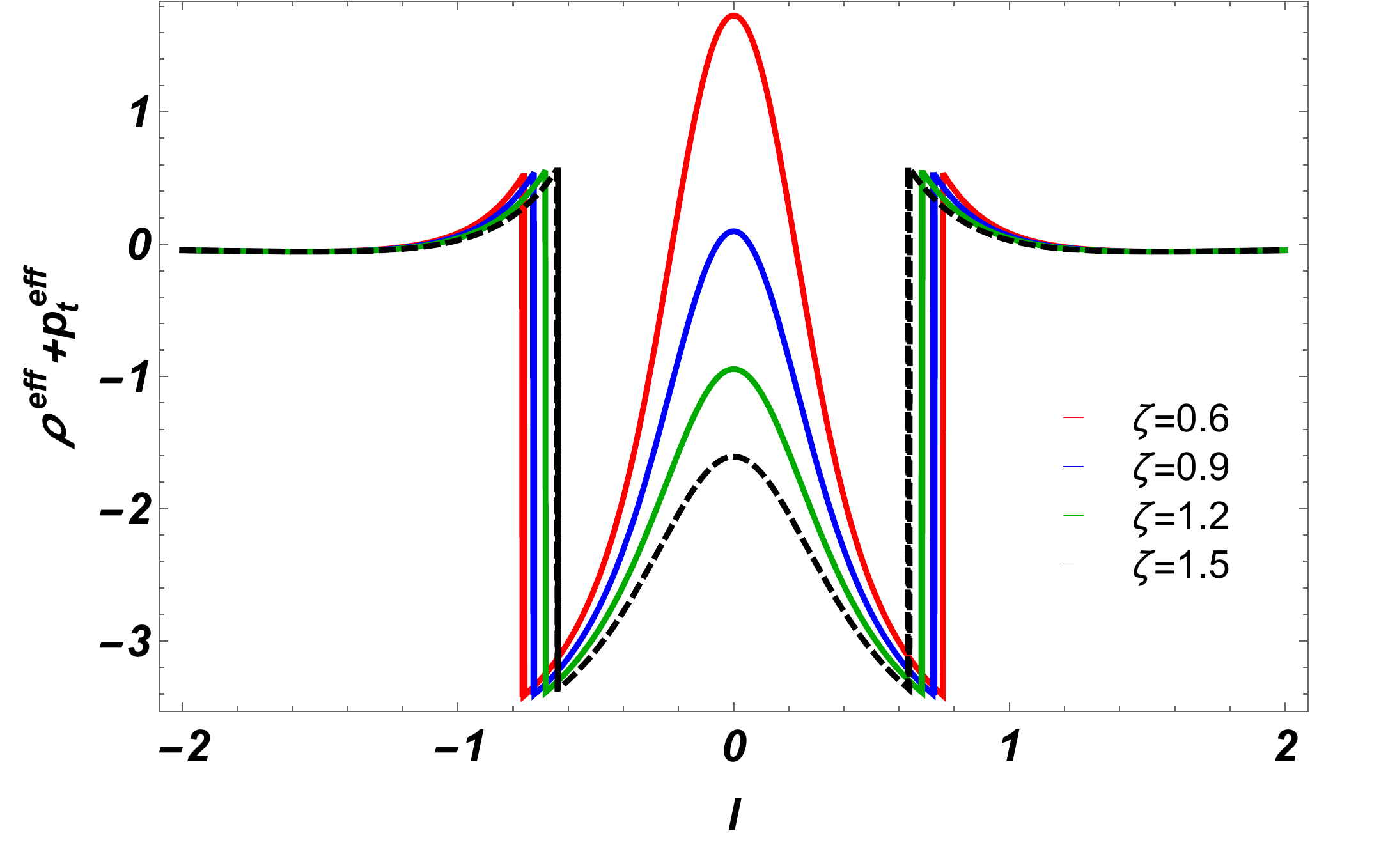} \includegraphics[width=7.6cm,height=6.0cm]{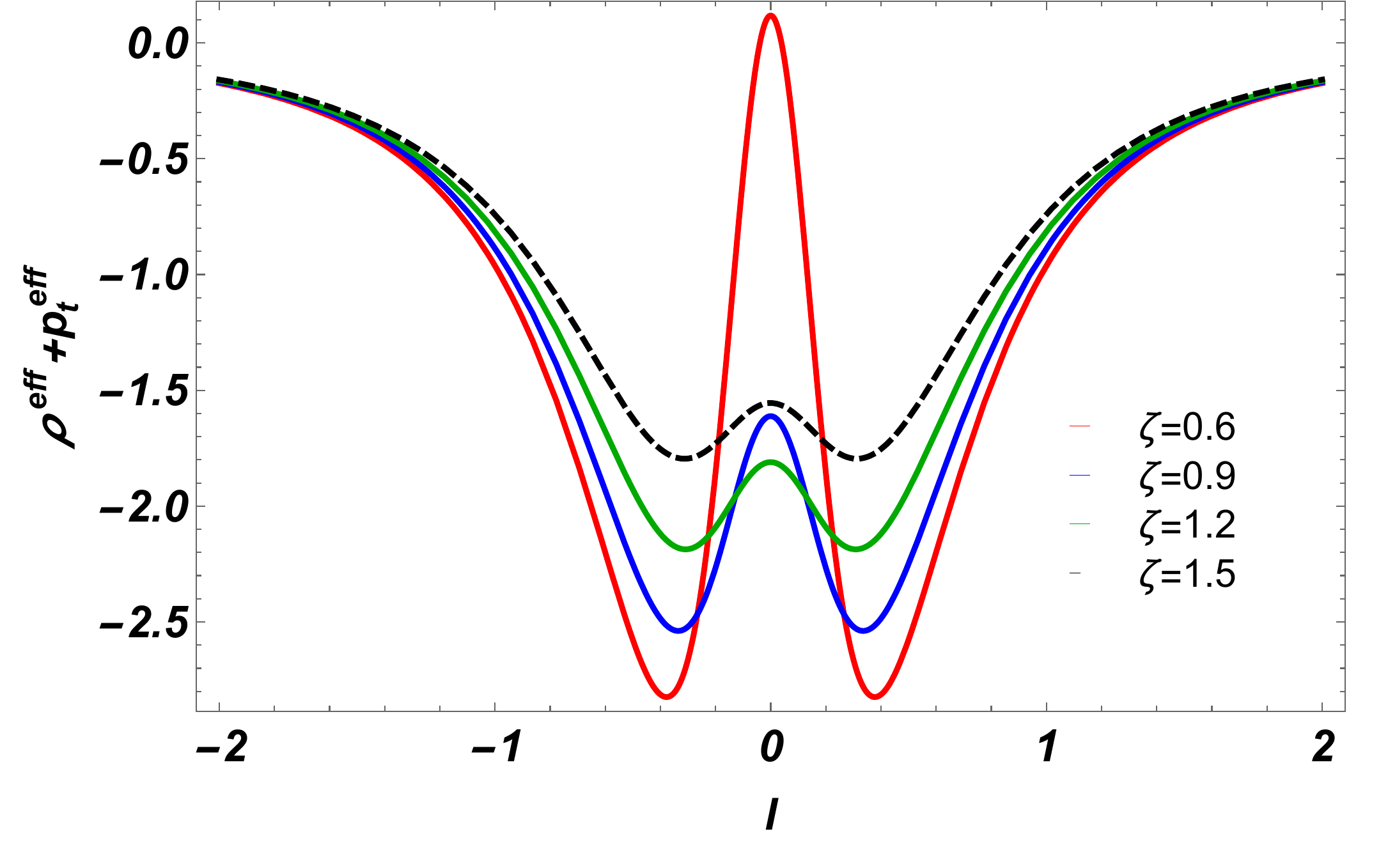}
\caption{Shows the variation of $\rho^{eff}+p^{eff}_{t}$ with Model-I (left) and Model-II (right) generalized embedded wormhole solutions under the effect of Polytropic EOS.}\label{F11}
\end{figure}

\begin{figure}[htb!]
\centering 
\includegraphics[width=7.6cm,height=6.0cm]{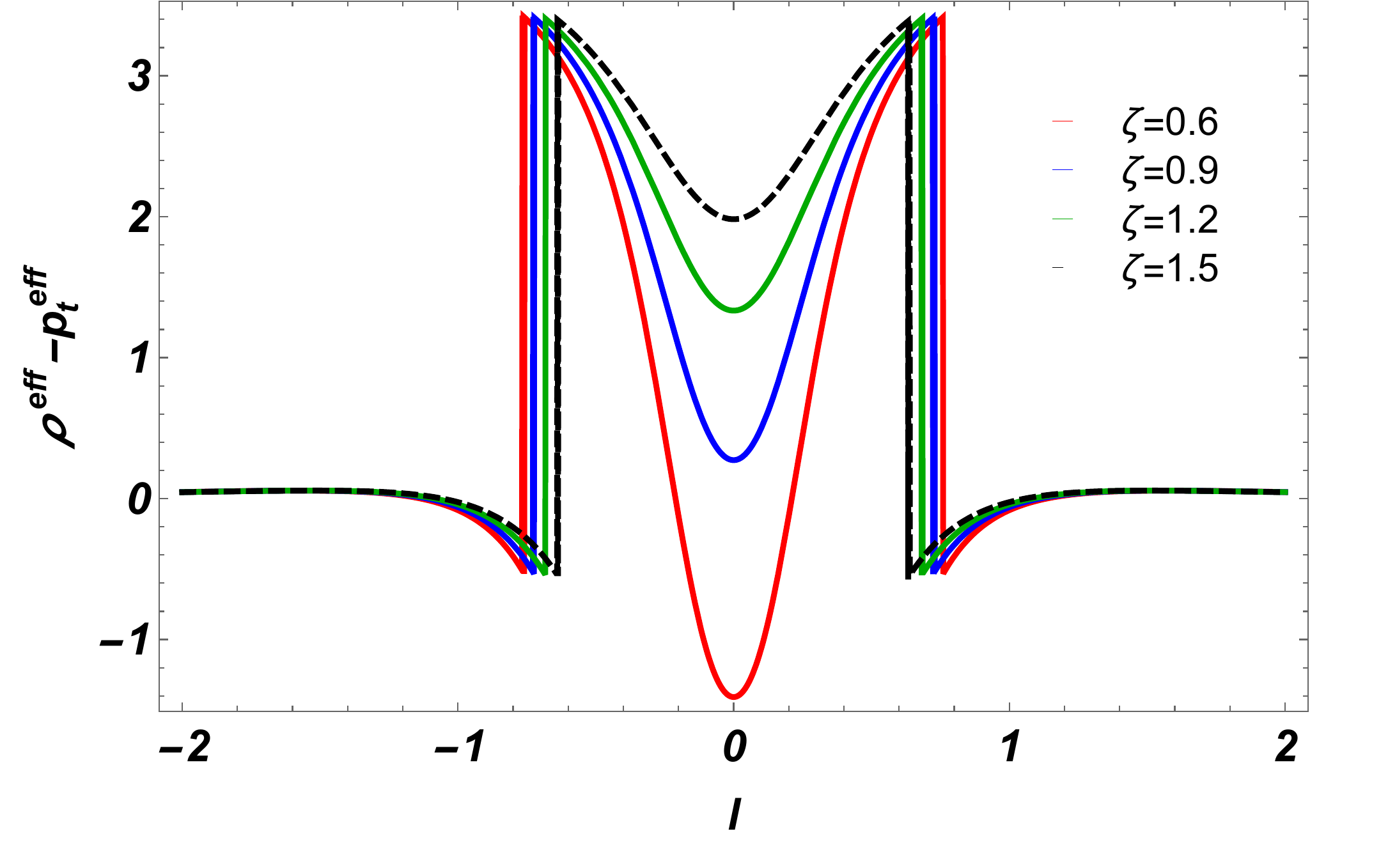} \includegraphics[width=7.6cm,height=6.0cm]{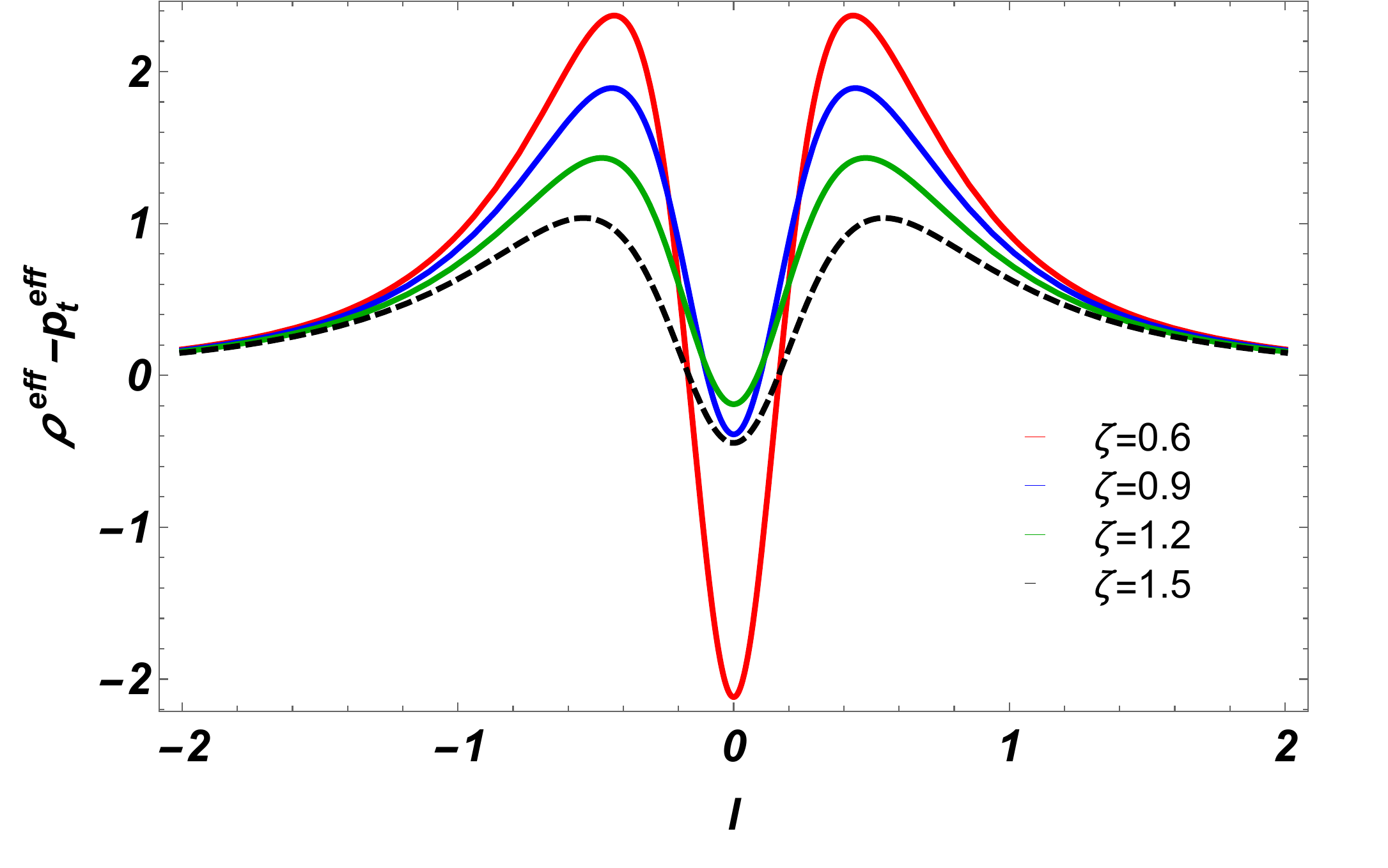}
\caption{Shows the variation of $\rho^{eff}-p^{eff}_{t}$ with Model-I (left) and Model-II (right) generalized embedded wormhole solutions under the effect of Polytropic EOS.}\label{F12}
\end{figure}


\section{Energy Conditions}\label{sec5}

The energy conditions of the physically realistic matter configuration can be generated from the Raychaudhuri equations. The Raychaudhuri equations~\cite{rrr1} express the temporal evolution of the scalar expansion ($\psi$) for timelike ($u^{i}$) and null ($\zeta_{i}$) geodesics as congruences which can be provided as
\begin{equation}
\label{r1a}
\frac{d\psi}{d\tau}-\mathcal{W}_{ij}\,\mathcal{W}^{ij}+\sigma_{ij}\sigma^{ij}+\frac{1}{3}\psi^2+R_{ij}u^i\,u^j=0,
\end{equation}

\begin{equation}
\label{r2a}
\frac{d\psi}{d\tau}-\mathcal{W}_{ij}\,\mathcal{W}^{ij}+\sigma_{ij}\sigma^{ij}+\frac{1}{2}\psi^2+R_{ij}\zeta^i\zeta^j=0,
\end{equation}
where $\sigma^{ij}$ and $\mathcal{W}_{ij}$ are the shear and the rotation associated with the vector field $u^i$ respectively. For the attractive nature of gravity ($\psi<0$) and neglecting the quadratic terms, the Raychaudhuri Eqs. (\ref{r1a}) and (\ref{r2a}) satisfy the following conditions:
\begin{equation}
\label{3a}
R_{ij}u^i\,u^j\geq0,
\end{equation}

\begin{equation}
\label{4a}
R_{ij}\zeta^i\zeta^j\geq0.
\end{equation}

Because we are dealing with anisotropic fluid matter distribution, the energy conditions recovered for GR case as\\
$\bullet$ Strong energy conditions (SEC) if $\rho^{eff}+p_k^{eff}\geq0$, $\rho^{eff}+\sum_k p_k^{eff}\geq0$, $\forall k$.\\
$\bullet$ Dominant energy conditions (DEC) if $\rho^{eff}\geq0$, $\rho^{eff}\pm p_k^{eff}\geq0$, $\forall k$.\\
$\bullet$ Weak energy conditions (WEC) if $\rho^{eff}\geq0$, $\rho^{eff}+p_k^{eff}\geq0$, $\forall k$.\\
$\bullet$ Null energy condition (NEC) if $\rho^{eff}+p_k^{eff}\geq0$, $\forall k$.\\
$\bullet$ Trace energy conditions (TEC) if $\rho^{eff}-p_k^{eff}\geq0$, $\rho^{eff}-\sum_k p_k^{eff}\geq0$, $\forall k$,
where $\rho^{eff}$ and $p^{eff}$ describe the energy density and pressure, respectively and $k=r,t$.

Here, we are going to provide a detailed discussion regarding our calculated results through energy conditions in the current study. In this analysis, our motive is to calculate the energy conditions under the effect of the cloud of string parameter and quintessence field by plugging two different and physically viable embedded wormhole shape functions with a specific choice of the red shift function. In order to calculate the involved quintessence energy density, we use the most famous and novel EOS like van der Waals EOS involving three unknown parameters. Later, we make use of another famous EOS, say polytropic EOS. Finally, we have two specific shape functions and two different EOS. 
 
Let us now discuss the graphic analysis of all the energy conditions including the trace energy conditions. The positive behavior of energy density is a very necessary component for the existence of wormhole geometry. From Figs. \ref{F1} and \ref{F8}, the behavior of energy density for both the embedded shape functions in the framework of two well-known EOS, like the van der Waals and polytropic EOS, under the effect of a cloud of string parameter and quintessence field can be perceived. In all cases, the energy density, say $\rho^{eff}$, remains positive throughout the wormhole configurations for the radial distance ranging $-2\leq l\leq2$. It is interesting to note here that $\rho^{eff}$ is seen as denser for Model-II as compared to Model-I for both the van der Waals and polytropic EOS.

The graphic development of NEC, i.e. $\rho^{eff}+p^{eff}_{r}$ and $\rho^{eff}+p^{eff}_{t}$, is provided in Figs. \ref{F2} and \ref{F4}, respectively for van der Waals EOS and in Figs. \ref{F9} and \ref{F11}, respectively for polytropic EOS. For Model-I it  is noticed that the energy condition strongly violated due to its negative behavior for $-2\leq l\leq2$, while for Model-II, it is satisfied with the positive behavior except one specific value of $\zeta=0.6$ near the wormhole throat $S_{0}$ for polytropic EOS. 

\begin{figure}[htb!]
\centering 
\includegraphics[width=7.6cm,height=6.0cm]{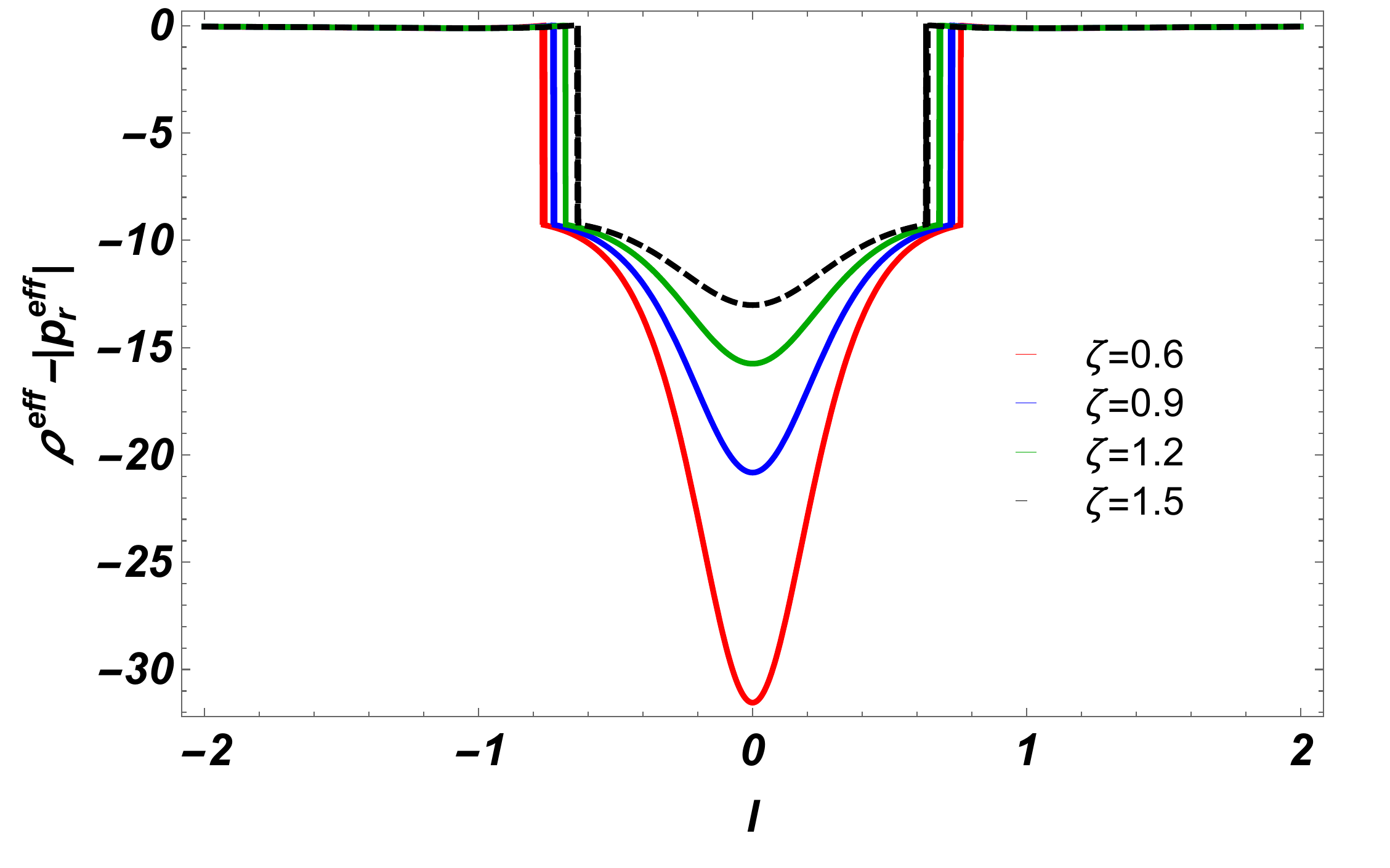} \includegraphics[width=7.6cm,height=6.0cm]{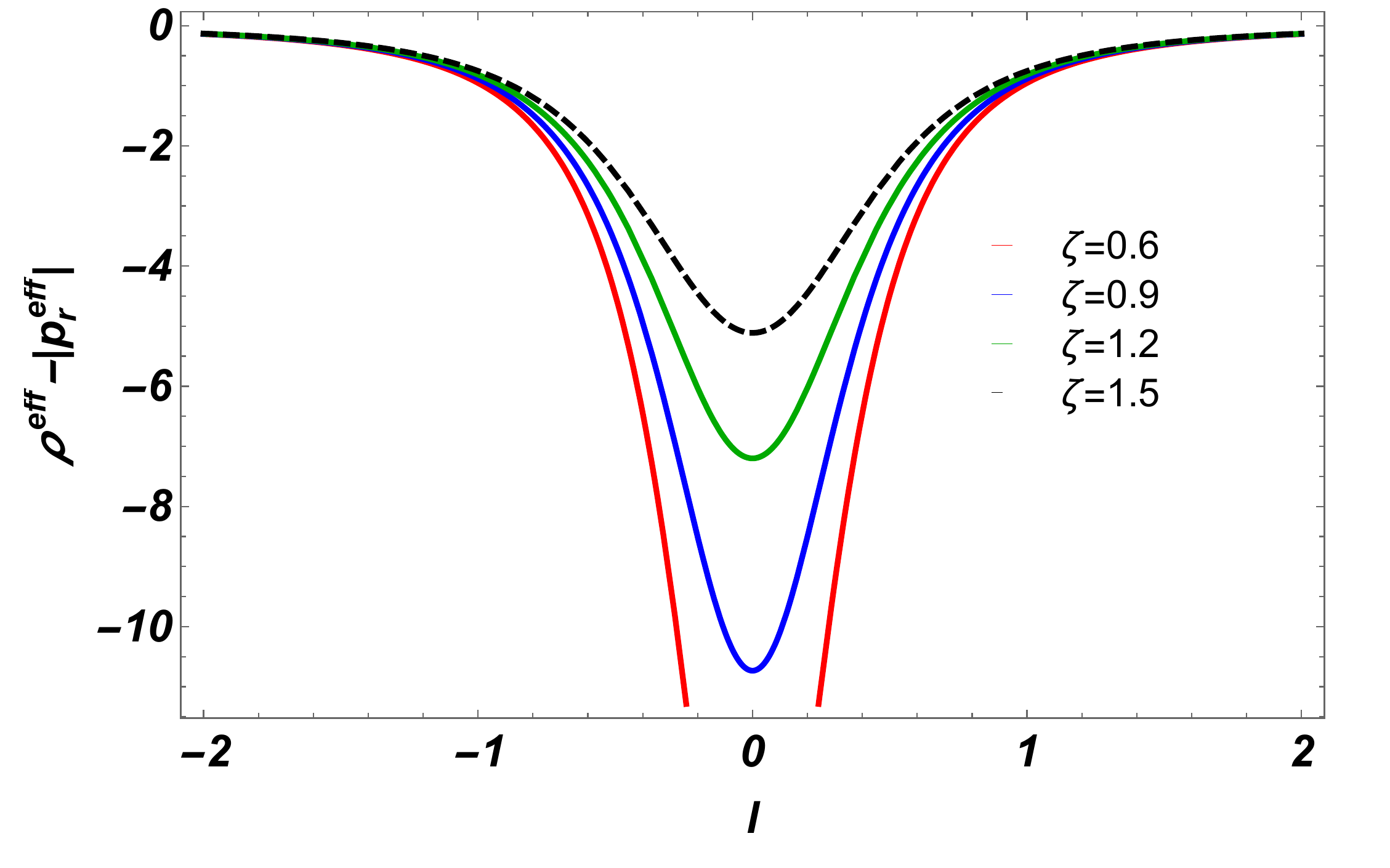}
\caption{Shows the variation of $\rho^{eff}-\mid p^{eff}_{r}\mid$ with Model-I (left) and Model-II (right) generalized embedded wormhole solutions under the effect of Polytropic EOS.}\label{rF12}
\end{figure}

\begin{figure}[htb!]
\centering 
\includegraphics[width=7.6cm,height=6.0cm]{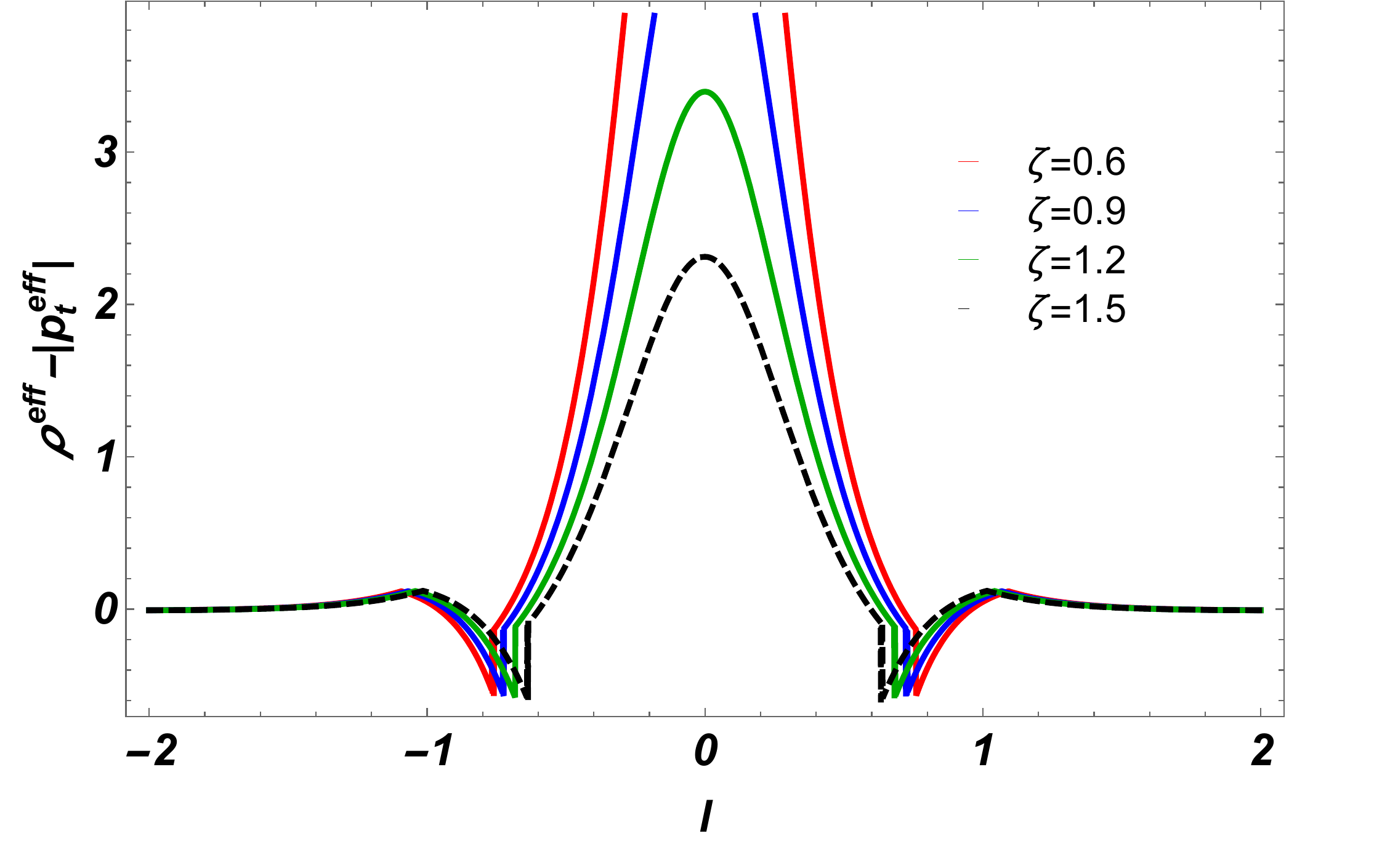} \includegraphics[width=7.6cm,height=6.0cm]{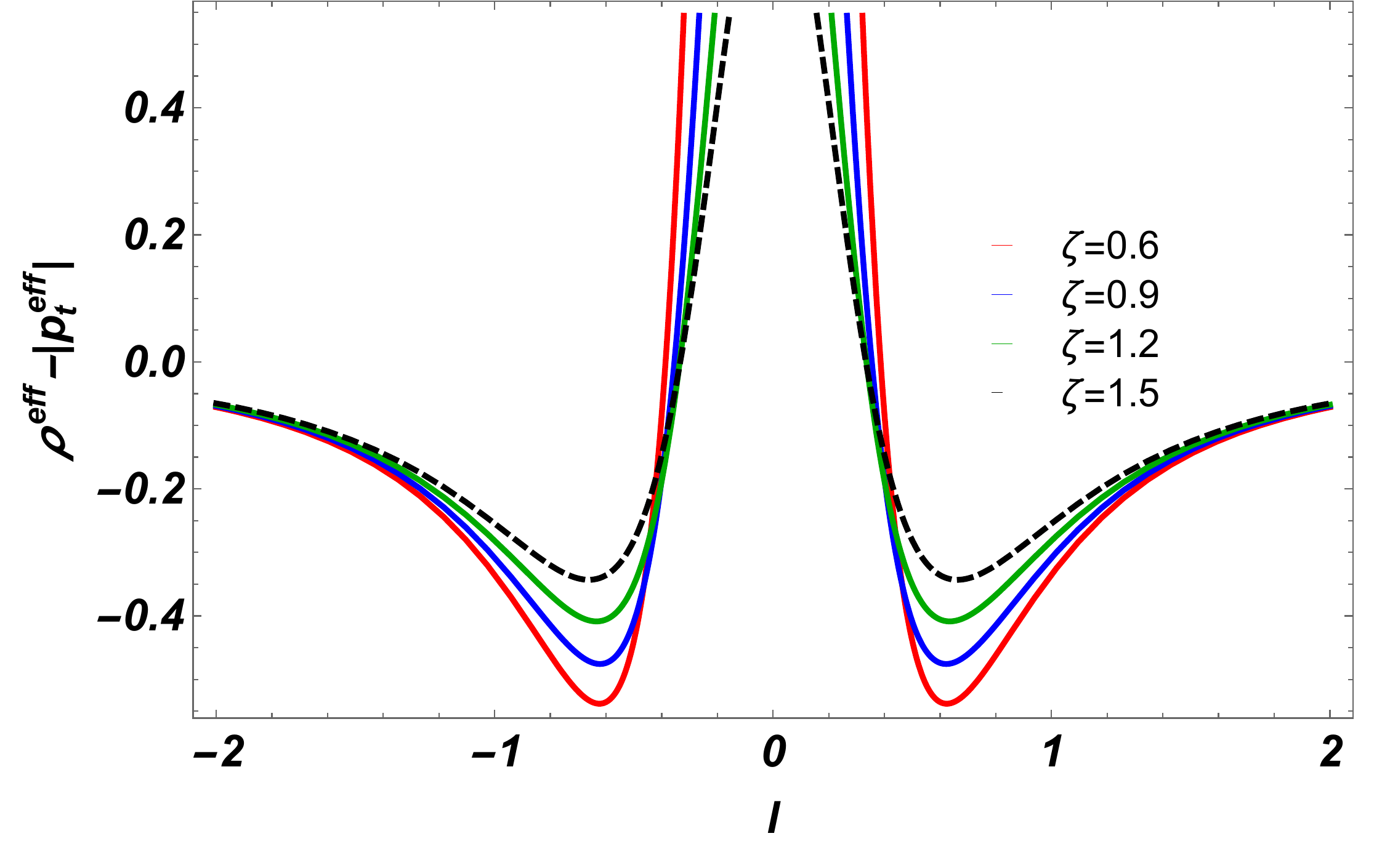}
\caption{Shows the variation of $\rho^{eff}-\mid p^{eff}_{t}\mid$ with Model-I (left) and Model-II (right) generalized embedded wormhole solutions under the effect of Polytropic EOS.}\label{rrF12}
\end{figure}

The DEC, like $\rho^{eff}-p^{eff}_{r}$, $\rho^{eff}-p^{eff}_{t}$ within $\rho^{eff}-\mid p^{eff}_{r}\mid$ and $\rho^{eff}-\mid p^{eff}_{t}\mid$ can be shown graphically in Figs. \ref{F3}, \ref{F5}, \ref{rF5} and \ref{rrF5}, respectively for van der Waals EOS and in Figs. \ref{F10}, \ref{F12}, \ref{rF12} and \ref{rrF12} for Polytropic EOS. It is noted from Fig. \ref{F3} for Van der Waals EOS in Fig. \ref{F10} for Polytropic EOS that $\rho^{eff}-p^{eff}_{r}$ remains negative with decreasing behavior in $-2\leq l\leq-1$ and $1\leq l\leq2$ while it is seen positive with the increasing behavior around the wormhole throat $-1\leq l \leq S_{0}$ and $S_{0}< l \leq 1$ for Model-I. Further, for Model-II, $\rho^{eff}-p^{eff}_{r}$ remains negative in $-2\leq l\leq2$. Also, $\rho^{eff}-\mid p^{eff}_{r}\mid$ and $\rho^{eff}-\mid p^{eff}_{t}\mid$ remain negative for Van der Waals EOS and polytropic EOS against both the models.

\begin{figure}[htb!]
\centering 
\includegraphics[width=7.6cm,height=6.0cm]{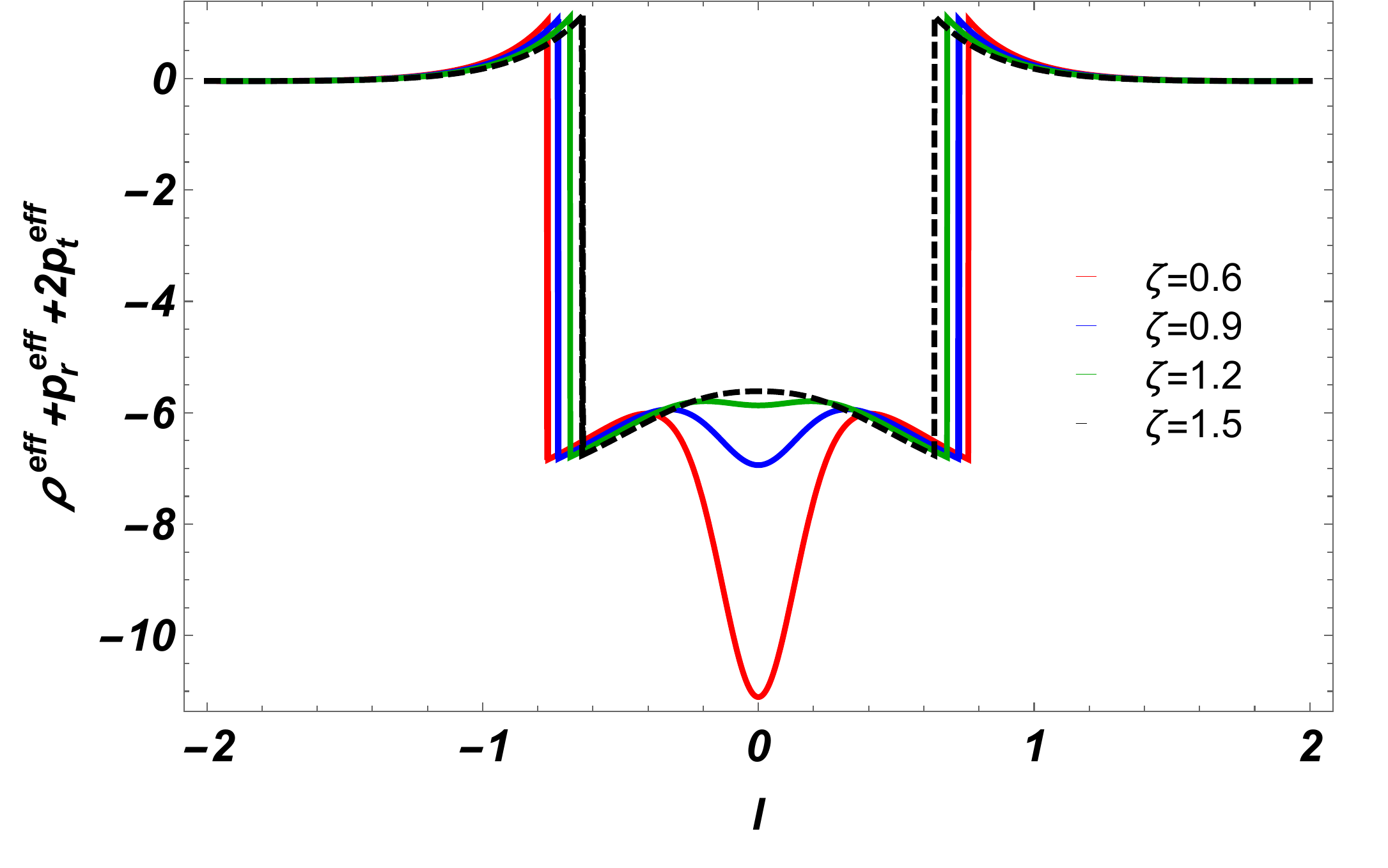} \includegraphics[width=7.6cm,height=6.0cm]{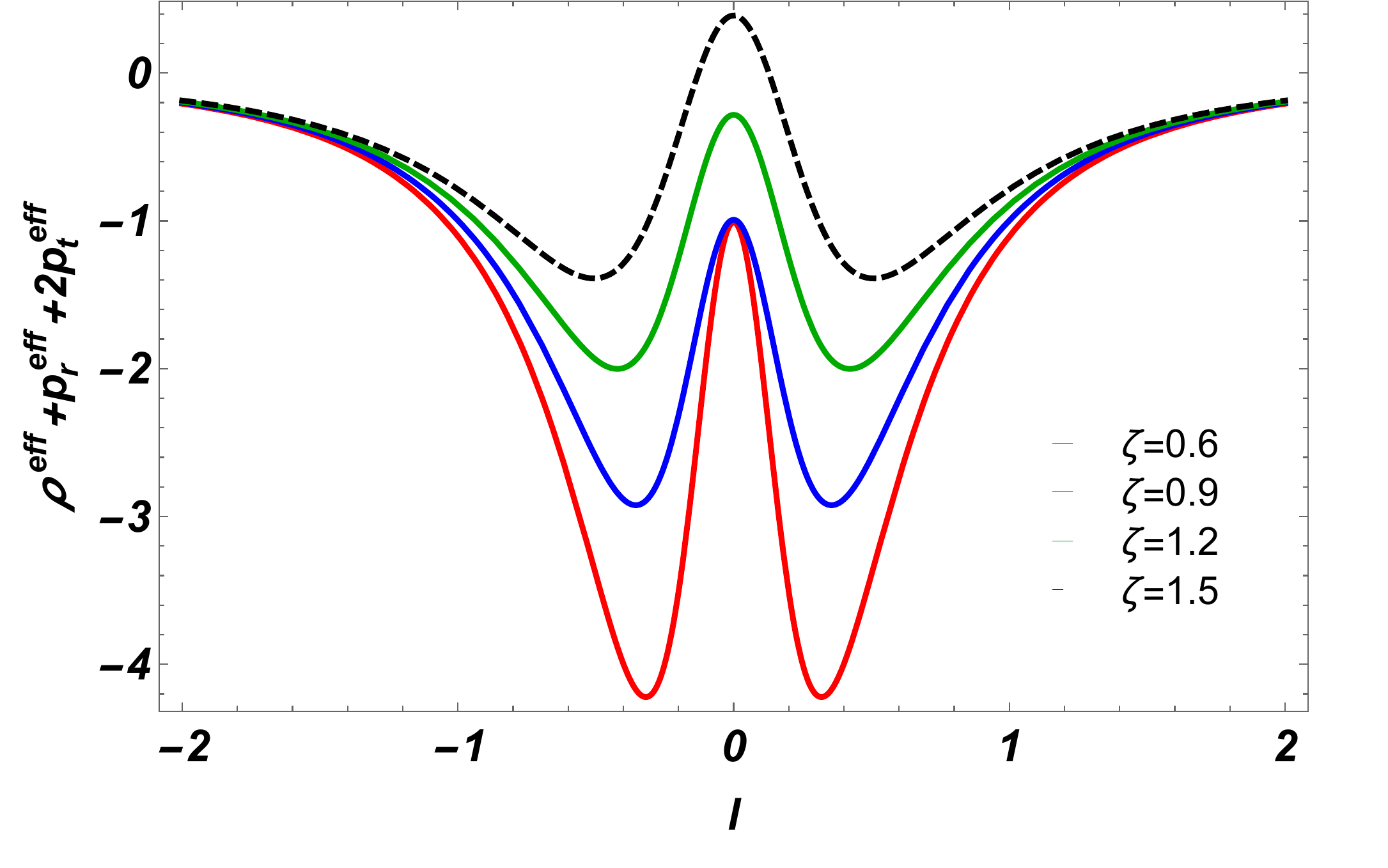}
\caption{Shows the variation of $\rho^{eff}+p^{eff}_{r}+2p^{eff}_{t}$ with Model-I (left) and Model-II (right) generalized embedded wormhole solutions under the effect of Polytropic EOS.}\label{F13}
\end{figure}

\begin{figure}[htb!]
\centering 
\includegraphics[width=7.6cm,height=6.0cm]{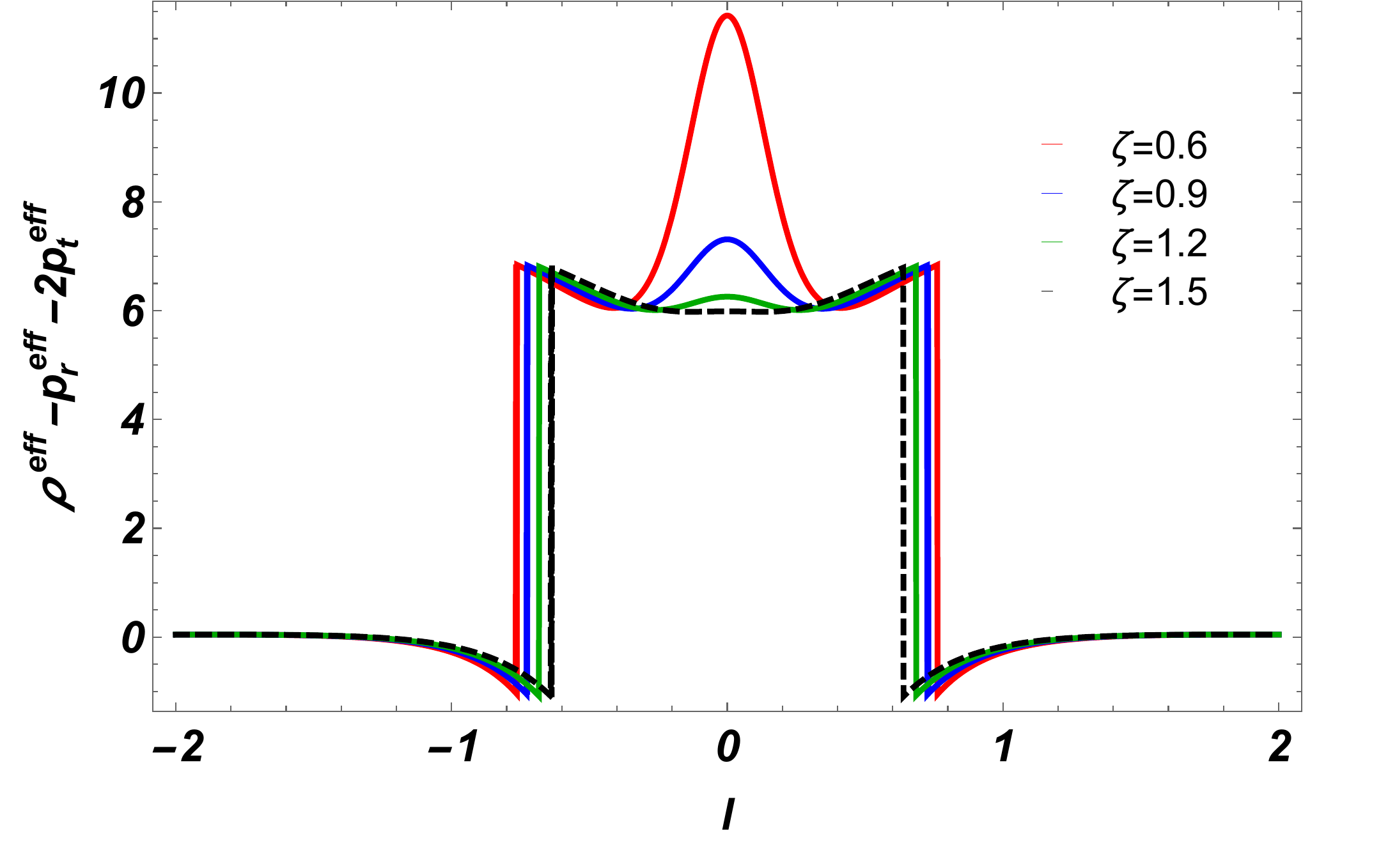} \includegraphics[width=7.6cm,height=6.0cm]{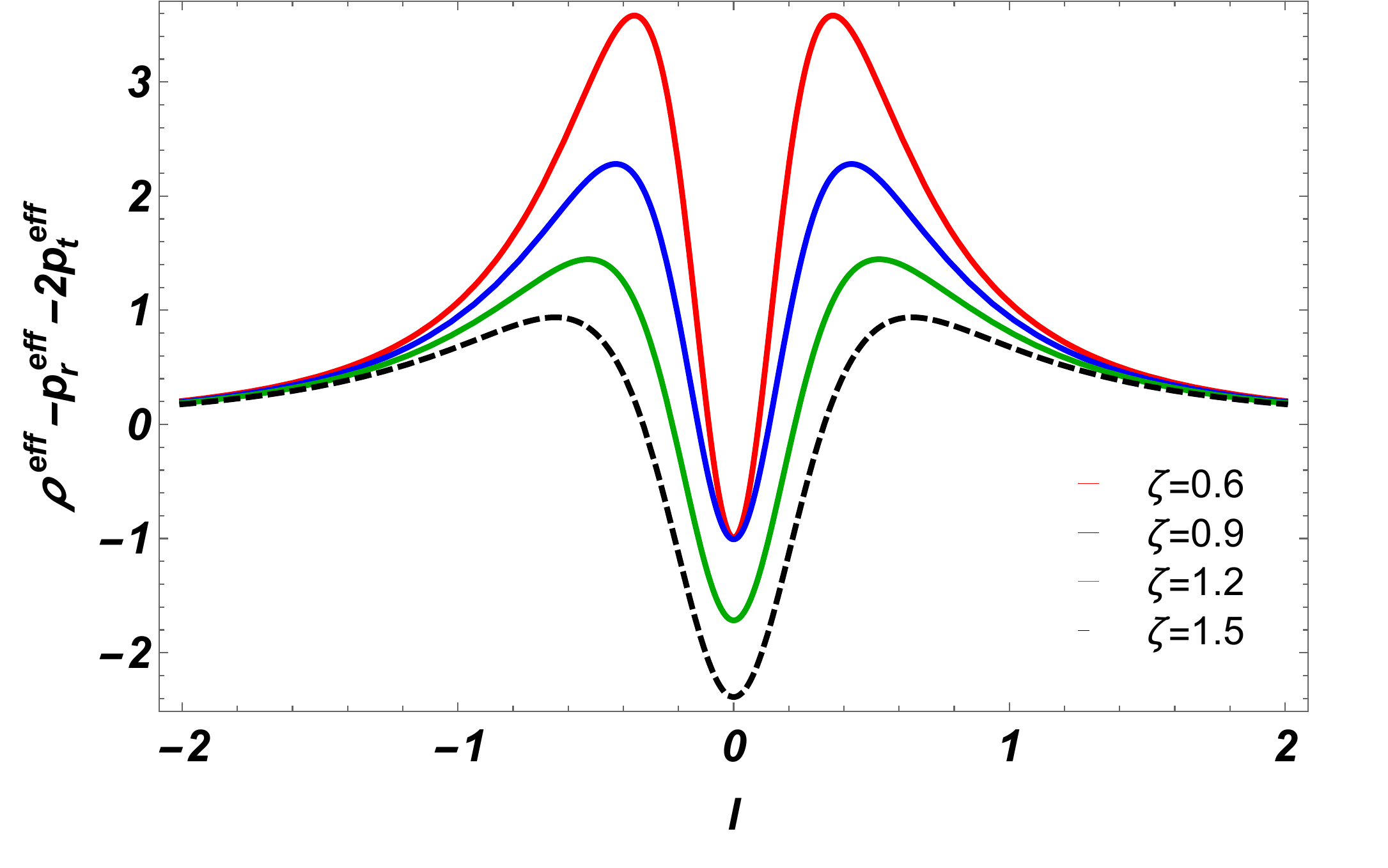}
\caption{Shows the variation of $\rho^{eff}-p^{eff}_{r}-2p^{eff}_{t}$ with Model-I (left) and Model-II (right) generalized embedded wormhole solutions under the effect of Polytropic EOS.}\label{F14}
\end{figure}

{The analysis of WEC, i.e. $\rho^{eff}+p^{eff}_{t}$, $\rho^{eff}+p^{eff}_{r}$ and $\rho^{eff}$ can be perceived from Figs. \ref{F4}, \ref{F9} and \ref{F1} for van der Waals EOS and in Figs. \ref{F11}, \ref{F9}, and \ref{F8} for polytropic EOS for both Model-I (left) and Model-II. For Model-I $\rho^{eff}+p^{eff}_{t}$ is seen positive for Model-I while for the model-II $\rho^{eff}+p^{eff}_{t}$ is observed negative for van der Waals EOS.  Further, polytropic EOS $\rho^{eff}+p^{eff}_{t}$ is seen as strongly negative for both Model-I and Model-II, which can be confirmed from the Fig. \ref{F11}. The graphical behavior of an energy condition like $\rho^{eff}-p^{eff}_{t}$ can be verified from Fig. \ref{F5} for van der Waals EOS in Fig. \ref{F12} for polytropic EOS for both the Model-I (left) and Model-II. For Model-I $\rho^{eff}-p^{eff}_{t}$ is seen negative while for Model-II $\rho^{eff}-p^{eff}_{t}$ is seen positive for van der Waals EOS.  Further, for polytropic EOS $\rho^{eff}-p^{eff}_{t}$ is seen as almost positive for both the Model-I and Model-II in some particular regions, It is also seen as negative in a very small region, which can be confirmed from the Fig. \ref{F11}.\\
It is really necessary to mention that the SEC, i.e. $\rho^{eff}+p^{eff}_{r}+2p^{eff}_{t}$ within $\rho^{eff}+p^{eff}_{t}$, $\rho^{eff}+p^{eff}_{r}$ has an important role in the wormhole study to investigate the exact nature of the matter. For the current analysis, the graphical behavior of $\rho^{eff}+p^{eff}_{r}+2p^{eff}_{t}$ can be verified from Fig. \ref{F6} for van der Waals EOS and in Fig. \ref{F13} for polytropic EOS for both the Model-I (left) and Model-II. The SEC like $\rho^{eff}+p^{eff}_{t}$ is observed positive for Model-I while for Model-II $\rho^{eff}+p^{eff}_{r}+2p^{eff}_{t}$ is shown to be negative for van der Waals EOS. Such kind of behavior can be also checked from Fig. \ref{F6}. For the polytropic EOS case, $\rho^{eff}+p^{eff}_{r}+2p^{eff}_{t}$ is seen strongly negative for both the Model-I and Model-II, which can be confirmed from Fig. \ref{F13} for $-2\leq l\leq2$. \\
Now, we are going to discuss TEC, i.e., $\rho^{eff}-p^{eff}_{r}-2p^{eff}_{t}$ within $\rho^{eff}-p^{eff}_{r}$, $\rho^{eff}-p^{eff}_{t}$ under the effect of both the van der Waals EOS and polytropic EOS. The behavior of $\rho^{eff}-p^{eff}_{r}-2p^{eff}_{t}$ is observed strongly negative for Model-I while for Model-II $\rho^{eff}-p^{eff}_{r}-2p^{eff}_{t}$ is seen positive in the background of van der Waals EOS (the variation with graphical analysis is provided in Fig. \ref{F7}). In Fig. \ref{F14}, the graphical development of $\rho^{eff}-p^{eff}_{r}-2p^{eff}_{t}$ is given under the effect of polytropic EOS, which is seen positive in some regions and it is also observed negative in some smaller part of the configurations. }

The behavior of all the energy conditions is also summarized in Table \ref{tab1} (for the van der Waals EOS) and in Table \ref{tab2} (for polytropic EOS). 

\begin{table*}
\centering
\caption{\label{tab1}{Outlook for van der Waals EOS case via energy conditions}}
\scalebox{0.80}{\begin{tabular}{|c|c|c|c|c|c|c|c|c|}
\hline

Energy conditions                 & Model-I:                                                           & Model-II:   \\                                   
{}                                    & with $S_{0}=0.9$, $\Psi =0.5$, $\alpha =0.8$, $\beta=4$,                       & with $S_{0}=0.9$, $\alpha =0.8$, $\beta=4$,  \\
{}                                    & $\gamma=0.8$, $w_q=-2/3$  and $m=2$,                         & $\gamma=0.8$, $w_q=-2/3$  and $m=2$\\
\hline
    & \multicolumn{2}{|c|}{Behavior of Energy conditions} \\
\hline
$\rho^{eff}$ ~~~~~~~~~~~~~~~~~~~~~~~~~(Fig. \ref{F1})            & $\rho^{eff}>0$ for $-2\leq l\leq2$                                 & $\rho^{eff}>0$ for $-2\leq l\leq2$  \\
\hline
$\rho^{eff}+p^{eff}_{r}$ ~~~~~~~~~~~~~~~\;(Fig. \ref{F2})           & $\rho^{eff}+p^{eff}_{r}<0$ for $-2\leq l\leq2$                  &$\rho^{eff}+p^{eff}_{r}>0$ for $-2\leq l\leq2$     \\
\hline
$\rho^{eff}-p^{eff}_{r}$ ~~~~~~~~~~~~~~~\;\;(Fig. \ref{F3})           & $\rho^{eff}-p^{eff}_{r}>0$ for $-2\leq l\leq2$                &$\rho^{eff}-p^{eff}_{r}<0$ for $-2\leq l\leq2$    \\
\hline
$\rho^{eff}+p^{eff}_{t}$ \;~~~~~~~~~~~~~~~~(Fig. \ref{F4})        & $\rho^{eff}+p^{eff}_{t}>0$ for $-2\leq l\leq2$                    &$\rho^{eff}+p^{eff}_{t}<0$ for $-2\leq l\leq2$   \\
\hline
$\rho^{eff}-p^{eff}_{t}$~~~~~~~~~~~~~~~\;\;\;(Fig. \ref{F5})            & $\rho^{eff}-p^{eff}_{t}<0$ for $-2\leq l\leq2$              &$\rho^{eff}-p^{eff}_{t}>0$ for $-2\leq l\leq2$    \\
\hline
$\rho^{eff}-\mid p^{eff}_{r}\mid$~~~~~~~~~~~~\;\;\;(Fig. \ref{rF5})     & $\rho^{eff}-\mid p^{eff}_{r}\mid<0$ for $-2\leq l\leq2$ & $\rho^{eff}-\mid p^{eff}_{r}\mid<0$ for $-2\leq l\leq2$    \\
\hline
$\rho^{eff}-\mid p^{eff}_{t}\mid$~~~~~~~~~~~~\;\;\;(Fig. \ref{rrF5})  & $\rho^{eff}-\mid p^{eff}_{t}\mid<0$ for $-2\leq l\leq2$  & $\rho^{eff}-\mid p^{eff}_{t}\mid<0$ for $-2\leq l\leq2$    \\
 \hline
$\rho^{eff}+p^{eff}_{r}+2p^{eff}_{t}$~~~\;\;\;\;(Fig. \ref{F6})  & $\rho^{eff}+p^{eff}_{r}+2p^{eff}_{t}>0$ for $-2\leq l\leq2$ &$\rho^{eff}+p^{eff}_{r}+2p^{eff}_{t}<0$ for $-2\leq l\leq2$    \\
\hline
$\rho^{eff}-p^{eff}_{r}-2p^{eff}_{t}$~~~\;\;\;\;(Fig. \ref{F7})  & $\rho^{eff}-p^{eff}_{r}-2p^{eff}_{t}<0$ for $-2\leq l\leq2$ &$\rho^{eff}-p^{eff}_{r}-2p^{eff}_{t}>0$ for $-2\leq l\leq2$    \\
\hline
\end{tabular}}
\end{table*}

\begin{table*}
\centering
\caption{\label{tab2}{Outlook for Polytropic EOS case via energy conditions}}
\scalebox{0.79}{\begin{tabular}{|c|c|c|c|c|c|c|c|c|}
\hline

Energy conditions                                                         & Model-I:                                                                             & Model-II:   \\
{}                                                                        & with $S_{0}=0.9$, $\Psi =0.5$, $\alpha =0.8$, $\beta=4$,                             & with $S_{0}=0.9$, $\alpha =0.8$, $\beta=4$,  \\
{}                                                                        & $\gamma=0.8$, $w_q=-2/3$  and $m=2$,                                                 & $\gamma=0.8$, $w_q=-2/3$  and $m=2$\\
\hline
    & \multicolumn{2}{|c|}{Behavior of Energy conditions} \\
\hline
$\rho^{eff}$ ~~~~~~~~~~~~~~~~~~~~~~~~~(Fig. \ref{F8})            & $\rho^{eff}>0$ for $-2\leq l\leq2$                    & $\rho^{eff}>0$ for $-2\leq l\leq2$   \\
\hline
$\rho^{eff}+p^{eff}_{r}$ ~~~~~~~~~~~~~~~\;(Fig. \ref{F9})           & $\rho^{eff}+p^{eff}_{r}<0$ for $-2\leq l\leq2$     &$\rho^{eff}+p^{eff}_{r}>0$ for $-2\leq l\leq2$     \\
\hline
$\rho^{eff}-p^{eff}_{r}$ ~~~~~~~~~~~~~~~\;\;(Fig. \ref{F10})           & $\rho^{eff}-p^{eff}_{r}>0$ for $-2\leq l\leq2$  &$\rho^{eff}-p^{eff}_{r}<0$ for $-2\leq l\leq2$    \\
\hline
$\rho^{eff}+p^{eff}_{t}$ \;~~~~~~~~~~~~~~~~(Fig. \ref{F11})        & $\rho^{eff}+p^{eff}_{t}>0$ for $-2\leq l\leq2$      &$\rho^{eff}+p^{eff}_{t}<0$ for $-2\leq l\leq2$    \\
\hline
$\rho^{eff}-p^{eff}_{t}$~~~~~~~~~~~~~~~\;\;\;(Fig. \ref{F12})            & $\rho^{eff}-p^{eff}_{t}<0$ for $-2\leq l\leq2$  &$\rho^{eff}-p^{eff}_{t}>0$ for $-2\leq l\leq2$     \\
\hline
$\rho^{eff}-\mid p^{eff}_{r}\mid$~~~~~~~~~~~~\;\;\;(Fig. \ref{rF12})   & $\rho^{eff}-\mid p^{eff}_{r}\mid<0$ for $-2\leq l\leq2$  & $\rho^{eff}-\mid p^{eff}_{r}\mid<0$ for $-2\leq l\leq2$ \\
\hline
$\rho^{eff}-\mid p^{eff}_{t}\mid$~~~~~~~~~~~~\;\;\;(Fig. \ref{rrF12})  & $\rho^{eff}-\mid p^{eff}_{t}\mid<0$ for $-2\leq l\leq2$ & $\rho^{eff}-\mid p^{eff}_{t}\mid<0$ for $-2\leq l\leq2$  \\
 \hline
$\rho^{eff}+p^{eff}_{r}+2p^{eff}_{t}$~~~\;\;\;\;(Fig. \ref{F13}) & $\rho^{eff}+p^{eff}_{r}+2p^{eff}_{t}>0$ for $-2\leq l\leq2$  &$\rho^{eff}+p^{eff}_{r}+2p^{eff}_{t}<0$ for $-2\leq l\leq2$ \\
\hline
$\rho^{eff}-p^{eff}_{r}-2p^{eff}_{t}$~~~\;\;\;\;(Fig. \ref{F14})  & $\rho^{eff}-p^{eff}_{r}-2p^{eff}_{t}<0$ for $-2\leq l\leq2$  &$\rho^{eff}-p^{eff}_{r}-2p^{eff}_{t}>0$ for $-2\leq l\leq2$\\
\hline
\end{tabular}}
\end{table*}

\section{Matter Contents of Thin-Shell around Wormhole Geometry}\label{sec6}

Here, we use two different kinds of shape functions in the context of altered gravity to build a thin shell surrounding WH solutions. For this purpose, the interior manifold is represented by a WH geometry, and the external geometry is represented by a Schwarzschild BH surrounded by cloud and quintessence-type fluid distribution. The outer manifolds can be given as
\begin{equation}\label{1aa}
ds^{2}_+=-\mathcal{F}_+(r_+)^{-1}dr^{2}_+-r^{2}_+
d\theta^{2}_+-r^{2}_+\sin^2{\theta}_+
d\phi^{2}_++\mathcal{F}_+(r_+)dt^{2}_+,
\end{equation}
the lapse function of exterior (+) manifold can be written
\begin{equation}\label{2aa}
\mathcal{F}_+(r_+)=-a-\frac{q}{r^{3 \omega _q+1}_+}-\frac{2 M}{r_+}+1.
\end{equation}

Here $M$, $q$, and $\omega _q$ represent the mass of Schwarzschild BH, normalization factor, and quintessence parameter. For simplification, we can consider  $r_+=\left(\ell^m+1\right)^{1/m}$, where $\ell$ is length parameter and $m$ is a constant.   
By considering WH as an interior manifold (-) as
\begin{equation}\label{2aaa}
ds^2_-=-e^{2\phi(r_-)}dt^2_-+r^2_-d\theta^2_-+\frac{dr^2_-}{\mathcal{F}_-(r_-)}+r^2_-sin^2\theta_-
d\phi^2_-,
\end{equation}
where
\begin{equation}\label{2aaaa}
\mathcal{F}_-(r_-)=-\frac{b(r_-)}{r_-}+1.
\end{equation}

The geometry of thin-shell enveloping WH spacetime is developed by using the cut and paste approach in the current paper. This approach generates a unique regular manifold that can be mathematically written as $\mathcal{W}=\mathcal{W}^{-}\cup \mathcal{W}^{+}$. The shell radius must be greater than $r_h$ and the resulting structure becomes non-singular.

\begin{figure}[htb!]
\centering 
\includegraphics[width=11.6cm,height=8.0cm]{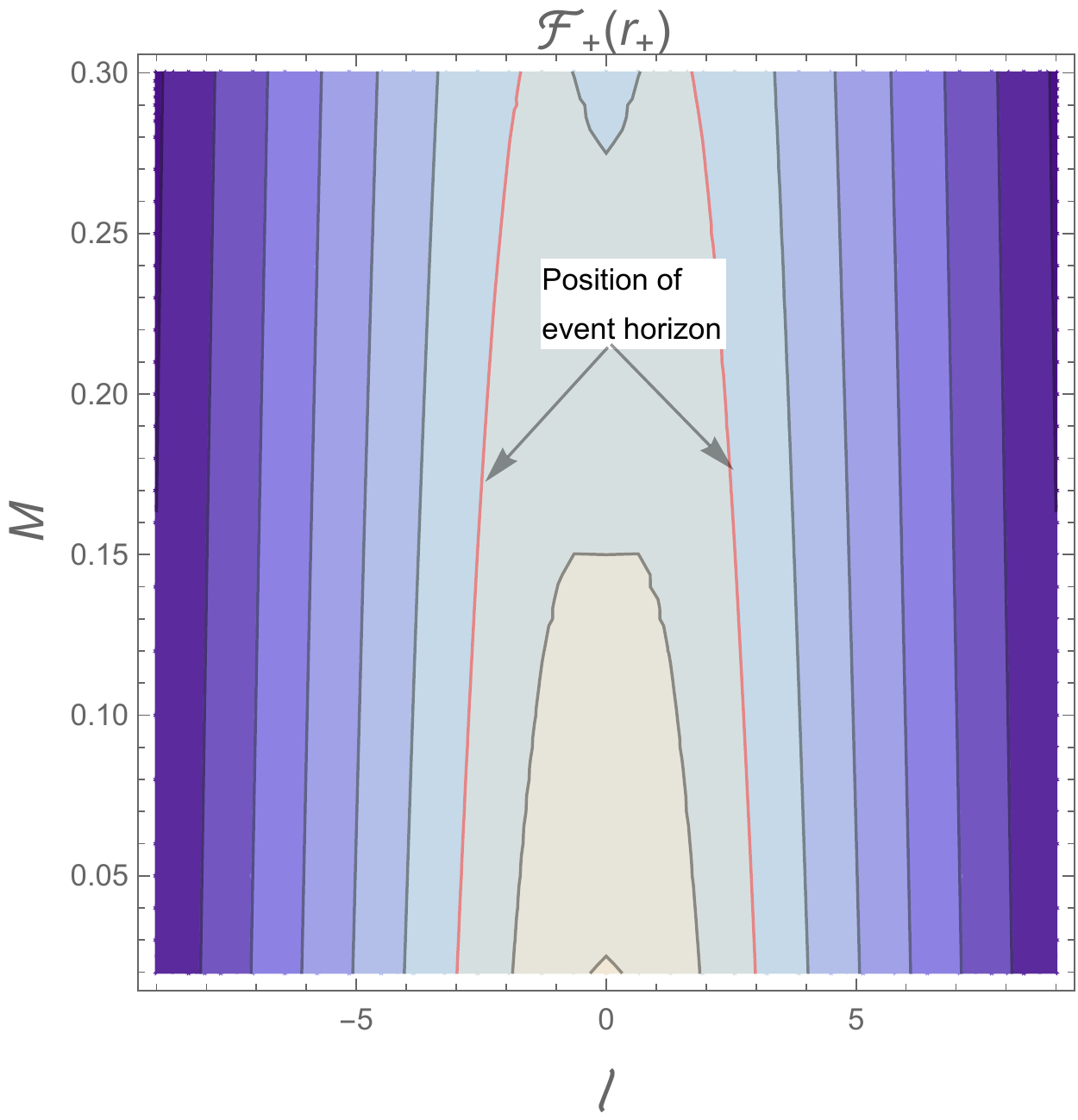} \includegraphics[width=1.5cm,height=8.0cm]{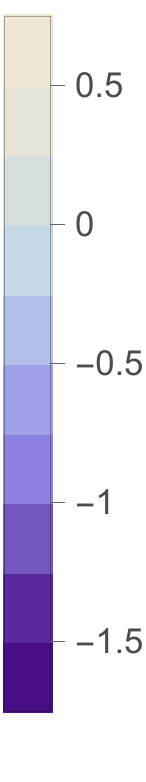}
\caption{ Contour plots of metric function versus $\ell$ and $M$ with $\omega_q=-2/3$, $a=0.2$, $m=2$ and $q=0.125.$ }\label{Fr6a}
\end{figure}

For this purpose, we observe the behavior of the metric function of the exterior BH spacetime versus $M$ and $\ell$, to determine the position of the event horizon as shown in Fig. (\ref{Fr6a}). Here, red lines represent the position of the event horizon versus $M$ and $\ell$. It is found that the position of the event horizon is located in the range of $-3.029\leq\ell\leq3.029$. Hence, thin shell radius must follow either $\ell\leq-3.029$ or  $\ell\geq3.029$.

According to the Israel formalism, the coordinates of considered manifolds and hypersurfaces are in the following form:
$z^{\gamma}_\pm=(t_\pm,r_\pm,\theta_\pm,\phi_\pm)$ and $\eta^{i}=(\tau,\theta,\phi)$, respectively. Here $\tau$ represents the proper time over the hypersurface. These coordinate systems are
related to one another by using the following coordinate transformation:
\begin{eqnarray}\label{5aa}
g_{ij}=\frac{\partial
z^{\gamma}_\pm}{\partial\eta^{i}}\frac{\partial z^{\beta}_\pm}{\partial\eta^{j}}g_{\gamma\beta}^\pm.
\end{eqnarray}

The hypersurface's parametric equation is defined as
\begin{eqnarray}\nonumber 
\mathcal{H}: R(r,\tau)=r-y(\tau)=0.
\end{eqnarray}

At hypersurface, the Lanczos equations, which are the simplified form of Einstein field equations, are used to analyze the physical characteristics of matter distributions provided as
\begin{equation}\label{6aa}
S_{\beta}^{\alpha}=\frac{1}{8\pi}(\delta_{\beta}^{\alpha}
\zeta_{\gamma}^{\gamma}-\zeta_{\beta}^{\alpha}),
\end{equation}
where $\zeta_{\alpha\beta}=K^{+}_{\alpha\beta}-K^{-}_{\alpha\beta}$ and $K^{-}_{\alpha\beta}$ represents the components of extrinsic curvature. For perfect fluid distribution, the stress-energy tensor is expressed as ${S^{\alpha}}_{\beta}=diag(\mathfrak{S},\mathfrak{P},\mathfrak{P})$. The surface density and pressure at $\mathcal{H}$ are denoted with $\mathfrak{S}$ and $\mathfrak{P}$, respectively. The Interior and outer geometries extrinsic curvature is defined as
\begin{equation}\label{7aa}
{K_{\alpha\beta}^{\pm}}= -n_{\mu}^\pm \left[\frac{\partial^2 z^{\mu}_\pm}{\partial \eta^{\alpha} \eta^{\beta}}+\Gamma^{\mu}_{\lambda\nu}\left(\frac{\partial
z^{\lambda}_\pm}{\partial\eta^{\alpha}}\right)\left(\frac{\partial z^{\nu}_\pm}{\partial\eta^{\beta}}\right)\right].
\end{equation}

Additionally, the unit normals can be defined as follows:
\begin{equation}\label{8aa}
n_{\pm}^{\mu}=\left(\frac{\dot{y}}{\mathcal{F}_\pm(y)},\sqrt{\mathcal{F}_\pm(y) +\dot{y}^2},0,0\right),
\end{equation}
where the overdot represents the proper time derivative. 

By using Lanczos equations, we get
\begin{eqnarray}\label{9aa}
\mathfrak{S}(y)&=&-\frac{1}{4 \pi y}\left(\sqrt{\dot{y}^2-a-\frac{q}{y^{3 \omega _q+1}}-\frac{2 M}{y}+1}-\sqrt{-\frac{b (y)}{y}+\dot{y}^2+1}\right),
\\\label{10aa}
\mathfrak{P}(y)&=&\frac{1}{8 \pi  y}\Bigg(\frac{y^{-3 \omega _q-1} \left(2 y^{3 \omega_q} \left(y \left(-a+\dot{y}^2+\ddot{y} y+1\right)-M\right)+3 q \omega
_q-q\right)}{\sqrt{-a+\dot{y}^2-\frac{q y^{-3 \omega_q}+2 M}{y}+1}} \nonumber\\&& +\frac{y \left(b'(y)-2 \left(c^2+d y+1\right)\right)+b(y)}{y \sqrt{-\frac{b(y)}{y}+\dot{y}^2+1}}\Bigg).
\end{eqnarray}

At equilibrium shell radius $y_0$, it is now assumed that the thin shell of the developed geometry does not move in a radial direction. As a result, it is noteworthy to point out that the
proper time derivative of shell radius vanishes, i.e., $\dot{y_0}=0=\ddot{y_0}$. Hence, we have
\begin{eqnarray}\label{11aa}
\mathfrak{S}(y_0)&=&-\frac{1}{4 \pi y_0}\left(\sqrt{-a-\frac{q}{y^{3 \omega _q+1}_0}-\frac{2 M}{y_0}+1}-\sqrt{-\frac{b (y_0)}{y_0}+1}\right),\\\label{12aaa}\mathfrak{P}(y_0)&=&\frac{1}{8
\pi y^2_0}\left(\frac{y^{-3 \omega _q}_0 \left(-2 ((a-1) y_0+M) y^{3 \omega _q}_0+3 q \omega _q-q\right)}{\sqrt{-a-\frac{q y_0^{-3 \omega _q}+2 M}{y_0}+1}}+\frac{y_0 \left(b'(y_0)-2\right)+b(y_0)} {\sqrt{1-\frac{b(y_0)}{y_0}}}\right),~~~~~~~
\end{eqnarray}
where density and pressure at equilibrium position are denoted with $\mathfrak{S}(y_0)$ and $\mathfrak{P}(y_0)$, respectively.

\section{Stability Analysis Using Linearized Radial Perturbation}\label{sec7}

We now want to study the stable configuration of a developed thin-shell around WH configuration using linearized radial perturbation at $y=y_ 0$. We get the equation of motion of the shell
from Eq. (\ref{9aa}) as follows:
\begin{equation}\label{13aa}
\dot{y}^2+\Pi(y)=0,
\end{equation}
where $\mathcal{V}(y)$ is denoted with effective potential function given as
\begin{equation}\label{14aa}
\Pi(y)=-\frac{(b(y)+y (\mathcal{F}(y)-1))^2}{64 \pi ^2 y^4 \mathfrak{S} (y)^2}+\frac{-b(y)+y \mathcal{F}(y)+y}{2 y}-4 \pi ^2 y^2 \mathfrak{S}(y)^2.
\end{equation}


\begin{figure}[htb!]
\centering 
\includegraphics[width=7.5cm,height=6.0cm]{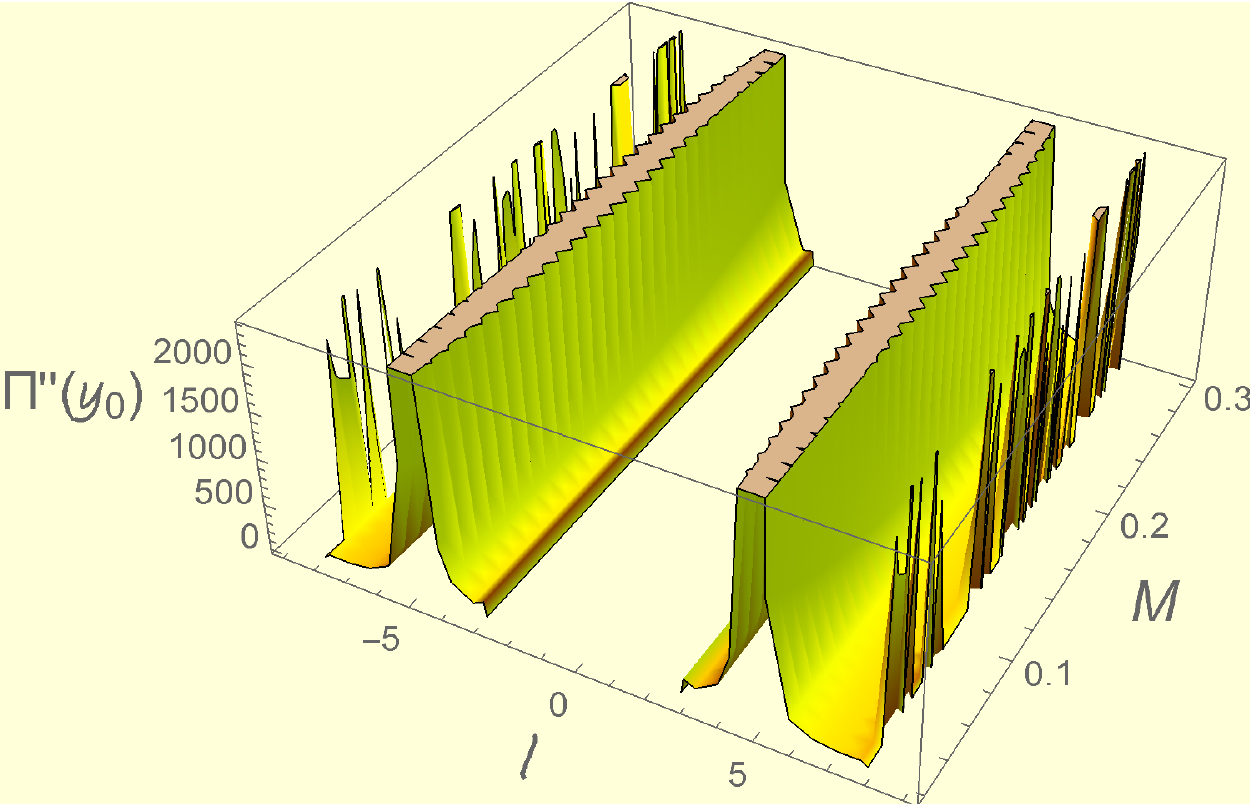}~~ \includegraphics[width=7.5cm,height=6.0cm]{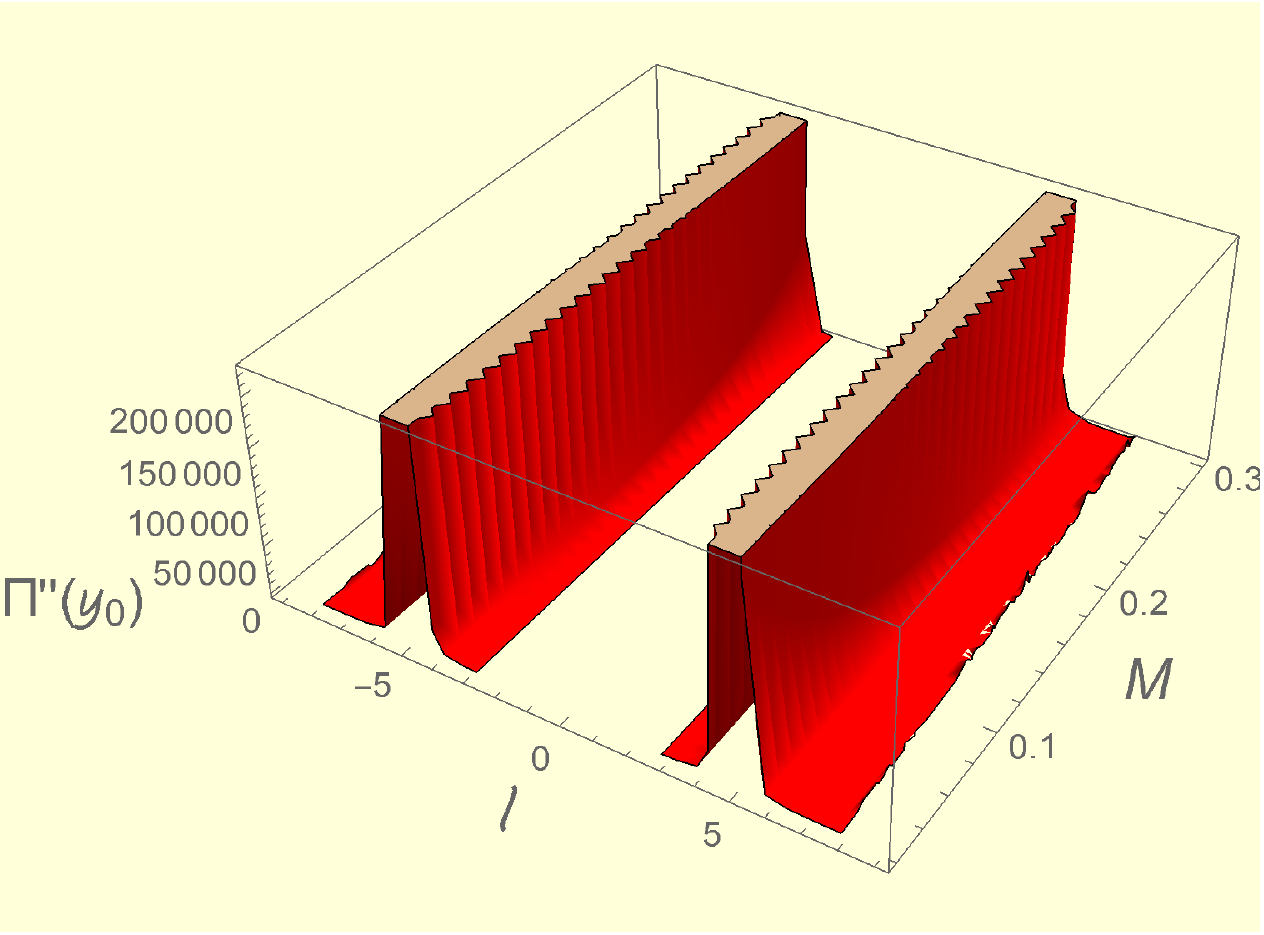}
\includegraphics[width=7.6cm,height=6.0cm]{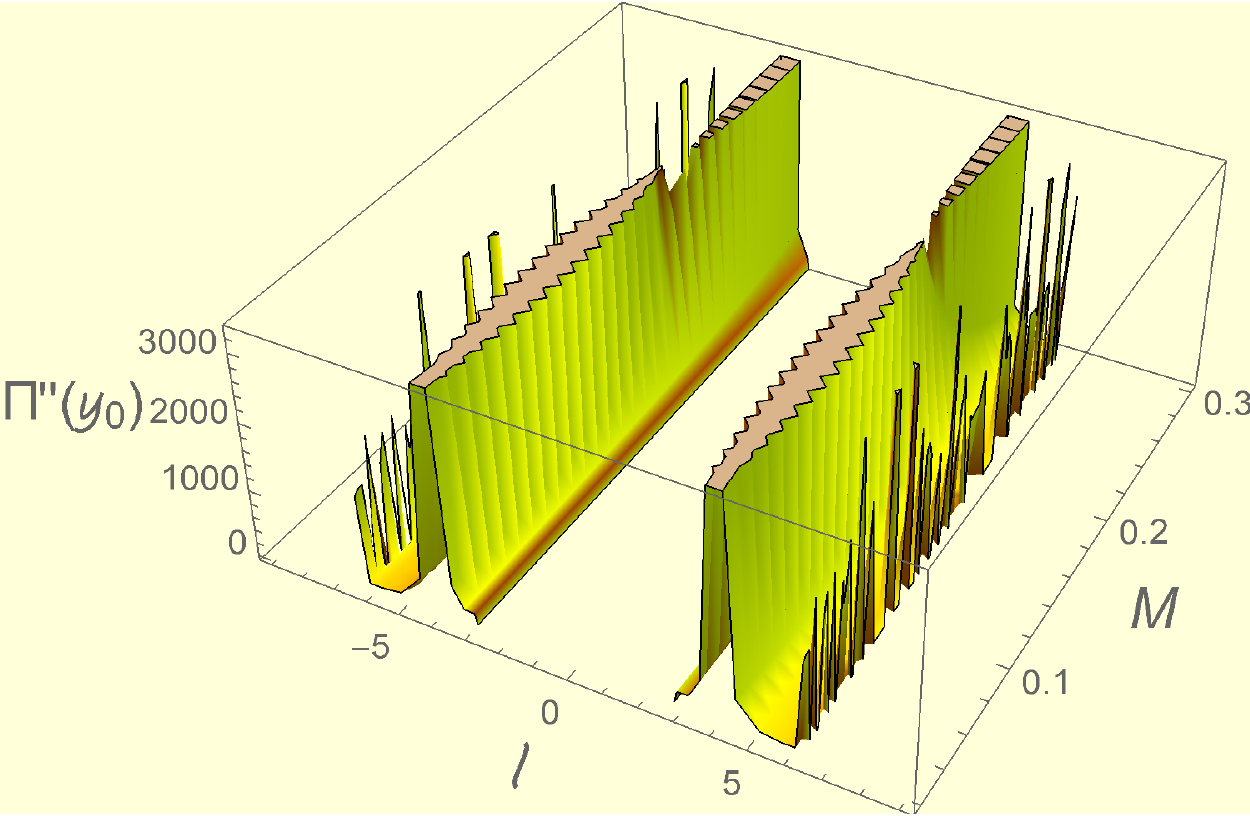} \includegraphics[width=7.6cm,height=6.0cm]{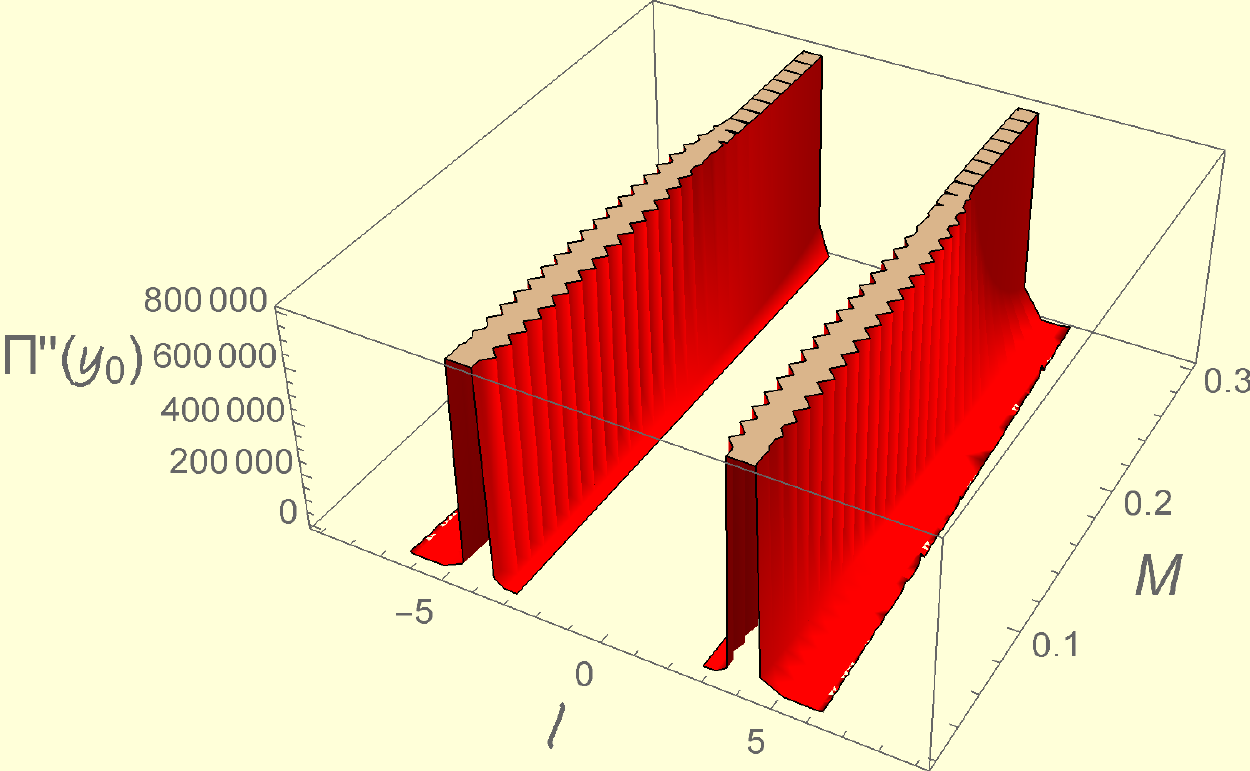}
\caption{Plots of $\Pi''(y)$ of thin shell around WH geometry for $b_1(y)$ (left plots) and $b_2(y)$ (right plots) for $q = 0.125$ with different values of $a$, i.e., $a=0$ (upper panel) and $a=0.2$ (lower panel).}\label{F6a}
\end{figure}

\begin{figure}[htb!]
\centering 
\includegraphics[width=7.5cm,height=6.0cm]{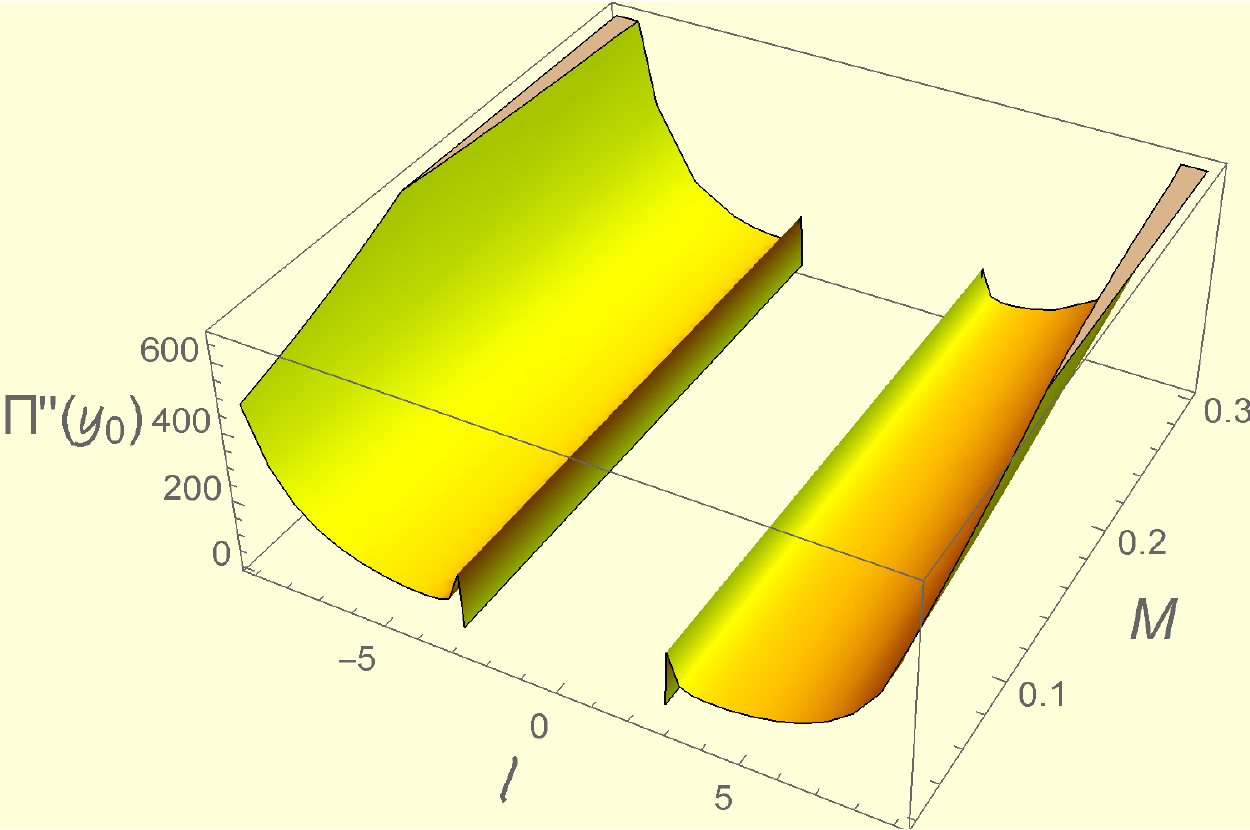}~~ \includegraphics[width=7.5cm,height=6.0cm]{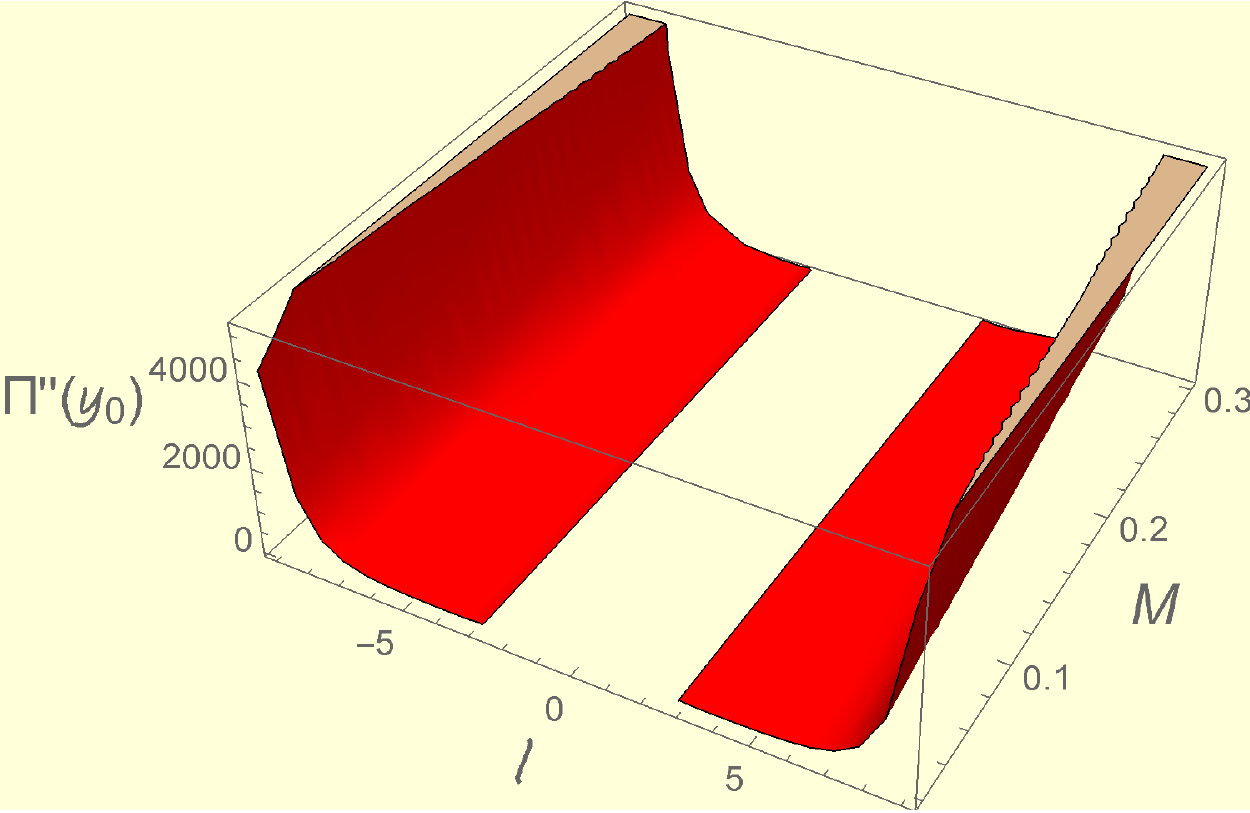}
\includegraphics[width=7.6cm,height=6.0cm]{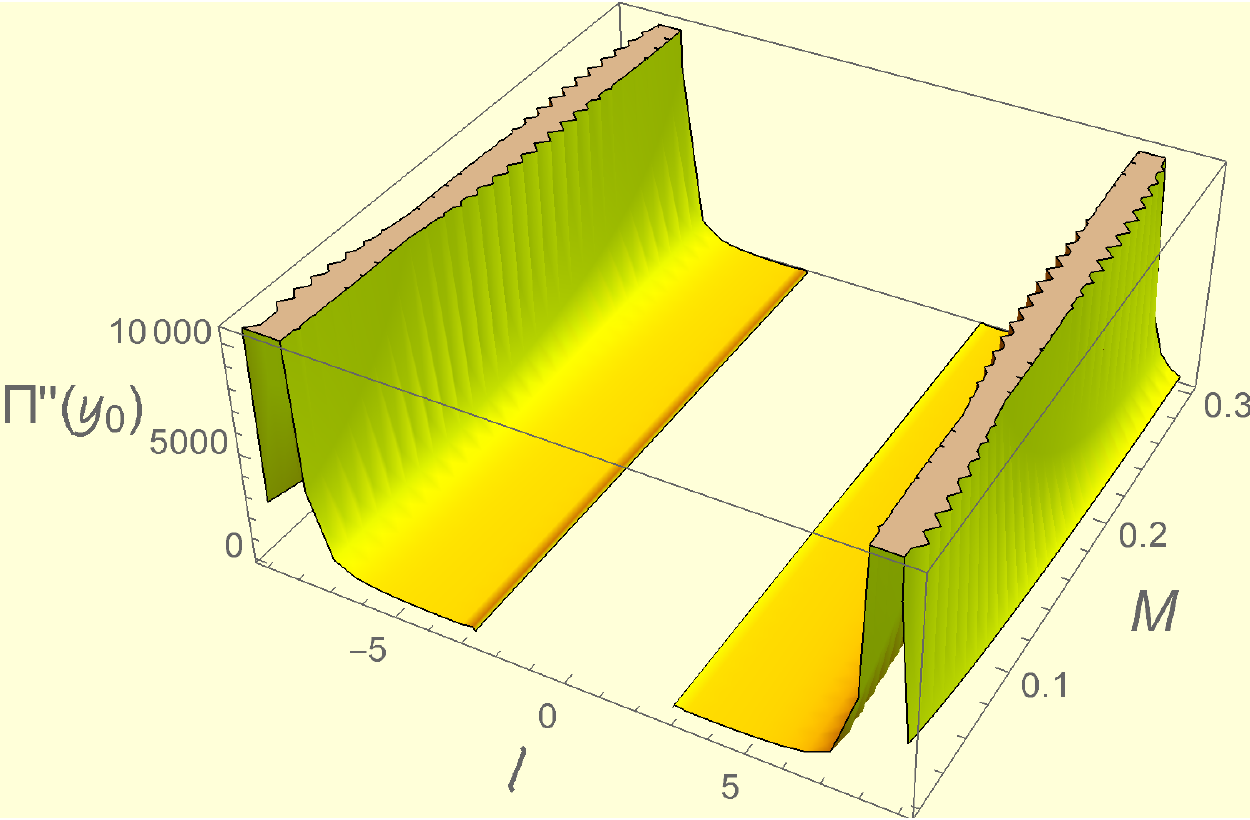} \includegraphics[width=7.6cm,height=6.0cm]{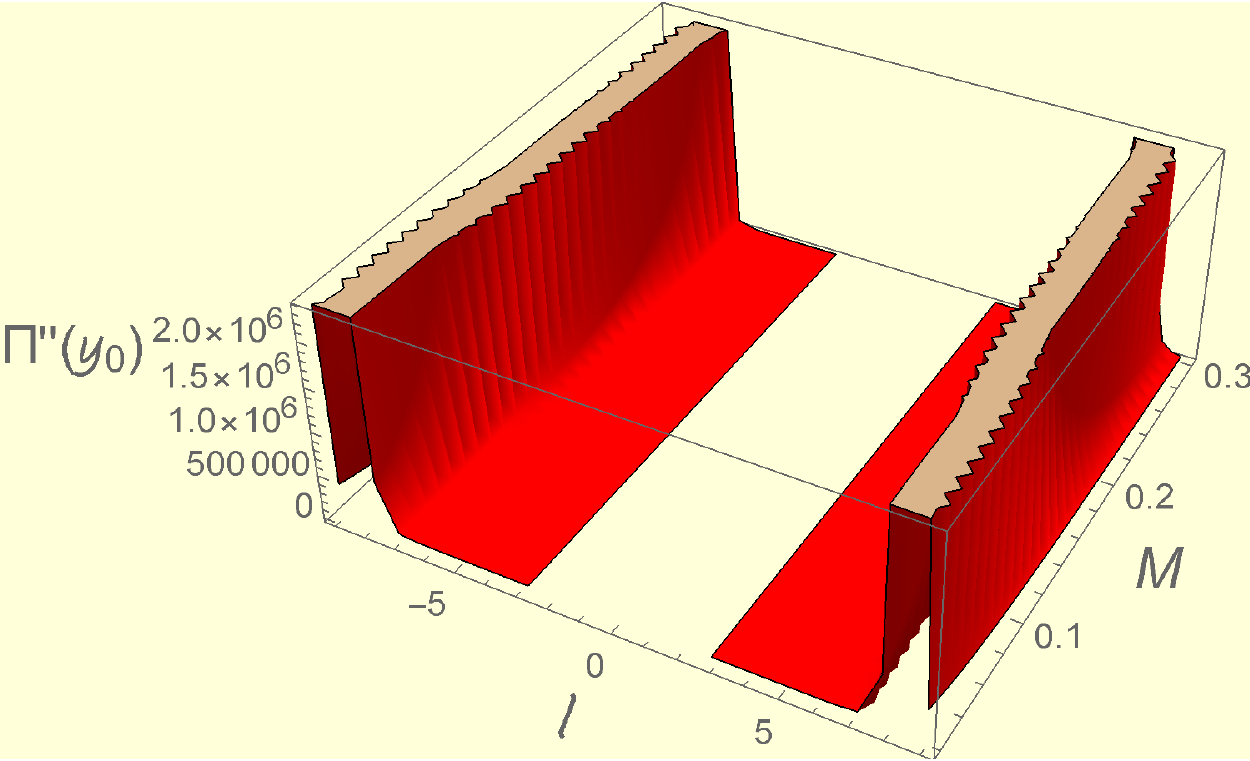}
\caption{Plots of $\Pi''(y)$ of thin shell around WH geometry for $b_1(y)$ (left plots) and $b_2(y)$ (right plots) with different values of $q$, i.e., $q=0.02$ (upper panel) and $q=0.04$ (lower panel).}\label{F6b}
\end{figure}

\begin{figure}[htb!]
\centering 
\includegraphics[width=7.5cm,height=6.0cm]{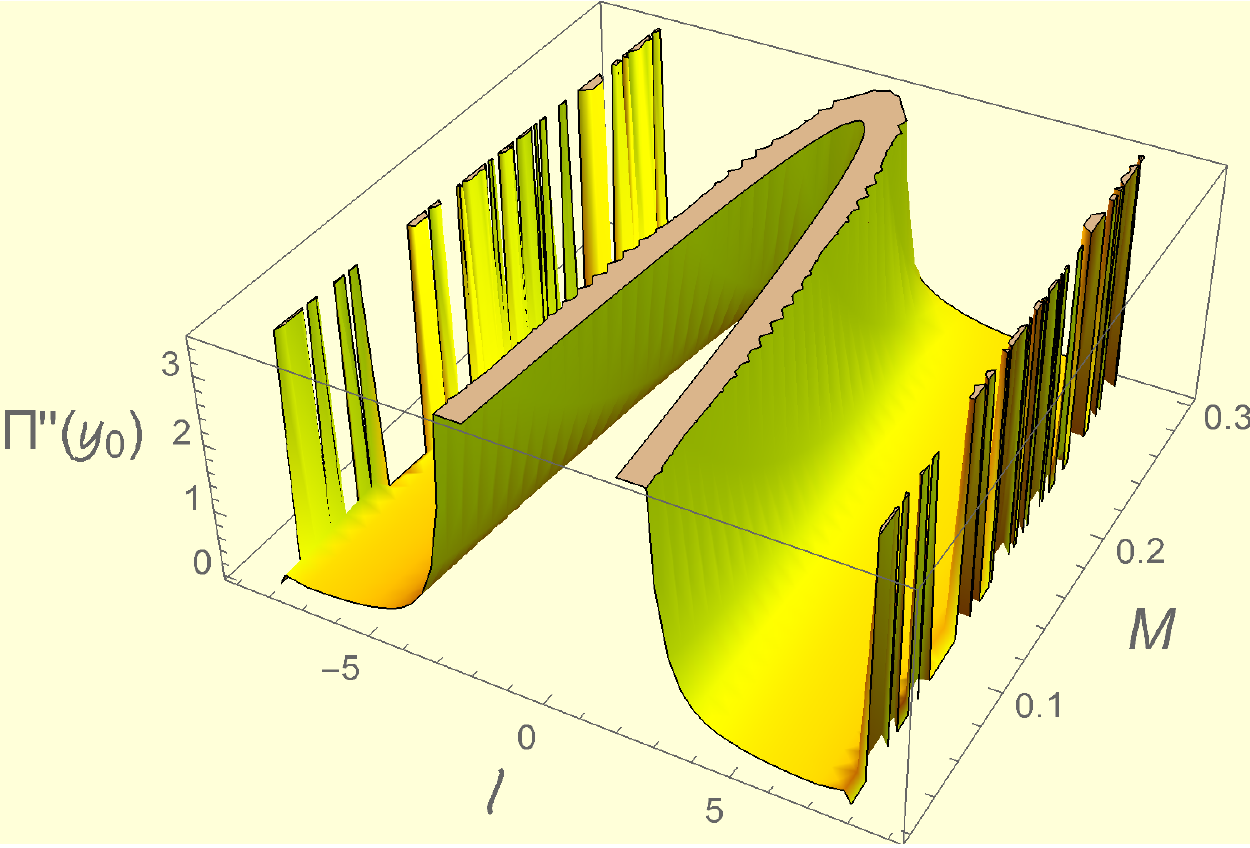}~~ \includegraphics[width=7.5cm,height=6.0cm]{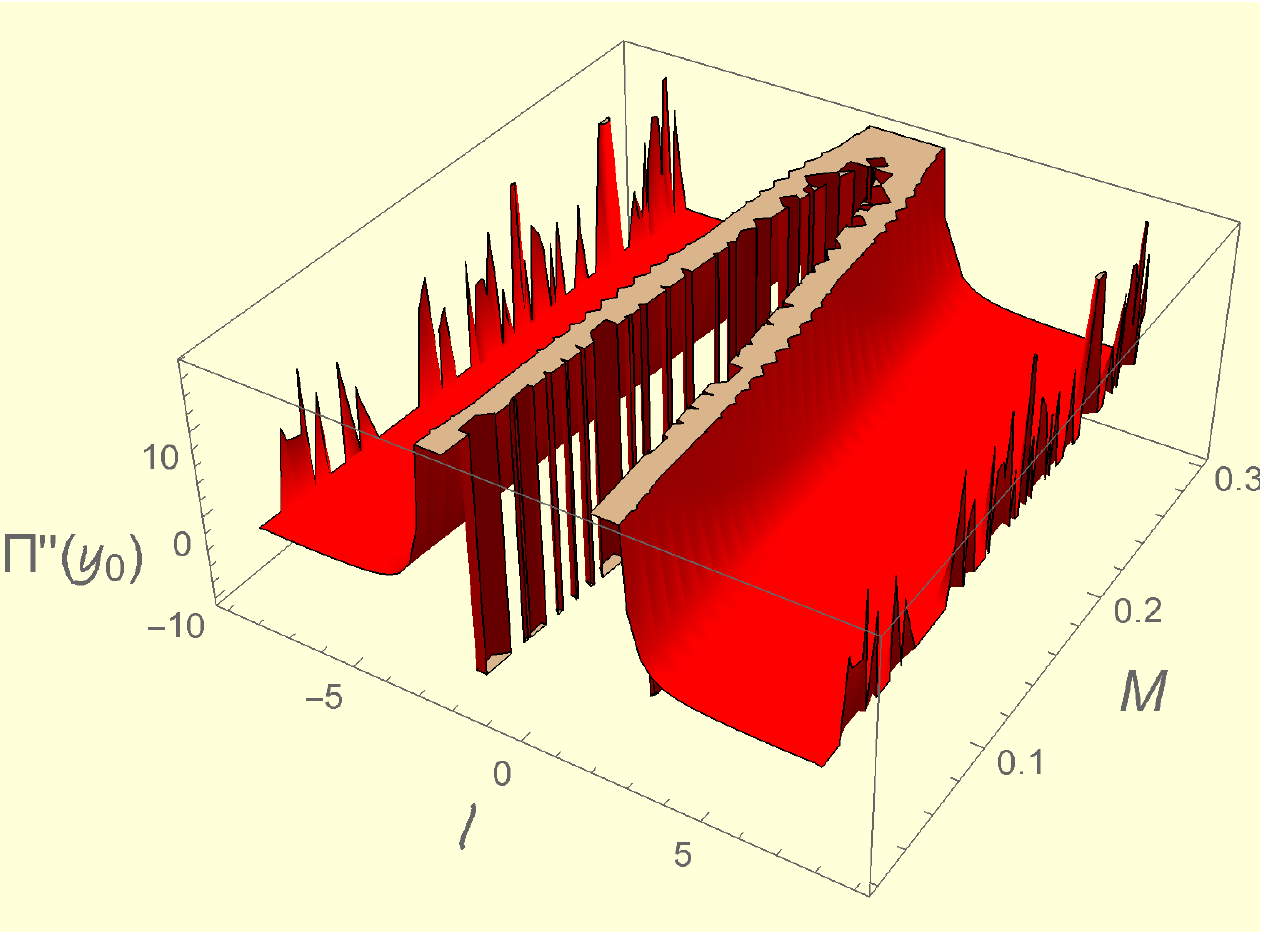}
\includegraphics[width=7.6cm,height=6.0cm]{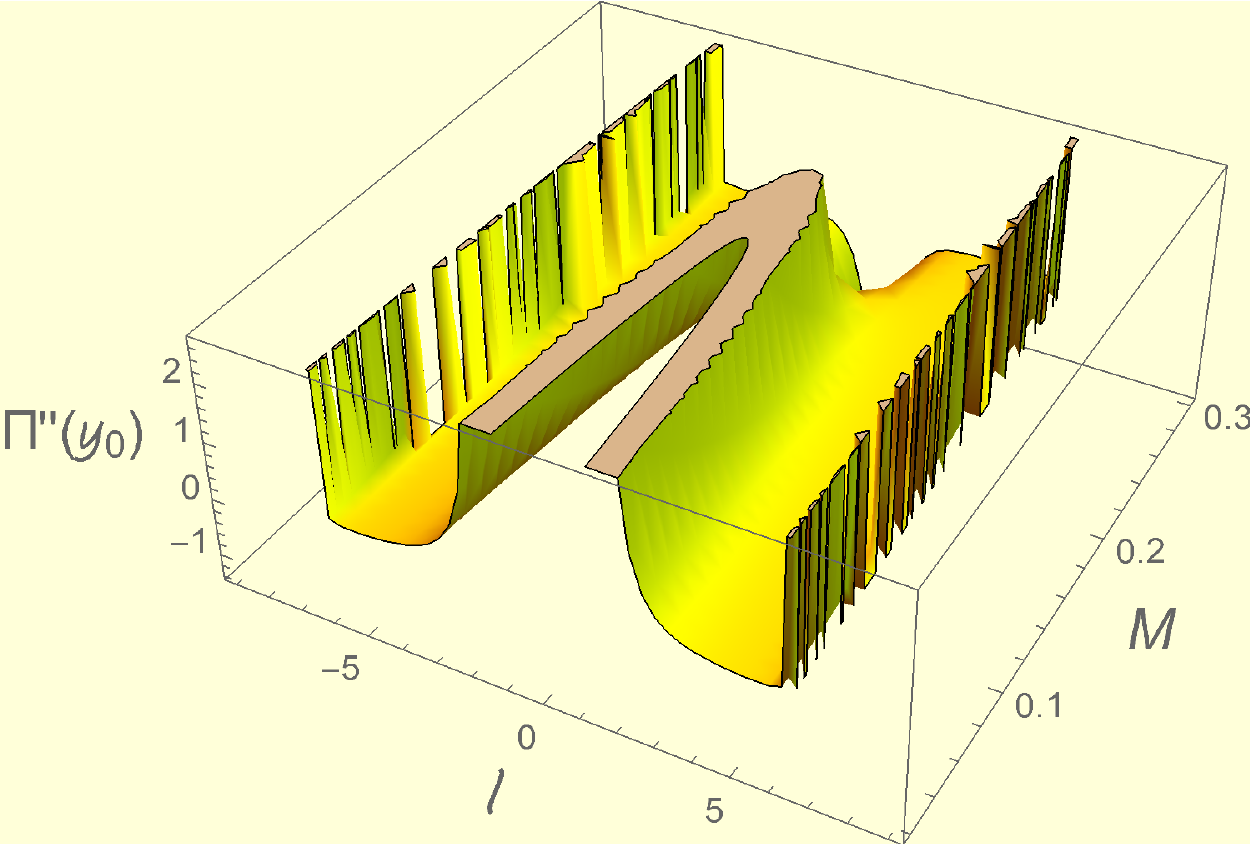} \includegraphics[width=7.6cm,height=6.0cm]{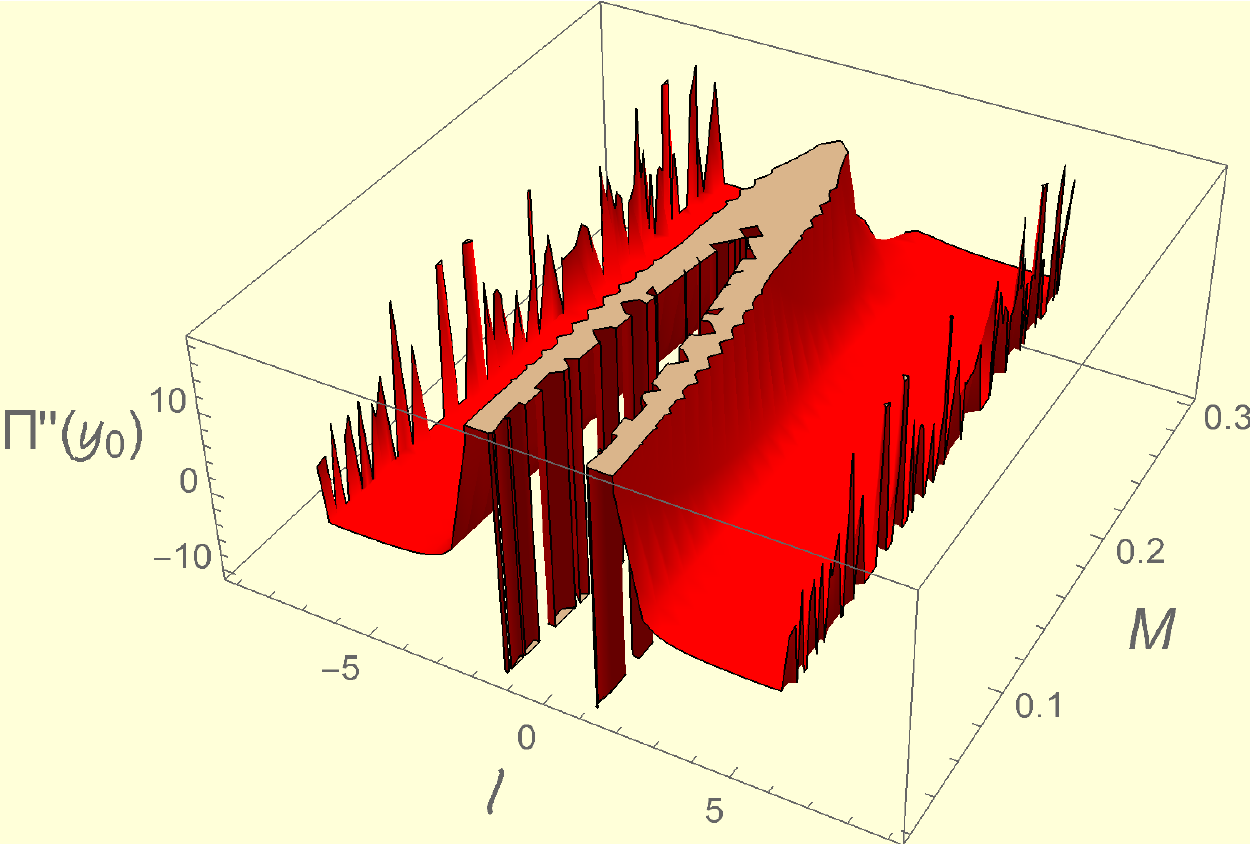}
\caption{Plots of $\Pi''(y)$ of thin shell around WH geometry for $b_1(y)$ (left plots) and $b_2(y)$ (right plots) for $q = 0.125$ with different values of $a$, i.e., $a=0$ (upper panel) and $a=0.2$ (lower panel).}\label{F6c}
\end{figure}

\begin{figure}[htb!]
\centering 
\includegraphics[width=7.5cm,height=6.0cm]{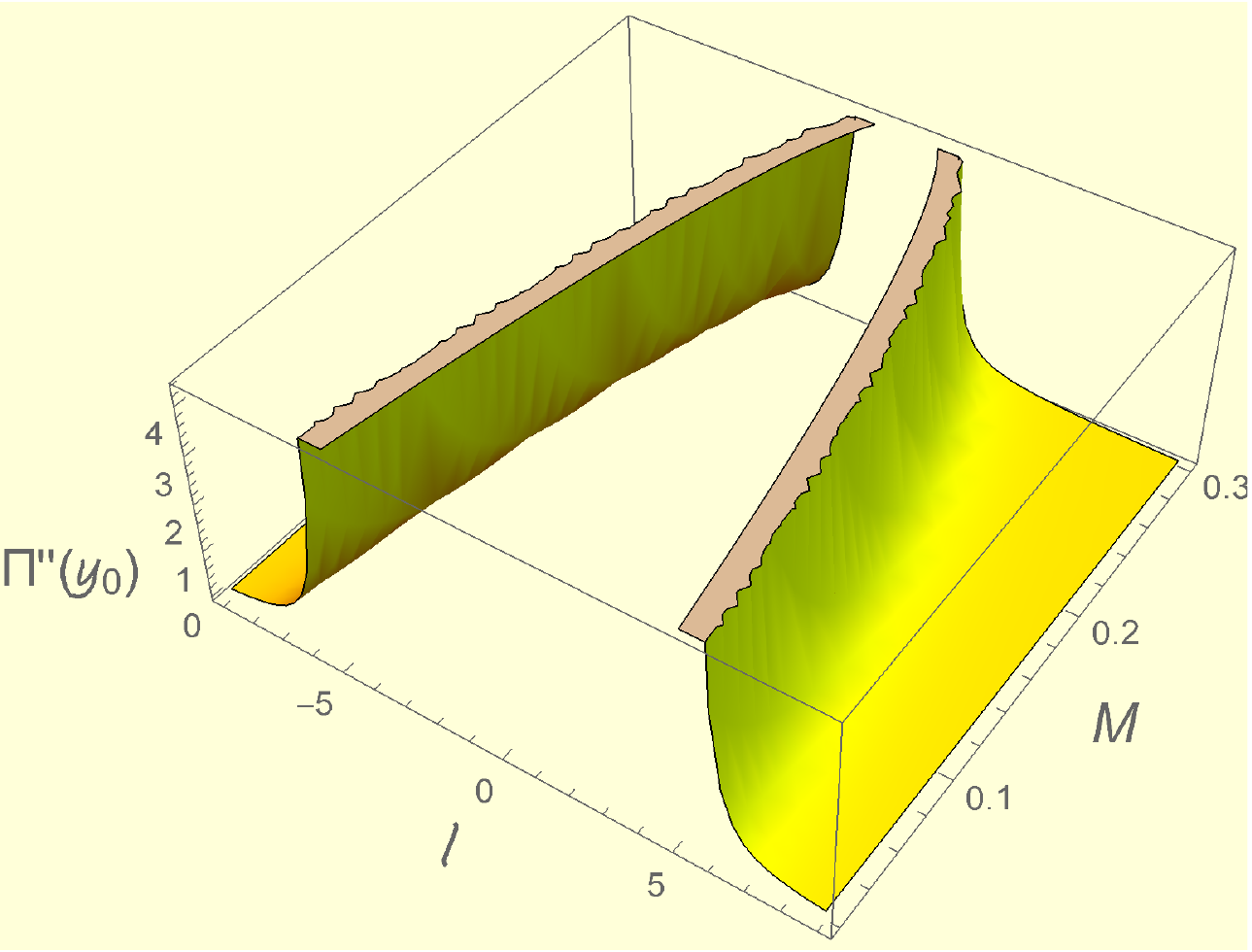}~~ \includegraphics[width=7.5cm,height=6.0cm]{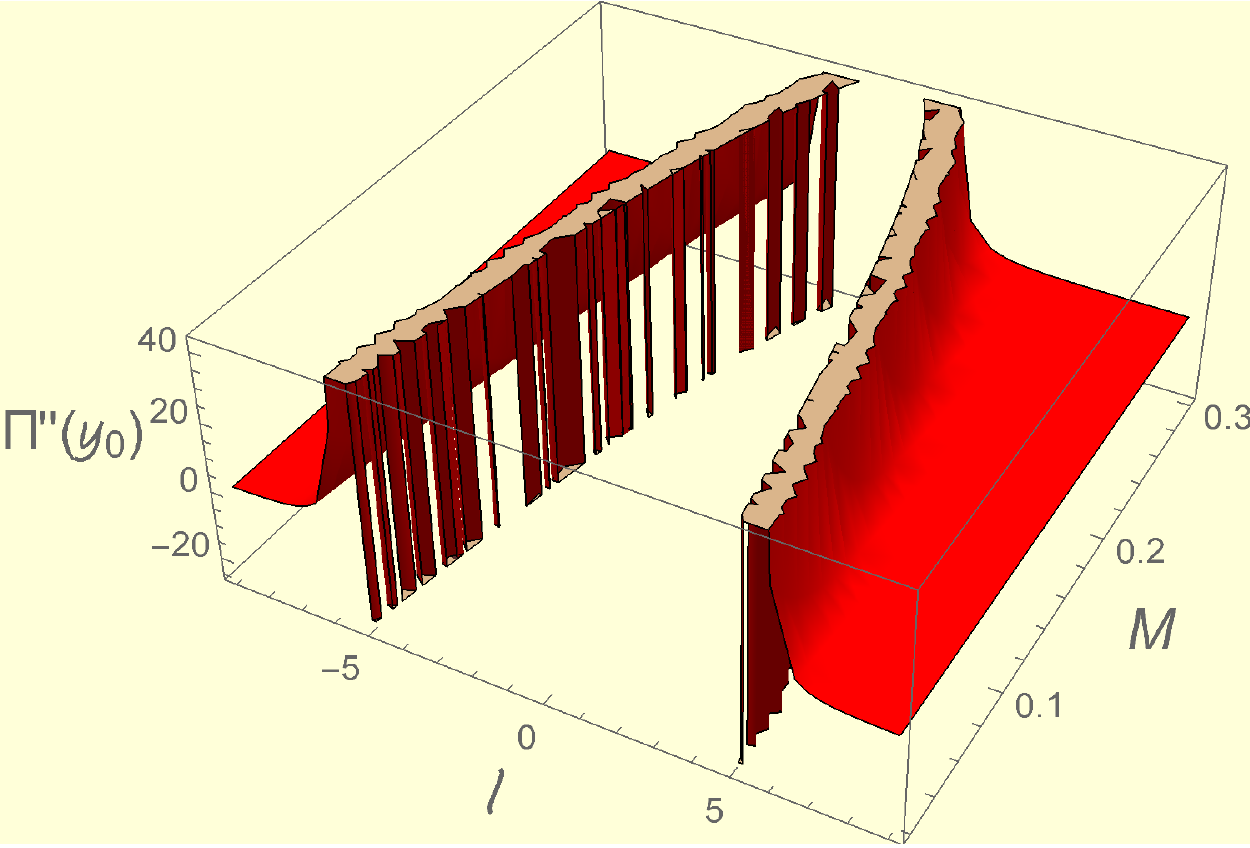}
\includegraphics[width=7.6cm,height=6.0cm]{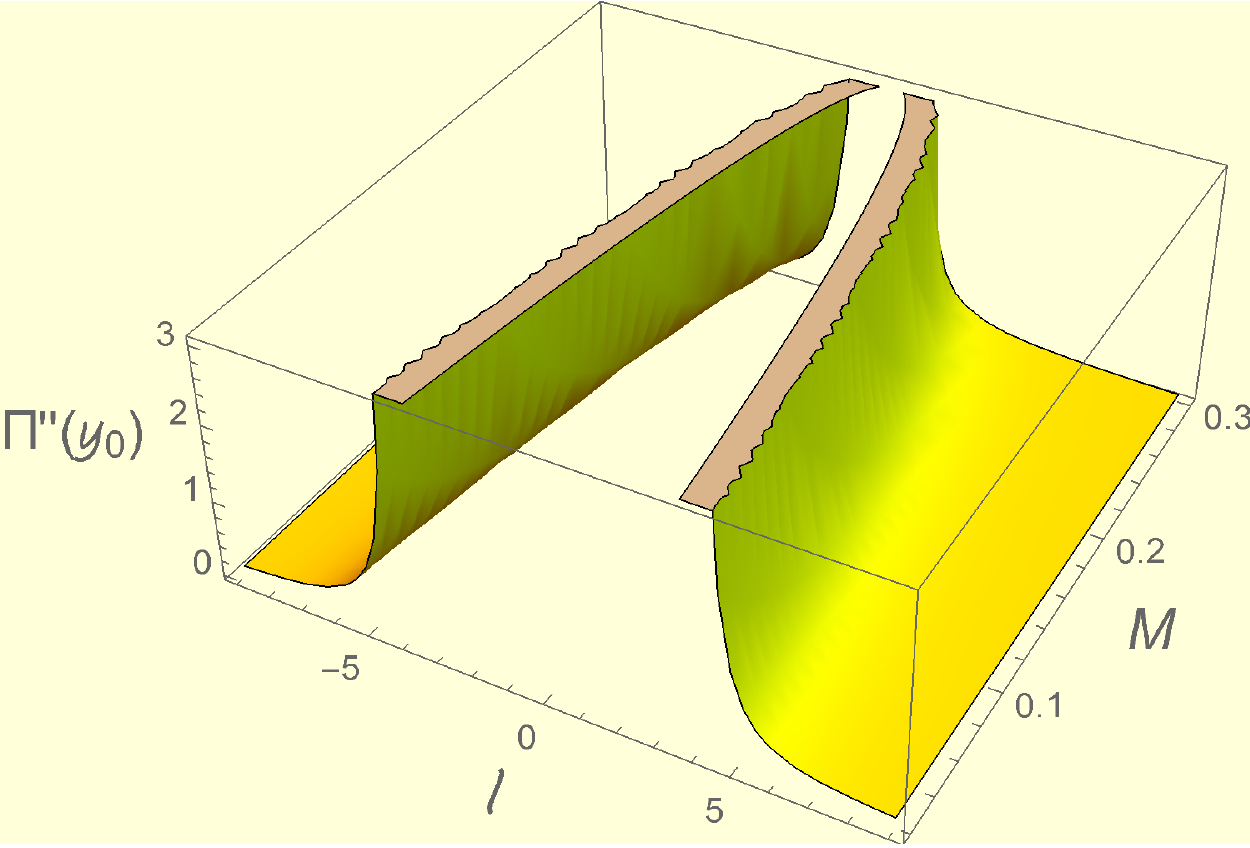} \includegraphics[width=7.6cm,height=6.0cm]{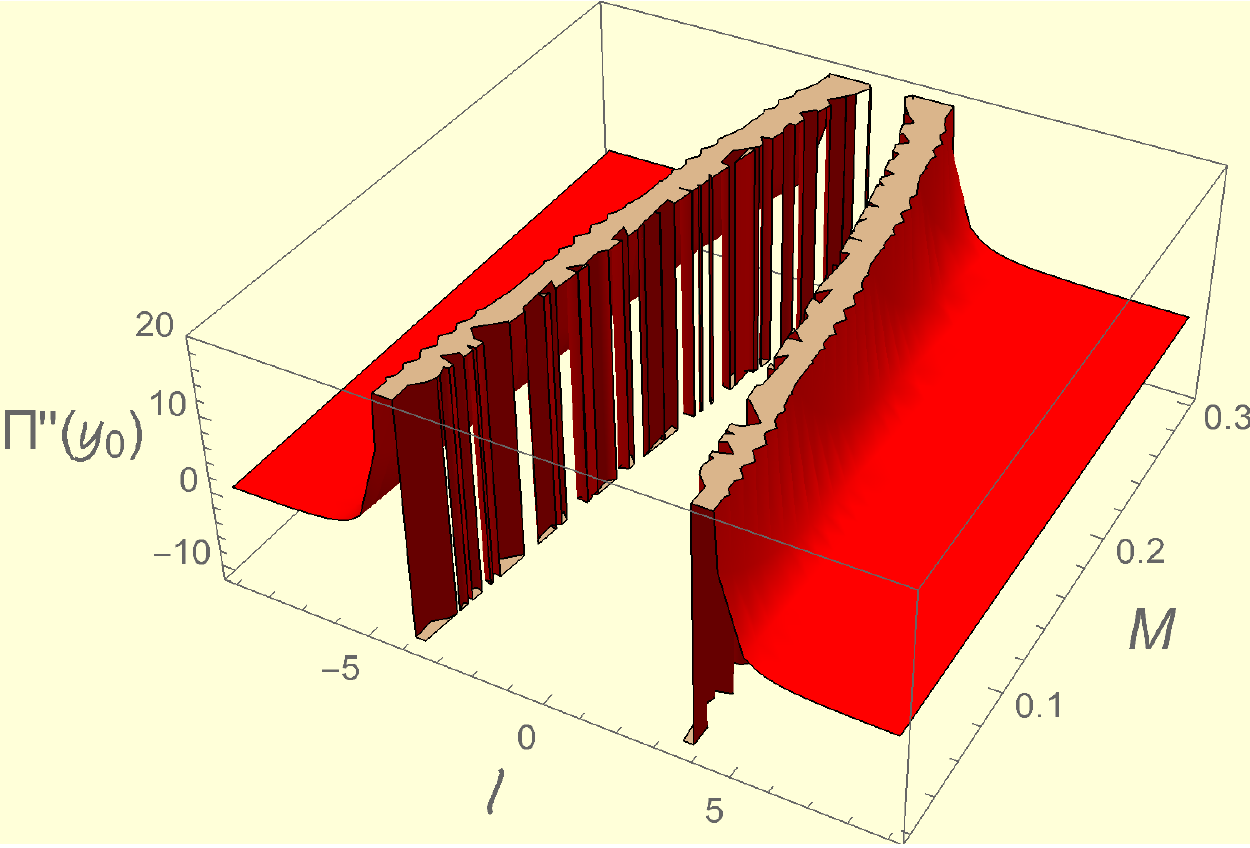}
\caption{Plots of $\Pi''(y)$ of thin shell around WH geometry for $b_1(y)$ (left plots) and $b_2(y)$ (right plots) with different values of $q$, i.e., $q=0.02$ (upper panel) and $q=0.04$ (lower panel)..}\label{F6d}
\end{figure}

It is worthwhile that the components of the stress-energy tensor follow the energy conservation constraints as
\begin{equation}\label{15aa}
\mathfrak{P}(y) \frac{d}{d\tau}(4\pi y^2)+\frac{d}{d\tau}(4\pi y^2\mathfrak{S}(y))=0,
\end{equation}
which turns out to be
\begin{equation}\label{16aa}
\mathfrak{S}'(y)=-\frac{2(\mathfrak{S}(y)+\mathfrak{P}(\mathfrak{S}(y)))}{y}.
\end{equation}

For stable configuration, we employ the Taylor series to expand the effective potential up to second-order terms around the equilibrium shell radius as:
\begin{equation}\nonumber
\Pi(y)=\Pi(y_{0})+(y-y_{0})\Pi'(y_{0})+\frac{1}{2}
(y-y_{0})^2\Pi''(y_{0})+O[(y-y_{0})^3].
\end{equation}

The developed structure become stable if $\Pi(y_0)=0=\Pi'(y_0)$. Hence, we get
\begin{equation}\label{17aa}
\Pi(y)=\frac{1}{2}(y-y_{0})^2\Pi''(y_{0}).
\end{equation}

To demonstrate thin-shell stability, the second derivative of the potential function at $y=y_0$ can be applied as follows:
\begin{itemize}
\item For stable $\Rightarrow$\quad $\Pi''(y_{0})>0$,
\item For unstable $\Rightarrow$\quad $\Pi''(y_{0})<0$,
\item Unpredictable $\Rightarrow$\quad  $\Pi''(y_{0})=0$.
\end{itemize}

The features of thin-shell surrounding WH geometry are investigated using the second derivative of the potential function for both shape functions given as:
\begin{eqnarray}\nonumber
b_1(y)&=&-\frac{y^5}{y^4+r_0^4 (r_0-n)}+y+n,\\\nonumber
b_2(y)&=&y-y^{3-2 n} \left(y^n-r_0^n\right)^{2-\frac{2}{n}}.
\end{eqnarray}

In the following, we explore the effects of two types of matter contents located at thin-shell which ensure to satisfy the Van der Walls EOS as well as phantom-like EOS.

\begin{table}[ht]\centering
\caption{\label{tab3}{Stable configuration of thin-shell with van
der Waals EOS and by using both shape functions. Here, shortcut keys
are used as SF (shape function)}}
\begin{tabular}{|c| c| c| c |c| c|c|}
\hline
\multicolumn {1}{ |c| }{ BH geometries } & SF& $q,a$&$\Pi''(y)$ &Stability&Fig \\
\hline  BH surrounded  & $b_1(y)$&$q =
0.125,a=0$&$0<\Pi''(y)<2\times 10^3$&Stable&\ref{F6a}\\\cline {2 -6}
by quintessence & $b_2(y)$&$q = 0.125,a=0$&$0<\Pi''(y)<2\times
10^5$&Stable&\ref{F6a}
\\\hline  BH surrounded by  &
$b_1(y)$&$q = 0.125, a=0.2$&$0<\Pi''(y)<3\times
10^3$&Stable&\ref{F6a}\\\cline {2 -6} cloud and quintessence &
$b_2(y)$&$q = 0.125, a=0.2$&$0<\Pi''(y)<8\times
10^5$&Stable&\ref{F6a}\\\cline {2 -6} ($a\neq0, q\neq0$) &
$b_1(y)$&$ q = 0.02= a$&$0<\Pi''(y)<6\times
10^2$&Stable&\ref{F6b}\\\cline {2 -6}  & $b_2(y)$&$ q = 0.02=
a$&$0<\Pi''(y)<4\times 10^3$&Stable&\ref{F6b}\\\cline {2 -6}  &
$b_1(y)$&$ q = 0.04, a=0.02$&$0<\Pi''(y)<1\times
10^4$&Stable&\ref{F6b}\\\cline {2 -6}  & $b_2(y)$&$ q = 0.04,
a=0.02$&$0<\Pi''(y)<2\times 10^6$&Stable&\ref{F6b}
\\\hline
\end{tabular}
\end{table}

\begin{table}[ht]\centering
\caption{\label{tab4}{Stable configuration of thin-shell with
polytropic EOS and by considering both shape functions. Here,
shortcut keys are used as SF (shape function)}}
\begin{tabular}{|c| c| c| c |c| c|c|}
\hline
\multicolumn {1}{ |c| }{ BH geometries } & SF& $q,a$&$\Pi''(y)$ &Stability&Fig \\
\hline  BH surrounded  & $b_1(y)$&$q =
0.125,a=0$&$0<\Pi''(y)<3$&Stable&\ref{F6c}\\\cline {2 -6} by
quintessence & $b_2(y)$&$q =
0.125,a=0$&$-10<\Pi''(y)<10$&Unstable&\ref{F6c}
\\\hline  BH surrounded by  &
$b_1(y)$&$q = 0.125,
a=0.2$&$-2<\Pi''(y)<2$&Unstable&\ref{F6c}\\\cline {2 -6} cloud and
quintessence & $b_2(y)$&$q = 0.125,
a=0.2$&$-10<\Pi''(y)<10$&Unstable&\ref{F6c}\\\cline {2 -6} ($a\neq0,
q\neq0$) & $b_1(y)$&$ q = 0.02=
a$&$0<\Pi''(y)<4$&Stable&\ref{F6d}\\\cline {2 -6}  & $b_2(y)$&$ q =
0.02= a$&$-20<\Pi''(y)<40$&Unstable&\ref{F6d}\\\cline {2 -6} &
$b_1(y)$&$ q = 0.04, a=0.02$&$0<\Pi''(y)<3$&Stable&\ref{F6d}\\\cline
{2 -6}  & $b_2(y)$&$ q = 0.04,
a=0.02$&$-10<\Pi''(y)<20$&Unstable&\ref{F6d}
\\\hline
\end{tabular}
\end{table}

\subsection{van der Waals EOS}\label{subsec7.1}

First, we consider the developed structure filled with matter distribution which follows the van der Waals EOS. Mathematically, it can be defined in terms of shell radius as
\begin{eqnarray}\label{21aa}
\mathfrak{P}(\mathfrak{S}(y))=\alpha (\mathfrak{S}(y))^2+\frac{\beta \mathfrak{S}(y)}{1+\gamma \mathfrak{S}(y)},
\end{eqnarray}
and we have
\begin{eqnarray}\label{22aa}
\frac{d\mathfrak{P}(\mathfrak{S}(y))}{d\mathfrak{S}(y)}=2 \alpha \mathfrak{S}(y) +\frac{\beta}{\gamma \mathfrak{S}(y) +1}-\frac{\beta \gamma \mathfrak{S}(y) }{(\gamma \mathfrak{S}(y) +1)^2}.
\end{eqnarray}

By using the final expression of the second derivative of the potential function for van der Waals EOS. To analyze the stable configuration of the thin shell, we plot the $\Pi''(y)$ for suitable values of the physical parameter for both shape functions at equilibrium shell radius $y_0$. For simplicity, we choose
$y_0=\left(\ell^m+1\right)^{1/m}$.

Figs. \ref{F6a} and \ref{F6b} show the effects of cloud parameter $a$ and quintessence field on the stable configuration of the developed structure with van der Waals EOS for both choices of shape functions. Here, the left plots show the stability of the developed thin shell in the background of the first shape function ($b_1(y)$) and the right plots represent the stable structure ($\Pi''(y)>0$) with second shape function ($b_2(y)$). It is very interesting to mention that for such type of matter contents, thin-shell shows a stable structure for both choices of shape functions for suitable values of physical parameters as $\omega=-\frac{2}{3},q\in(0,0.25),\beta =2,\alpha =0.3,\gamma 
=10,n=\frac{2}{3},r_0=0.002$. {It is noted that the stable/unstable regions are formed in the desired range which follows the basic condition of the thin shell as $y_0>r_h$ (vide Fig. \ref{Fr6a}). Hence, thin shell regions follow the condition $\ell\leq-3.029$ or  $\ell\geq3.029$. The developed structure shows symmetric behavior for the values of the length parameter. The more stable configuration is found for higher values of cloud parameter and normalization factor (Figs. \ref{F6a} and \ref{F6b}). It is noted that the stability of the developed structure for the case of $b_2(y)$ is much greater than $(b_1(y))$. The complete overview of the stability of thin-shell around wormhole geometry with van der Waals EOS and both shape functions is given in Table \ref{tab3}}.

\subsection{Polytropic EOS}\label{subsec7.2}

Here, we choose the polytropic type matter distribution to discuss the stable configuration of thin-shell around WH geometry through the second derivative of the effective potential. Similarly, we can write as
\begin{eqnarray}\label{23aa}
\mathfrak{P}(\mathfrak{S}(y))=\omega (\mathfrak{S}(y))^\chi, \quad \chi=1+\frac{1}{n},\quad n\neq0,
\end{eqnarray}
and we have
\begin{eqnarray}\label{24aa}
\frac{d\mathfrak{P}(\mathfrak{S}(y))}{d\mathfrak{S}(y)}=\omega \chi (\mathfrak{S}(y))^{\chi-1}.
\end{eqnarray}

As mentioned in the above case, also one can observe the stability of thin-shell at the equilibrium of shell radius choosing $y_0=\left(l^m+1\right)^{1/m}$ with suitable physical parameters as $\omega=-\frac{2}{3}, \chi =\frac{5}{2}, n=\frac{2}{3}, m= 2,q\in(0,0.25)$. For such matter contents, we find some interesting behavior of the developed structure for different values of cloud parameter and quintessence field as shown in Figs. \ref{F6c} and \ref{F6d}. Similarly, the stable/unstable regions are formed in the desired range as $\ell\leq-3.029$ or  $\ell\geq3.029$. In the absence of a cloud parameter, it is noted that the developed structure shows stable behavior for the choice of $b_1(y)$ (first plot in the upper panel of Fig. \ref{F6c}) while the unstable structure is observed for $b_2(y)$ (second plot in the upper panel of Fig. \ref{F6c}) with $q=0.125$. Furthermore, for non-zero values of cloud parameters, both the cases show the possibility of unstable configuration ($\Pi''(y)<0$) as shown in the lower panel of Fig. \ref{F6c}. For smaller values of cloud and normalization factor, thin-shell shows completely stable behavior for the first shape function while the unstable configuration for the second shape function. For such matter distribution, the first shape function is more suitable as compared to the second one for smaller values of the cloud and normalization factor (Figs. \ref{F6c} and \ref{F6d}). The stable and unstable configurations of thin-shell with Polytropic EOS for both the shape functions is given in Table \ref{tab4}.

\section{Conclusion} \label{sec8}

In this paper, firstly we have  presented the EFE within the context of a modified matter source, thus finding out the solutions for the WH. We have used the van der Waals and polytropic EOS to the EFE in order to find characteristic solutions set. Then the entire analysis has been done under the embedding class for the system to obtain WH solutions. Secondly, we constructed thin shell around the obtained WH solutions by using the Schwarzschild BH surrounded by the cloud and quintessence-type fluid distribution through cut and paste approach.  Also, we have examined the stability of thin shell with the linearized radial perturbation method for the same van der Waals and polytropic EOS. Several interesting features have been exhibited via the graphical plots some of which are in demand to discuss below:\\ 

(i) Energy Conditions 

From Fig. (\ref{F1}) and Fig. (\ref{F14}), the behavior of energy density for both embedded shape functions in the framework of two well-known EOS like the van der Waals and Polytropic EOS under the effect of a cloud of string parameter and quintessence field can be observed where the energy density remains positive throughout the wormhole configurations against the radial distance ranging $-2\leq l\leq2$. 

NEC, i.e., $\rho^{eff}+p^{eff}_{r}$ and $\rho^{eff}+p^{eff}_{t}$ is provided in Fig. \ref{F2} and \ref{F4} respectively for van der Waals EOS and in Fig. \ref{F9} and Fig. \ref{F11} respectively for Polytropic EOS. For Model-I it is noticed strongly violated due to its negative behavior for $-2\leq l\leq2$, while for Model-II, it is seen satisfied with positive behavior expect one value of $\zeta=0.6$ near the wormhole throat $S_{0}$ for Polytropic EOS. 

It is noted from Fig. (\ref{F3}) for van der Waals EOS in Fig. (\ref{F10}) for Polytropic EOS that expression $\rho^{eff}-p^{eff}_{r}$ remains negative with decreasing behavior in $-2\leq l\leq-1$ and $1\leq l\leq2$, while it is seen positive with the increasing behavior around the wormhole throat $-1\leq l \leq S_{0}$ and $S_{0}< l \leq 1$ for Model-I. Further, for Model-II, $\rho^{eff}-p^{eff}_{r}$ remains negative in $-2\leq l\leq2$. The DEC like $\rho^{eff}-p^{eff}_{r}$, $\rho^{eff}-p^{eff}_{t}$ within $\rho^{eff}-\mid p^{eff}_{r}\mid$ and $\rho^{eff}-\mid p^{eff}_{t}\mid$ can be seen graphically in Fig. \ref{F3}, Fig. \ref{F5}, Fig. \ref{rF5}, and Fig. \ref{rrF5} respectively for van der Waals EOS and in Fig. \ref{F10}, Fig. \ref{F12}, Fig. \ref{rF12}, and Fig. \ref{rrF12} for Polytropic EOS.

WEC, i.e., $\rho^{eff}+p^{eff}_{r}$ and $\rho^{eff}+p^{eff}_{t}$ can be perceived from Fig. \ref{F2} and \ref{F4} respectively for van der Waals EOS and in Fig. \ref{F9} and Fig. \ref{F11} respectively for Polytropic EOS for both Model-I (left) and Model-II. For model-I $\rho^{eff}+p^{eff}_{t}$ is seen positive for Model-I, while for Model-II, $\rho^{eff}+p^{eff}_{t}$ is observed negative for Van der Waals EOS.  Further, polytropic EOS $\rho^{eff}+p^{eff}_{t}$ is seen as strongly negative for both Model-I and model-II, which can be confirmed from the Fig. (\ref{F11}). 

The graphical behavior of an energy condition like $\rho^{eff}-p^{eff}_{t}$ can be verified from Fig. (\ref{F5}) for van der Waals EoS in Fig. (\ref{F12}) for polytropic EOS for both Model-I (left) and Model-II. For model-I $\rho^{eff}-p^{eff}_{t}$ is seen negative for Model-I, while for the model-II, $\rho^{eff}-p^{eff}_{t}$ is seen positive for van der Waals EOS.  Further, for polytropic EOS $\rho^{eff}-p^{eff}_{t}$ is seen as almost positive for both Model-I and Model-II in some particular regions, It is also seen as negative in a very small region, which can be confirmed from the Fig. (\ref{F11}).

SEC, i.e., $\rho^{eff}+p^{eff}_{r}+2p^{eff}_{t}$ with $\rho^{eff}+p^{eff}_{r}$ and $\rho^{eff}+p^{eff}_{t}$ can be verified from Fig. (\ref{F6}) for van der Waals EOS and in Fig. (\ref{F13}) for polytropic EOS for both Model-I (left) and Model-II which is observed positive for Model-I, while for the Model-II, it is shown negative for van der Waals EOS. Regarding two other conditions, discussion is already provided. 

The behavior of $\rho^{eff}-p^{eff}_{r}-2p^{eff}_{t}$ is observed strongly negative for Model-I, while for Model-II, it is seen as positive in the background of van der Waals EOS as can be verified in Fig. (\ref{F7}). In Fig. (\ref{F14}), it is given under the effect of polytropic EOS, which is seen as positive in some regions and it is also observed as negative in some small regions of the configurations. 

The behavior of all the energy conditions is also summarized in Table (\ref{tab1}) for van der Waals EOS and in Table (\ref{tab2}) for polytropic EOS. \\

(ii) Stability analysis using Linearized Radial Perturbation

Figs. \ref{F6a} and \ref{F6b} show the effects of the cloud parameter $a$ and normalization factor on the stable configuration of the developed structure with the van der Waals EOS for both the choices of the shape functions. It is very interesting to mention that for such matter contents, thin-shell shows a stable structure for both the choices of the shape functions for suitable values of physical parameters. The more stable configuration can be found for the higher values of the cloud parameter and normalization factor (Figs. \ref{F6a} and \ref{F6b}) also discussed in Table \ref{tab3}.

For polytropic EOS, we found some interesting behavior of the developed structure for different values of the cloud parameter and quintessence field as shown in Figs. \ref{F6c} and \ref{F6d}. The first shape function is more suitable as compared to the second one for the smaller values of the cloud and normalization factor (Figs. \ref{F6c} and \ref{F6d}) also discussed in Table \ref{tab4}. In both the cases, the stable/unstable regions are formed in the desired range as $\ell\leq-3.029$ or  $\ell\geq3.029$.

Hence, it is concluded that the van der Walls EOS plays a remarkable role to stabilize the dynamical behavior of thin-shell around developed WH geometries for both the chosen shape functions as compared to polytropic type matter distribution.

\end{document}